\documentstyle[12pt]{article}

\def\thefootnote{\fnsymbol{footnote}}
\def\bea{\begin{eqnarray}}
\def\eea{\end{eqnarray}}
\def\beq{\begin{equation}}
\def\btr{{\bf Tr}}
\def\eeq{\end{equation}}
\def\W{\overline{W}}
\def\G{{\cal G}}
\def\cW{{\cal W}}
\def\cbW{{\cal \W}}
\def\cJ{{\cal J}}
\def\tL{{\tilde L}}
\def\tX{{\widetilde X}}

\def\tF{{\tilde F}}
\def\pr{{1\over -p^2}}
\def\prr{{1\over p^4}}

\def\prrrr{{1\over p^8}}
\def\notp{\not{\hspace{-.03in}p}}

\def\hcA{\hat{{\cal A}}}
\def\hN{\hat{N}}
\def\hh{\hat{h}}
\def\hL{\hat{L}}
\def\hA{\hat{A}}

\def\notcD{\not{\hspace{-.05in}{\cal D}}}
\def\nottD{\not{\hspace{-.05in}{\tilde D}}}
\def\notF{\not{\hspace{-.05in}F}}

\def\notL{\not{\hspace{-.05in}L}}
\def\notD{\not{\hspace{-.05in}D}}
\def\notA{\not{\hspace{-.05in}A}}
\def\lll{{\ln\Lambda^2\over 32\pi^2}}
\def\ibar{\bar{\imath}}

\def\ba{\bar{\alpha}}
\def\[{\left [}
\def\]{\right ]}
\def\({\left (}
\def\){\right )}
\def\lbr{\left\{}
\def\rbr{\right\}}
\def\pp{\partial}
\def\M{\bar{M}}

\def\y{\bar{y}}
\def\z{\bar{z}}

\def\R{{\cal{R}}}

\def\STr{{\rm STr}}
\def\Tr{{\rm Tr}}
\def\cA{{\cal A}}
\def\K{{\cal K}}
\def\f{\bar{f}}
\def\F{{\cal F}}
\def\tcF{{\tilde\F}}

\def\L{{\cal L}}
\def\D{{\cal D}}
\def\bl{\bar{\lambda}}

\def\hz{\hat{z}}
\def\hM{\widehat{M}}

\def\hF{\hat{F}}
\def\hA{\hat{A}}

\def\hD{\hat{D}}
\def\hgg{\hat{g}}
\def\tgg{\tilde{g}}
\def\hG{\hat{G}}
\def\hV{\hat{V}}
\def\bell{\bar{\ell}}
\def\n{\bar{n}}
\def\m{\bar{m}}
\def\s{\bar{s}}
\def\cM{{\cal{M}}}

\def\bc{\bar{\chi}}

\def\aa{\bar{a}}
\def\A{\bar{A}}
\def\bc{\bar{\chi}}
\def\bps{\bar{\psi}}
\def\notgg{\not{\hspace{-.04in}{\cal J}}}

\def\notG{\not{\hspace{-.05in}G}}

\def\Z{{\bar{Z}}}

\def\bj{\bar{\jmath}}

\def\sA{\backslash{\hspace{-.1in}{A}}}
\def\sj{\backslash{\hspace{-.1in}{{\cal J}}}}

\def\sB{\backslash{\hspace{-.1in}{B}}}

\def\tj{\backslash{\hspace{-.1in}\tilde{{\cal J}}}}

\def\stB{\backslash{\hspace{-.1in}\tilde{B}}}

\def\stA{\backslash{\hspace{-.1in}\tilde{A}}}
\def\tG{{\tilde G}} 
\def\tD{{\tilde D}}
\def\tC{{\tilde C}}
\def\tB{{\tilde B}}
\def\tA{{\tilde A}}

\def\tbM{{\widetilde{\M}}}
\def\tM{{\widetilde M}}
\def\fM{{\cal M}_4}
\def\eM{{\cal M}_8}

\def\theequation{\ksection.\arabic{equation}}

\catcode`\@=11

\def\thesubsection{\Alph{subsection}.}
\def\thesubsubsection{\Roman{subsubsection}.}
 
\begin{document}

\begin{titlepage}
\begin{center}

\hfill LBL-34948 \\
\hfill UCB-PTH-93/37 \\
\hfill ITP-SB-95-38 \\ 
\hfill hep-th/9606052 \\
\hfill June, 1996\\

\vskip .3in

{\large \bf SUPERGRAVITY AT ONE LOOP II: CHIRAL AND YANG-MILLS
MATTER}\footnote{This 
work was supported in part by
the Director, Office of Energy Research, Office of High Energy and Nuclear
Physics, Division of High Energy Physics of the U.S. Department of Energy under
Contract DE-AC03-76SF00098 and in part by the National Science Foundation under
grants PHY-95-14797, PHY-90-21139 and PHY-93-09888.}

Mary K. Gaillard$^{\em a}$, Vidyut Jain${\em ^b}$ {\em and} 
Kamran Saririan$^{\em a}$

{\em $^a$Physics Department and Theoretical Physics Group,
 Lawrence Berkeley Laboratory, 
 University of California, Berkeley, California 94720}\\

{\em $^b$Institute for Theoretical Physics, SUNY at Stony Brook, NY 11794}

\end{center}

\begin{abstract}

We present the full calculation of the divergent one-loop contribution to the 
effective boson Lagrangian for supergravity, including the Yang-Mills
sector and the helicity-odd operators that arise from integration over fermion
fields. The only restriction is on the Yang-Mills kinetic energy normalization 
function, which is taken diagonal in gauge indices, as in 
models obtained from superstrings.

\end{abstract}
\end{titlepage}
\newpage
\renewcommand{\thepage}{\roman{page}}
\setcounter{page}{2}
\mbox{ }

\vskip 1in

\begin{center}
{\bf Disclaimer}
\end{center}

\vskip .2in

\begin{scriptsize}
\begin{quotation}
This document was prepared as an account of work sponsored by the United
States Government. While this document is believed to contain correct 
 information, neither the United States Government nor any agency
thereof, nor The Regents of the University of California, nor any of their
employees, makes any warranty, express or implied, or assumes any legal
liability or responsibility for the accuracy, completeness, or usefulness
of any information, apparatus, product, or process disclosed, or represents
that its use would not infringe privately owned rights.  Reference herein
to any specific commercial products process, or service by its trade name,
trademark, manufacturer, or otherwise, does not necessarily constitute or
imply its endorsement, recommendation, or favoring by the United States
Government or any agency thereof, or The Regents of the University of
California.  The views and opinions of authors expressed herein do not
necessarily state or reflect those of the United States Government or any
agency thereof, or The Regents of the University of California.
\end{quotation}
\end{scriptsize}

\vskip 2in

\begin{center}
\begin{small}
{\it Lawrence Berkeley Laboratory is an equal opportunity employer.}
\end{small}
\end{center}

\newpage
\renewcommand{\thepage}{\arabic{page}}
\setcounter{page}{1}
\def\thefootnote{\arabic{footnote}}
\setcounter{footnote}{0}

\section{Introduction}\indent

Understanding the structure of the divergences in supergravity 
is a necessary step in determining the 
counterterms~\cite{dixon},~\cite{anomalies},~\cite{tom} that are needed to 
fully restore modular invariance in an effective supergravity theory from 
superstrings.  The determination of
these loop corrections may also provide a guide to the construction of an
effective theory for a composite chiral multiplet that is a bound state of
strongly coupled Yang-Mills superfields, which in turn could shed light on
gaugino condensation as a mechanism for supersymmetry breaking.

In a recent paper~\cite{us} (hereafter referred to as I), we gave the
divergent contributions to the
bosonic Lagrangian in a general supergravity theory coupled to chiral matter, in
a general bosonic background, averaged over quantum fermion helicities. That 
work extended and completed the results of several earlier 
calculations~\cite{josh}--\cite{sigma}. In particular, using specific
choices of the gauge fixing and of the expansion of the action, we were able
to cast the results in an especially simple form in which most of the one-loop
corrections can be
interpreted in terms of renormalizations. In the present paper we extend these
results to incorporate the Yang-Mills sector~\cite{noncan}, including
helicity-odd operators that arise from integration over quantum fermions. 
Our results are completely general, except that we assume
that the tree-level gauge kinetic energy normalization function $f(z)
$~\cite{cremmer}, where $z$ represents the complex scalar
fields of the theory, is proportional to the unit matrix. This is the case for 
all known theories derived from superstrings, up to possible
multiplicative constants
for different factor gauge groups that correspond to higher affine
levels~\cite{gins}. This modification is easily incorporated into our
formalism, as explained in Section 5.

The generalization of the results of I to the more general case considered here
can be summarized as follows.  We define an operator of dimension $d$ as a 
K\"ahler invariant operator whose term of lowest dimension is $d$, where 
scalar and Yang-Mills fields are assigned the canonical dimension of unity.  
Then, among the ultra-violet divergent terms generated at one
loop, all operators of dimension 6 or less (as well as many operators of 
dimension 8) that involve neither the K\"ahler curvature nor derivatives of 
the gauge kinetic function can be absorbed by field redefinitions, interpreted 
as renormalizations of the K\"ahler potential, or take the form 
$F_{ab}(z,\z)\(W^aW^b\)_F + $ h.c., where $W^a$ is a chiral Yang-Mills
supermultiplet, the subscript denotes the F-component, and the matrix-valued
function $F_{ab}(z,\z)$ is not in general holomorphic.  The remaining terms of 
dimension 8 and higher must be interpreted as arising from higher order 
spinorial derivatives of superfield operators.

As noted in I, the effective cut-off for effective theories derived from 
superstrings is field dependent~\cite{tom},~\cite{bg},~\cite{mk};
moreover the field dependence is different for loop corrections arising from 
different sectors of the theory~\cite{tom},~\cite{mk}. As in I we use here a 
single cut-off and neglect its 
derivatives; terms involving derivatives of the cut-off have a
different dependence on the moduli and must be considered together with terms
that are one-loop finite. Our results, some of which are collected
in the appendix, are presented in such a way that the contributions from
different sectors can be isolated and the corresponding Pauli-Villars
contributions can easily be evaluated. 

In Section 2 we discuss gauge fixing and the definition of the action 
expansion and in 
Section 3 we evaluate the helicity-odd fermion loop contributions.
Our result for the one-loop corrected effective action is given in Section 4,
and applied to generic models from string theory in Section 5.  We summarize
our results and discuss applications in Section 6.

In I we included appendices that define our conventions and list the 
operators that appear in the quantum action as defined by our gauge fixing
and expansion prescriptions, as well as the traces of products of
these operators that determine the divergent terms in the effective one loop
action. Appendix C of this paper extends that compilation to include
operators involving the Yang-Mills background field and new operators arising
from integration over Yang-Mills quantum fields. Additional conventions and
techniques used in
the evaluation of helicity-odd fermion traces are included in Appendix A.
In Appendix B we specify our Yang-Mills sign conventions and list relations 
among the covariant scalar derivatives of the 
K\"ahler potential $K$, the superpotential $W$ and the gauge field normalization
function $f$ that follow from gauge invariance of these functions and that are
useful in evaluating traces.  Corrections to I are included in footnotes to the
text.

\section{Gauge Fixing and the Expansion of the Action}
\setcounter{equation}{0}\indent

Our gauge fixing procedure is described in I.  Here we generalize the
formalism of I to the case $x\ne$ constant, where $x =$ Re$f(z)$ is the 
inverse squared gauge coupling.  In the general supergravity 
Lagrangian~\cite{cremmer}, the function $f_{ab}(z)$, where $a,b$ are gauge 
indices, that determines the inverse squared gauge coupling constant, 
is matrix-valued.  Throughout this paper we set 
$$ f_{ab}(z) = \delta_{ab}f(z) \equiv \delta_{ab}\(x + iy\).$$
The Yang-Mills gauge fixing prescription is modified when $x\ne$ 
constant, and, since we are now including background as
well as quantum Yang-Mills fields, gauge-graviton ghost mixing must be included.
We discuss only gauge fixing of 
the bosonic sector in this section.  The fermion sector gauge 
fixing is unchanged\footnote{There are some sign errors in the fermionic part 
of the Lagrangian and gauge fixing terms given in I that are corrected in
Appendix C of this paper; they do not affect the results of I.} from that 
defined in I, and is summarized in Appendix C.2. Our gauge sign conventions 
are those of~\cite{cremmer} and are defined in Appendix B.

The gauge-fixed Lagrangian is defined by\footnote{There is a factor 2 missing in
the last term in (2.6) of I.}
\beq\L\to \L + \L_{gf}, \;\;\;\; \L_{gf} = -{\sqrt{g}\over 2}C_AZ^{AB}C_B, 
\;\;\;\;
Z = \pmatrix{\delta^{ab}&0\cr0& -g^{\mu\nu}\cr}, \;\;\;\; C = \pmatrix{C_a\cr
C_\mu\cr}, $$
$$ C^a = \D''^\mu\hcA^a_\mu + {i\over\sqrt{x}}K_{i\m}\[(T^a\z)^{\m}\hz^i
- (T^az)^i\hz^{\m}\],$$
$$ \sqrt{2}C_\mu = \(\nabla^\nu h_{\mu\nu}
- {1\over 2}\nabla_\mu h^\nu_\nu - 2\D_\mu z^IZ_{IJ}\hz^J + 
2\F^a_{\mu\nu}\hcA_a^\nu \),
\end{equation}
where hatted variables refer to quantum fields and unhatted ones refer to
background fields, $h_{\mu\nu}$ is the quantum part of the space-time metric
whose classical part is $g_{\mu\nu}$, and $K_{i\m}$ is the K\"ahler metric, 
which here is a function of the background fields. 
Following~\cite{noncan} we have introduced canonically normalized Yang-Mills 
fields:
\beq \cA_\mu = \sqrt{x}A_\mu, \;\;\;\; \hcA_\mu = \sqrt{x}\hA_\mu, \;\;\;\;
\F_{\mu\nu} = \sqrt{x}F_{\mu\nu}, \;\;\;\; 
\sqrt{x}\D_\mu A_\nu = \D'_\mu\cA_\mu,\eeq
and we have adopted the shorthand notation
\beq \D'_\mu = \D_\mu - {\pp_\mu x\over 2x},\;\;\;\;
\D''_\mu = \D_\mu + {\pp_\mu x\over 2x},\eeq
where $\D_\mu$ is the gauge and general coordinate invariant derivative.
Under a gauge transformation with parameter $\beta = T_a\beta^a$ and fixed
background fields we have, neglecting terms of order $\hz,\hA$:
\beq\delta\hz^i = -i(\beta z)^i,\;\;\;\;\delta\hz^{\m} = +i(\beta\z)^{\m}, \;
\;\;\; \delta\hcA^a_\mu = \sqrt{x}\D_\mu\beta^a.\eeq
If we implement the gauge fixing condition in the usual way, the ghost
determinant contains a factor Det$^{1\over2}x$ that translates into a 
quartically divergent term proportional to $\Tr\ln x$ in the effective action.
Note however that we have rescaled the quantum Yang-Mills fields~\cite{noncan} 
[see (2.2) above] and the quantum gaugino fields~\cite{josh} (see Appendix C.2
below) in order to canonically normalize their kinetic energy.  If we rescale 
the gauge parameter in the same way as the Yang-Mills
supermultiplet, and take, instead of $\beta$, the gauge parameter
$$\gamma = \sqrt{x}\beta, \;\;\;\; \sqrt{x}\D_\mu\beta = \D'_\mu\gamma, $$
we get
\begin{equation}
\delta\hcA_\mu  = \D'_\mu\gamma, \;\;\;\;\delta\hz^i = -{i\over\sqrt{x}}(\gamma
z)^i, \;\;\;\; \delta \hz^{\m} = +{i\over\sqrt{x}}(\gamma\z)^{\m},
\end{equation}
and no $\Tr\ln x$ term is generated in the ghost determinant.  We therefore
adopt the prescription (2.5).

Under a general coordinate transformation $x\to x' = x + \epsilon$, we have
$$ \delta \hz^i = \epsilon^\mu\pp_\mu z^i, \;\;\;\;
\delta \hcA_\nu = \sqrt{x}\(\epsilon^\sigma \nabla_\sigma A_\nu + A_\sigma
\nabla_\nu\epsilon^\sigma\),$$ 
which is general coordinate, but not gauge, covariant.  To obtain a manifestly 
gauge covariant result,
we add a compensating gauge transformation with parameter
$\gamma^a(\epsilon^\mu) = - \epsilon^\mu\cA^a_\mu$, giving
\beq \delta\hz^i = \epsilon^\mu\D_\mu z^i , \;\;\;\;
\delta \hcA_\nu = \epsilon^\sigma\F_{\sigma\nu}.\eeq

Then, relabelling the gauge parameter as $\epsilon_a\equiv\gamma_a$, the ghost 
determinant $M$ is obtained in the usual way as
\beq M^A_B = {\pp\over\pp\epsilon_A}\delta C_B, \eeq
where the variation $\delta C$ is determined from
\beq
\delta \hz^i = -{i\over\sqrt{x}}(T_b z)^i\epsilon^b + \epsilon^\mu\D_\mu z^i, 
\;\;\;\; \delta {\hz}^{\m} = 
{i\over\sqrt{x}}(T_b\z)^{\m}\epsilon^b + \epsilon^\mu\D_\mu\z^{\m}, $$ 
$$ \delta\hcA^a_\mu = \D'_\mu\epsilon^a + \epsilon^\sigma\F^a_{\sigma\mu},
 \;\;\;\; \delta h_{\mu\nu} = \nabla_\nu\epsilon_\mu + \nabla_\mu\epsilon_\nu
.\eeq
This gives a contribution to the gauge-fixed Lagrangian:
\bea g^{-{1\over2}} \L_{gh} &=& \bar{c}^B M^A_Bc_A^0 
\equiv \bar{c}Z\(\hD^2 + H_{gh}\)c \nonumber \\
&=& {\bar c}^b\[(\D''_\mu\D'^\mu)^a_b + q^a_Iq_b^I\]c_a - {\bar c}^\nu 
\sqrt{2}\[\D''^\mu\F^a_{\nu\mu} + q^a_I\(\D_\nu z^I\)\]c_a \nonumber \\
& & - {\bar c}^\mu\[\nabla^2g_{\mu\nu}  -r_{\mu\nu}- 2\(\D_\mu z^I\)Z_{IJ}
\(\D_\nu z^J\) + 2\F^a_{\mu\rho}\F_{a\nu}^{\;\;\;\;\rho}\]c^\nu \nonumber \\
& & - {\bar c}^a\sqrt{2}\[\(\D_\mu z^I\)q_{aI} - \F_{a\mu\nu}\D'^\nu\]c^\mu,
\;\;\;\; c_0^a = c^a, \;\;\;\;c_0^\mu = -\sqrt{2}c^\mu, \nonumber \\ 
q^a_i &=& {i\over \sqrt{x}}(T^a\z)^{\m}K_{i\m}, \;\;\;\;
q_a^i = -{i\over \sqrt{x}}(T_az)^i.\eea
The rescaling of the graviton ghost in order to canonically normalize the
ghost kinetic energy yields a factor Det$^{-{1\over 2}}2$ in the functional
integration that cancels a factor Det$^{1\over 2}2$ from the gravitino
auxiliary field~\cite{josh},~\cite{us}.  The matrix elements of $H_{gh}$ and 
of the covariant derivative $\hD$ are given in (2.11), (C.29) and (C.30).

Finally, as discussed in I, we modify the graviton propagator by
adding terms that are proportional to $\L_A = \pp\L/\pp\phi^A$, where $\phi^A$
is any field.  This modification, which is equivalent to a nonlinear
redefinition of the quantum variables,  does not change the S-matrix and can 
lead to simplifications as well as enhancing manifest covariance under the 
symmetries of the theory~\cite{gauge}.  We define the graviton propagator 
by\footnote{(2.21) of I should read: $\Delta^{-1}_{\mu\nu,\rho\sigma} \to
\Delta^{-1}_{\mu\nu,\rho\sigma} - 2P_{\mu\nu,\rho\sigma}\L^\lambda_\lambda
-{1\over 2}\[g_{\mu\nu}\L_{\rho\sigma} + g_{\rho\sigma}\L_{\mu\nu}\] 
+ {1\over 2}\[g_{\mu\rho}\L_{\nu\sigma} + g_{\nu\rho}\L_{\mu\sigma} + 
g_{\mu\sigma}\L_{\nu\rho} + g_{\nu\sigma}\L_{\mu\rho}\].$}
(2.20) and (2.21) of I, and by 
\begin{equation}
\Delta^{-1}_{\mu\nu,a\rho} = \L_{\mu\nu,a\rho}-{1\over 2}g_{\mu\nu}\L_{a\rho}
+ {1\over 2}g_{\mu\rho}\L_{a\nu} + {1\over 2}g_{\nu\rho}\L_{a\mu}
= \L_{\mu\nu,a\rho}+4P_{\mu\nu,\rho\sigma}\L_a^\sigma,$$
$$ \L_{\mu\nu,a\rho} = g_{\mu\mu'}g_{\nu\nu'}g_{\rho\rho'}
{\pp^2\over \pp g_{\mu'\nu'}\pp A^a_{\rho'}}\L, \;\;\;\; \L_a^\sigma =
g^{\sigma\rho}\L_{a\rho} = {\pp\over\pp A^a_\sigma}\L.\end{equation}
It should be emphasized that the propagator modifications that we use have been
chosen purely for convenience; they considerably simplify the matrix elements
that are listed in Appendix C.1, and are not necessarily derivable from a
generalized metric~\cite{gauge}.  A natural choice\footnote{This choice for
$G^{\mu\nu,\rho\sigma}$ coincides with that of Fradkin and Tseytlin~\cite{gauge}
for the case of supergravity with their parameter $t=1$, which corresponds to 
$\lambda = -1/2$ in their pure gravity case.} for this metric would be $G_{AB} 
= \sqrt{g}\(Z_\Phi\)_{AB}$, where $A,B$ run over all bose degrees of freedom 
and the metric $Z_\Phi$ is defined in (2.11) below.  Then defining
$ \Delta^{-1}_{AB} = \L_{AB} - \Gamma_{AB}^C\L_C$, where $\Gamma_{AB}^C$ is the
Christoffel connection derived from the metric $G_{AB}$, the propagator
corrections would be precisely half the ones used here (with additional
corrections to scalar propagator $\Delta^{-1}_{IJ}$ and the vector propagator
$\Delta^{-1}_{a\rho,b\sigma}$ proportional to $\L_{\mu\nu,\rho\sigma}$).
It is possible that the use of this generalized metric would reduce the
need for field redefinitions as described in Section 4 [see  (4.11-13)],
but its use would make the intermediate calculations more cumbersome.

Once the above prescriptions have been implemented, the quadratic quantum 
Lagrangian for the bosonic sector takes the general form:
\bea \L_{\rm bose} + \L_{gh} &=& -{1\over 2}\Phi^T\[Z_\Phi\(D^2 + M^2_\Phi\)
+ \{D_\mu,X_\Phi^\mu\}\]\Phi \nonumber \\ & &
+ {1\over 2}\bar{c}\[Z_{gh}\(\D^2 + M^2_{gh}\)+ \{\D_\mu,X_{gh}^\mu\}\]c, 
\nonumber \eea
where $\Phi = (h_{\mu\nu},\hcA^a,\hz^i,\hz^{\m})$, $D_\mu$ is covariant under
scalar field redefinitions as well as gauge and general coordinate
transformations, and the $X_\mu$ connect fields of different spin; in addition,
there is a vector-vector connection~\cite{noncan} in $X_\Phi^\mu$.
Following the procedure described in~\cite{noncan}, we introduce off-diagonal
connections in both the bosonic and ghost sectors, as well as an additional
connection for the gauge fields, so as to cast the quantum
Lagrangian for the full gauge-fixed bosonic sector in the form
\bea \L_{\rm bose} + \L_{gh} &=& -{1\over 2}\Phi^T Z_\Phi\(\hD_\Phi^2 + H_\Phi\)
\Phi + {1\over 2}\bar{c}Z_{gh}\(\hD^2_{gh} + H_{gh}\)c, \nonumber \\
\hD_\mu^\Phi &=& D_\mu + V_\mu, \;\;\;\; \(V_\mu\)_{a\rho,b\sigma} = -
\delta_{ab}\epsilon_{\rho\mu\sigma\nu}{\pp^\nu y\over2x},\nonumber \\
\(ZV_\mu\)_{\alpha\beta,a\nu} &=& \(V_\mu\)_{a\nu,\alpha\beta} = 
{1\over 4}\(\F_{a\beta\mu}g_{\alpha\nu} + \F_{a\alpha\mu}g_{\beta\nu}\),  
\nonumber \\
\(V_\mu\)_{a\nu,i} &=& \(V_\mu\)_{i,a\nu} = \[\(V_\mu\)_{\ibar,a\nu}\]^*=
{1\over 4x}f_i\(\F_{a\mu\nu} - i\tcF_{a\mu\nu}\) ,\nonumber \\
\hD_\mu^{gh} &=& \D_\mu + B_\mu, \;\;\;\; \(B_\mu\)_{a\nu} = 
\(B_\mu\)_{\nu a} = -{1\over\sqrt{2}}\F_{a\nu\mu}.\eea
This introduces corresponding shifts in the background field-dependent
``squared mass'' matrices:
\beq M^2_\Phi \to H_\Phi = M^2_\Phi - V_\mu V^\mu, \;\;\;\;
M^2_{gh} \to H_{gh} = M^2_{gh} - B_\mu B^\mu.\eeq
The elements of $M_\Phi^2$ were evaluated in~\cite{noncan}; here they are 
somewhat modified by the different Yang-Mills gauge fixing and action 
expansion.  These modified matrix elements are listed in Appendix C.1 below.

As explained in Section 3 and Appendix A, we evaluate the fermion determinant
by first writing it in two-component notation, separating it into helicity-even
and -odd contributions, and then recasting these two contributions in Lorentz
covariant four-component notation.  As discussed in~\cite{mk}, this separation 
is not uniquely defined.  The choice that respects supersymmetry as well as
manifest gauge and K\"ahler covariance allows a consistent Pauli-Villars
regulation.  We follow that choice here; the corresponding matrix elements are
given in the Appendix.  The contribution from fermion loops to the effective
action is evaluated (see Appendix A) by introducing~\cite{josh} the $8\times 8$ 
matrices 
\beq D_\mu = \pmatrix{ D_\mu^+ & 0\cr 0& D_\mu^-\cr},\;\;\;\;
M_\Theta = \pmatrix{ 0 & M \cr \M & 0 \cr},\;\;\;\;\notD = \gamma^\mu D_\mu \eeq
that operate on an eight component fermion $f^T = (f_L,\; f_R = f_L^c)$.
The helicity averaged contribution of the fermion determinant is then
\begin{equation}
-{i\over 4}\Tr\ln(-i\notD + M_\Theta)_+ = 
-{i\over 8}\Tr\ln\(\notD^2 + M^2_\Theta -i[\notD,M_\Theta] \),\end{equation}
Because the fermion mass matrix and connection contain the terms $ 
\sigma^{\mu\nu}M_{\mu\nu}$ and $iL_\mu\gamma_5$, respectively, 
they do not commute with $\gamma_\mu$; thus
\bea \notD^2 &=& D^2 + {1\over4}\[\gamma^\mu,\gamma^\nu\]G_{\mu\nu} +
{1\over2}\lbr D_\nu,\gamma^\mu\[D_\mu,\gamma^\nu\]\rbr -
{1\over2} \[D_\nu,\gamma^\mu\[D_\mu,\gamma^\nu\]\], \;\;\;\;
\nonumber \\ \[\notD,M_\Theta\] &=& {1\over2}\lbr\gamma_\mu,D^\mu M_\Theta\rbr +
{1\over2}\lbr D^\mu,\[\gamma_\mu,M_\Theta\]\rbr + {1\over2}[M_\Theta,[D^\mu,
\gamma_\mu]],
\nonumber \\ D^\mu M_\Theta &\equiv& \[D^\mu,M_\Theta\].\eea
Therefore, in analogy with the  boson case discussed above, we write
\beq -{i\over 4}\Tr\ln(-i\notD + M_\Theta)_+ = -{i\over8}\Tr\ln\(\hD_\Theta^2 + 
H_\Theta\), \eeq
\bea H_\Theta &=& M^2_\Theta - {i\over2}\{\gamma^\mu,D_\mu M_\Theta\} 
+ {1\over4}\[\gamma^\mu,M_\Theta\]\[\gamma_\mu,M_\Theta\] - 
{i\over2}[M_\Theta,[D^\mu,\gamma_\mu]] 
+ {1\over 4}[\gamma^\mu,\gamma^\nu]G_{\mu\nu} \nonumber \\ & & -
{1\over4}\gamma^\mu\[D_\mu,\gamma^\nu\]\gamma^\rho\[D_\rho,\gamma_\nu\] -
{1\over2}\[D_\nu,\gamma^\mu\[D_\mu,\gamma^\nu\]\] +
{i\over4}\{\[\gamma^\mu,M_\Theta\],\gamma^\nu\[D_\nu,\gamma_\mu\]\}, 
\nonumber \\ \hD_\mu^\Theta &=& D_\mu - {i\over2}\[\gamma_\mu,M_\Theta\]
+ {1\over2}\gamma^\nu\[D_\nu,\gamma_\mu\].\eea

\section{Helicity-Odd Fermion Loop Contributions}
\setcounter{equation}{0}\indent

In this section we determine the helicity-odd operators that arise from 
integration over fermionic degrees of freedom. 
They are particularly relevant to the
evaluation of anomalies~\cite{anomalies},~\cite{tom}, in effective 
supergravity theories, which is currently of
special interest in attempts to extract physics from string theory. 
We show that these terms are finite, except in the presence of a Yang-Mills
sector with a nontrivial kinetic normalization function $f(z)$, in which case
there are logarithmically divergent contributions that are invariant under 
chiral $U(1)_R$
transformations, {\it i.e.}, under K\"ahler (or modular) transformations 
up to a possible dependence of the cut-off on the K\"ahler potential.
We also indicate how the finite contributions
to the effective action can be obtained. 

\subsection{General formalism}

The fermion loop contribution is given by 
\beq \L_1 = -{i\over2}\Tr\ln\(-i\notD + M_\Theta\) \equiv 
-{i\over2}\Tr\ln\cM.\eeq
To evaluate the determinant (3.1), we write 
\beq T = \Tr\ln\cM = T_+ + T_-,\;\;\;\; T_{\pm} = 
{1\over 2}\[\Tr\ln\cM(\gamma_5) \pm \Tr\ln\cM(-\gamma_5)\].\eeq
Only $T_+$ has been calculated previously for 
supergravity~\cite{us}--\cite{sigma}.
Here we will evaluate the additional contribution, $T_-$: 
\beq T_- = -{1\over 2}\Tr\ln\cM(-\gamma_5)\cM^{-1}(\gamma_5) = 
- {1\over 2}\Tr\ln\{1 - \cM^{-1}[\cM(\gamma_5) - \cM(-\gamma_5)]\} $$ 
$$ = {1\over 2}\Tr\sum_{n=1}^{\infty}{1\over n}\{\cM^{-1}[\cM(\gamma_5) - 
\cM(-\gamma_5)]\}^n.\eeq 
Using the techniques described in~\cite{mkg},~\cite{josh}, 
we can write the trace in (3.3) as (see Appendix A)
\beq T_- = \int d^4xT(x), \;\;\;\; T(x)= \int{d^4p\over (2\pi)^4} T(p,x), \eeq
and then expand $T(p,x)$ as
\beq
T(p,x) = \Tr\sum_{n=1}^{\infty}{2^n\over 2n}\{\sum_{\ell = 0}^{\infty}
(-\R)^{\ell}\R_5\}^n, \eeq 
where $\R,\R_5$ are defined in (A.19--20):
\bea \R &=& \pr\[p^2 - T^{\mu\nu}\Delta_\mu\Delta_\nu + \hh + X
+ \(p^\nu + G^\nu\)P_{\mu\nu}\hM^\mu\], \nonumber \\
\R_5 &=& \pr\[\(p^\nu + G^\nu\)P_{\mu\nu}\hN^\mu\].\eea
The operators appearing in (3.5) are defined in Appendix A as power series of 
the form $\sum_n c_n(O)(D\cdot\pp/\pp p)^nO,$ where $D_\mu = D_\mu^+R +
D_\mu^-L$ is the fully covariant derivative defined in (A.8) of the Appendix, 
and the operator $O$ is a function of the background bosons.
The coefficients $c_n(O)$ are constants with, in particular, $c_0(G) = 0$ in 
the expansion of $G_\mu^{\pm}$; more specifically
\beq \notG^{\pm} = \gamma^\mu G_\mu^\pm \;\;\;\; G_\mu^\pm = 
{1\over2}G^{\pm}_{\nu\mu}{\partial\over \partial p_\nu}
+ O\({\pp^2\over\pp p\pp p}\), \;\;\;\;
G^{\pm}_{\mu\nu}  = -G^{\pm}_{\nu\mu} = [D_\mu^{\pm},D_\nu^{\pm}]\eeq 
Thus we have to evaluate the following contribution to the effective 
one-loop Lagrangian:
\beq \L_1 \ni -{i\over 2}T_- = -i\int{d^4p\over 4(2\pi)^4}
\Tr\sum_{n=1}^{\infty}{2^n\over n}\{\sum_{\ell = 0}^{\infty}
(-\R)^{\ell}\R_5\}^n, \eeq  
where now the trace is over only Dirac indices and internal quantum numbers (and
Lorentz indices for the gravitino). 

To keep the integrals finite, the integration should
be performed including Pauli-Villars regulator masses
$\mu_0$: $-p^{-2}\to (-p^2 +\mu_0^2)^{-1}$ in the derivative expansion.  
However, as shown below, $T_-$, when suitably
defined, contains no quadratically divergent terms. Once the integrals are 
properly regulated--including the appropriate definitions of 
$T_{\pm}$--the coefficients of log divergent terms are independent of the 
regularization scheme. On the other hand, if one wishes to evaluate finite 
terms, one has either to expand around an infrared regulator mass $\mu_0$ 
or, alternatively, to resum the derivative expansion~\cite{oren}~\cite{rey}.
In particular, the ultra-violet finite terms include 
the standard chiral anomaly. We
explicitly evaluated this term for the vector-vector-axial vertex induced by 
Dirac fermions with a common mass $\mu_0$, and
recovered the large mass limit of the Adler-Rosenberg formula \cite{adler}; the
complete expression for this formula requires a resummation of the derivative
expansion which will be presented elsewhere~\cite{rey}.  
We emphasize that, because of the anomaly, K\"ahler
invariance is broken at the quantum level. Classically, this 
invariance permits a 
choice~\cite{cremmer} of K\"ahler gauge such that the classical 
Lagrangian is derivable from only two functions of the scalar fields, the 
(in general matrix-valued) gauge
normalization function $f_{ab}(z)$ and the generalized K\"ahler potential
$\G(z,\z) = K(z,\z) + \ln|W(z)|^2$, where $K$ and $W$ are the K\"ahler potential
and the superpotential, respectively. For the purpose of calculating the
anomaly~\cite{anomalies},~\cite{tom}, one has to undo the K\"ahler rotation
of Cremmer {\it et al.}~\cite{cremmer}, by performing a phase 
transformation~\cite{kahler} on the fermion fields.  As in I we work throughout
in this K\"ahler covariant formalism.

As was discussed in~\cite{mk}, the separation (3.2) of $T$ into helicity-odd 
and -even parts is not uniquely defined because we can interchange terms that 
are even and odd in 
$\gamma_5$ using $\gamma_5 = (i/24)\epsilon^{\mu\nu\rho\sigma}\gamma_\mu
\gamma_\nu\gamma_\rho\gamma_\sigma$ and similar identities.  In most cases the
correct choice is dictated by gauge or K\"ahler covariance.  The remaining
ambiguities are resolved by supersymmetry.
A fully SUSY-invariant result for the quadratically divergent terms requires the
introduction of Pauli-Villars regulator fields~\cite{sigma},~\cite{casimir};
there is a unique definition of the matrix elements that allows a
supersymmetric Pauli-Villars regularization~\cite{mk}.  Specifically,
this fixes the forms of the fermion mass matrix and connection matrix: 
\bea M &=& m + \(\alpha_a F^a_{\mu\nu} + i\beta_a\gamma_5\tF^a_{\mu\nu}\)
\sigma^{\mu\nu}, \;\;\;\; \tF_{\mu\nu} =
{1\over2}\epsilon_{\mu\nu\rho\sigma}F^{\rho\sigma}, \nonumber \\ 
D_\mu &=& \D_\mu + i\Gamma_\mu\gamma_5 - {1\over24}L_\mu
\epsilon^{\lambda\nu\rho\sigma}\gamma_\lambda 
\gamma_\nu\gamma_\rho\gamma_\sigma, \eea
where $\Gamma_\mu,L_\mu,m,$ and $\alpha,\beta$ are proportional to the 
unit matrix in Dirac space. $\D_\mu$, which contains the spin connection, is 
the gauge and general coordinate covariant derivative, $\Gamma_\mu$ is the 
K\"ahler connection, $F_{\mu\nu}$ is
the Yang-Mills field strength, and $L_\mu$ is an additional axial connection 
for gauginos arising from the noncanonical form of the
kinetic energy term. $T_{\pm}$ are defined by (3.2) using the explicit
$\gamma_5$-dependence in (3.9).  Then the operators appearing in the derivative
expansion of (3.6) take the form:
\bea G_{\mu\nu}^{\pm} &=& \tG_{\mu\nu}^{\pm} + i\gamma_5 L^{\pm}_{\mu\nu} 
- [L_\mu,L_\nu],
\;\;\;\; \tG_{\mu\nu}^{\pm} = [\tD_\mu^{\pm},\tD_\nu^{\pm}], \nonumber \\
\tD_\mu^{\pm} &=& \D_\mu^{\pm} \pm i\Gamma_\mu + \Gamma'_\mu, \;\;\;\; 
L^{\pm}_{\mu\nu} = \tD^{\pm}_\mu L_\nu - \tD^{\pm}_\nu L_\mu, \quad
\tD^{\pm}_\mu L_\nu \equiv [\tD^{\pm}_\mu, L_\nu], \nonumber \\
\cJ_\mu &=& {i\over2}\(\tD^+_\mu - \tD^-_\mu\) = {i\over2}\(\D^+_\mu - 
\D^-_\mu\) -\Gamma_\mu, \;\;\;\; M_I = {1\over2}\(M - \M\), \nonumber \\
M &=& m + M_\sigma = m + M_{\mu\nu}\sigma^{\mu\nu}, \;\;\;\;
\M = \m + \M_\sigma = \m + \M_{\mu\nu}\sigma^{\mu\nu},   \nonumber \\
M_{\mu\nu} &=& \alpha F_{\mu\nu} - i\beta\tF_{\mu\nu}, 
\;\;\;\; \M_{\mu\nu} = \bar{\alpha}F_{\mu\nu} + i\bar{\beta}\tF_{\mu\nu} ,\eea
where $\Gamma_\mu$ is the K\"ahler connection and $\Gamma'_\mu$ is an
off-diagonal $\lambda$-$\psi$ connection. 
We consider only the case where the gauge field normalization function
$f(z)$ is diagonal in gauge indices; then, since $\Gamma_\mu$ is diagonal,
$L_\mu$ commutes with $\cJ_\nu$, 
and we have \bea  L_{\mu\nu}^+ &=& L_{\mu\nu}^- \equiv
\hL_{\mu\nu} = L_{\mu\nu} + [\Gamma'_\mu,L_\nu] - [\Gamma'_\nu,L_\mu], \nonumber
\\ L_{\mu\nu} &=& \nabla_\mu L_\nu - \nabla_\nu L_\mu, \;\;\;\;[L_\mu,L_\nu] = 0
.\eea
Note that the spin connection in $\tD_\mu$
[see eq. (A.12) of I] drops out of the covariant derivatives $\tD_\mu M$.  This 
is because we have taken the vierbein, and therefore $\gamma_\mu$, to be
covariantly constant~\cite{van}: $[\tD_\mu,\gamma_\nu] = 0.$
The spin connection is even in $\gamma_5$ and therefore contributes to 
$\tD_\mu M$
through the commutator which vanishes [see the definitions (3.27) below].

To identify the ultraviolet divergences, we have to study the large $p$
behavior of the integrand in (3.8) and keep terms up to $O(p^{-4})$. 
{\it A priori} $\R,\R_5 \sim p^{-1}$, so the ultraviolet
divergent part of (3.7) can occur only in terms with $n\le 4,\;\ell\le 4-n.$
Aside from terms
involving $L_\mu$, by construction, the integrand is odd in $\gamma_5$, 
and we need at least four $\gamma_\mu$'s to get a nonvanishing trace:
\beq T \propto \Tr\(A_{\mu\nu\rho\sigma}\gamma^\mu\gamma^\nu\gamma^\rho
\gamma^\sigma\gamma^5\) = -4i\epsilon^{\mu\nu\rho\sigma}\Tr 
A_{\mu\nu\rho\sigma}, \eeq
so $ \Tr \R_5 = 0.$  Finally, we note that $G^{\pm}_\mu$ in (3.7)
vanishes except when sandwiched between functions of
$p$, and is of order $p^{-1}$ in power counting. Once all
$p$-differentiations have been performed, surviving terms  must have at least 
three $\gamma_\mu$'s that are not contracted with
$p^\mu$ because of antisymmetry.  After integration over $p$, the tensor
$A_{\mu\nu\rho\sigma}$ in (3.12) can be constructed only from the four-vectors
${\cal J}_\mu$ and $L_\mu$, the tensors $M_{\mu\nu},G^{\pm}_{\mu\nu}$, the
Riemann tensor, and their covariant derivatives $\tD_\mu$.  Each factor of
$G^{\pm}_{\mu\nu}$ and of $D_\mu$ reduces the apparent divergence of a given
term by one power of $p$.  Furthermore, in the covariant derivative
expansions (A.19--20) of the operators $O$ appearing in (3.5) the indices
$\mu_i\cdots\mu_n$ in $D_{\mu_i}\cdots D_{\mu_n}O$ are automatically 
symmetrized, so at most one derivative of each operator can contribute to 
$A_{\mu\nu\rho\sigma}$ in (3.12).  

\subsection{Quadratically divergent contributions}

By construction, $T_-$ is antisymmetric under $\gamma_5\to-\gamma_5$. 
Therefore we can evaluate, instead of (3.5)
\bea T_- &\to& {1\over2}\[T_-(\gamma_5) - T_-(-\gamma_5)\], \eea
where  $T_-(-\gamma_5)$ is obtained from $T_-(\gamma_5)$ by the
substitutions 
$$ (D^+,D^-,M,\M,\cJ,M_I) \to (D^-,D^+,\M,M,-\cJ,-M_I). $$
The matrices $\R,\R_5$ are defined in (A.19--20). Since $\int d^4p\Tr\R_5 = 0,$
the potentially quadratically divergent contribution to $T_-$ is 
\beq \Tr\(\R_5^2 -\R\R_5\) \to {1\over p^4}\Tr\[\(p^\mu N_\mu - p^\mu M_\mu\)
p^\nu N_\nu\], \eeq
with $N_\nu, M_\nu$ given in (A.15).  Under Lorentz invariant integration, with
$M = m + \sigma_{\mu\nu}M^{\mu\nu}$, we have 
$$ \int d^4p\notp M\notp M'(1\pm \gamma_5)\propto\int d^4p\;p^2\gamma_\mu M
\gamma^\mu M'(1\pm \gamma_5) = 4\int d^4p\;m\notp M'(1\pm \gamma_5).$$
It follows that there are no quadratically divergent contribution involving the
mass matrix. The averaging procedure (3.13) eliminates a residual
spurious quadratic divergence proportional to Tr$\cJ_\mu\cJ^\mu$.
This divergence would vanish
identically if a Pauli-Villars regularization were used with P-V masses that 
leave all classical symmetries unbroken.  However this is not in general 
possible for the classical K\"ahler symmetry.\footnote{A detailed discussion of
Pauli-Villars regularization of $T_-$ will be given elsewhere~\cite{rey}.}  
Moreover, in the Pauli-Villars regularization described in~\cite{mk}, there are
no P-V fields that can regulate quadratic divergences proportional to 
$M_{\mu\nu}M^{\mu\nu}$, so the integrals, which are ill-defined unless they are
explicitly regulated, must be defined in such a way that these divergences
do not appear.   Note that no quadratically divergent contribution to $T_-$
arises if (3.3), as defined by (A.6), is expanded without performing the
the transformation (A.16) that makes use of partial integration, which is
ill-defined if the integrals are not finite.  However this transformation 
renders many terms explicitly covariant and thereby considerably simplifies the
derivative expansion.

\subsection{Logarithmically divergent contributions}

In the remainder of this section, $T_-$ is understood as the average (3.13).
Since we encounter only logarithmic divergences, after symmetric integration 
we may make the replacements:
\bea p_\mu p_\nu f(p^2) &\to& {p^2\over 4}g_{\mu\nu}f(p^2), \nonumber \\
p_\mu p_\nu p_\rho p_\sigma f(p^2) &\to& 
{p^4\over 24}\(g_{\mu\nu}g_{\rho\sigma} + g_{\mu\rho}g_{\nu\sigma} +
g_{\mu\sigma}g_{\nu\rho}\)f(p^2).\eea
To evaluate the terms with $p$-derivatives, we write 
\bea \pr p^\mu{\pp\over\pp p_\nu}&\to& -\pr A^{\mu\nu}, \;\;\;\;
\pr p^\mu G_{\nu\mu}{\pp\over\pp p_\nu} \to 0, \;\;\;\; A^{\mu\nu} 
= g^{\mu\nu} - {2\over p^2}p^\mu p^\nu \nonumber \\
{\pp\over\pp p_\nu}\pr p^\mu  &\to& \pr A^{\mu\nu} , \;\;\;\;
p^\mu G_{\nu\mu}{\pp\over\pp p_\nu}\pr p^\rho\to\pr p^\mu 
G_{\nu\mu}g^{\rho\nu} , \eea
where the first line is obtained by partial integration over $p$, and it is
understood that operators multiplying the first (second) line on the left 
(right) are independent of $p$. Similarly
\bea {\pp^2\over\pp p_\mu\pp p_\nu}\pr p^\rho &\to& 
{2\over p^4}\(g^{\rho\mu} p^\nu + g^{\rho\nu} 
p^\mu + g^{\mu\nu}p^\rho - {4\over p^2}p^\mu p^\nu p^\rho\),\nonumber \\
{\pp^2\over\pp p_\mu\pp p_\nu}\pr\notp p^\rho p^\sigma r_{\mu\rho\sigma\nu}
&\to& \pr\(r\notp - {2\over p^2}p^\mu p^\nu\notp r_{\mu\nu} + 2
r^{\mu\nu}\gamma_\mu p_\nu\), \nonumber \\
\pr{\pp^2\over\pp p_\mu\pp p_\nu}p^\rho p^\sigma r_{\mu\rho\sigma\nu}
&\to& - {2\over p^4}p^\mu p^\nu r_{\mu\nu}, ,\eea
where the last line is obtained by partial integration.

It is easy to see that the nonvanishing terms in $T_-$ involve the connection
$L_\mu$ and/or the off-diagonal mass $M_{\mu\nu}$. In the absence of these
contributions, since $\epsilon^{\mu\nu\rho\sigma}r_{\mu\nu\rho\tau} =0,$ the 
only helicity-odd terms are:
\beq \epsilon^{\mu\nu\rho\sigma}\Tr[(D_\mu^e{\cal J}_\nu)
{\cal J}_\rho{\cal J}_\sigma], \;\;\;\; 
\epsilon^{\mu\nu\rho\sigma}\Tr[G^A_{\mu\nu}
{\cal J}_\rho{\cal J}_\sigma] ,\;\;\;\; 
\epsilon^{\mu\nu\rho\sigma}\Tr[G^V_{\mu\nu}D^e_\rho\cJ_\sigma],\eeq
where $$ D^e_\mu = {1\over 2}\(D_\mu^+ + D_\mu^-\) \equiv 
\pp_\mu + \cJ'_\mu, \;\;\;\; 
G^{A(V)}_{\mu\nu} = {1\over 2}[G^+_{\mu\nu}- (+)G^-_{\mu\nu}].$$
The first term in (3.18) can be written
$$ {1\over 3}\epsilon^{\mu\nu\rho\sigma}
\Tr\[D^e_\mu\({\cal J}_\nu{\cal J}_\rho{\cal J}_\sigma\)\] 
= {1\over 3}\epsilon^{\mu\nu\rho\sigma}
\partial_\mu(\Tr[{\cal J}_\nu{\cal J}_\rho{\cal J}_\sigma]) ,$$  
where we used cyclic permutations in the trace together with the relation
\beq\Tr[D^e_\mu(\cJ\cJ\cJ)] = \Tr\{\partial_\mu(\cJ\cJ\cJ) + 
i[\cJ'_\mu,\cJ\cJ\cJ]\} = \partial_\mu\Tr(\cJ\cJ\cJ).\eeq 
Note that if a field-dependent ultraviolet regulator mass $\Lambda$ is present
one cannot drop the total derivative on the right hand side of (3.19), 
but integrating by parts gives
$\partial \ln\Lambda = \partial \Lambda/\Lambda$ which is finite for
$\Lambda\to\infty$. For the second term in (3.18), defining 
$D^{\pm}_\mu=\pp_\mu + \Gamma^{\pm}_\mu$, we have
\beq G^{\pm}_{\mu\nu} = 
\partial_\mu\Gamma^{\pm}_\nu - \partial_\nu\Gamma^{\pm}_\mu 
+ [\Gamma^{\pm}_\mu,\Gamma^{\pm}_\nu] = D_\mu\Gamma^{\pm}_\nu - 
D_\nu\Gamma^{\pm}_\mu - [\Gamma^{\pm}_\mu,\Gamma^{\pm}_\nu].\eeq
By the above argument the $D\Gamma$ terms give finite contributions, so we 
are left with
$$ \epsilon^{\mu\nu\rho\sigma}\Tr[(\Gamma^+_\mu\Gamma^+_\nu - 
\Gamma^-_\mu\Gamma^-_\nu)(\Gamma^+_\rho - \Gamma^-_\rho)
(\Gamma^+_\sigma - \Gamma^-_\sigma)] = 0,$$
again using cyclic permutations of the trace.  Since
$\epsilon^{\mu\nu\rho\sigma}D^e_\mu[D^e_\nu,D^e_\rho]$ vanishes by virtue of
the Bianchi identity, the third term in (3.18) reduces (up to a total
derivative) to the same form as the first term: $G^V\to [\cJ,\cJ]$.
 
First consider the terms quartic in $\R,\R_5$.  To obtain the logarithmically
divergent piece, we drop all $p$-derivatives:
\beq \R\to\pr p_\mu M^\mu,\;\;\;\; \R_5\to\pr p_\mu N^\mu.\eeq
We note that $F_a^{\mu\nu}F^b_{\nu\rho}F^{c\;\rho}_\mu$ and
$F_a^{\mu\nu}F^b_{\nu\rho}\tF^{c\;\rho}_\mu$ vanish if any two of the indices
$a,b,c$ are equal; there are therefore no terms cubic in $M_\sigma$.
Then using $\gamma_\mu M\gamma^\mu = 4m$, together with Eqs. (A.23) and
(B.12--13) and cyclic permutivity of the trace, we obtain:
\bea H(M_1,M_2) &\equiv&
\Tr\(\notp M_1\notp\notgg\notp M_2 \notp\notgg\gamma_5\) \to  {16i\over3}
p^4\Tr\(\tM_1^{\mu\nu}\cJ_\nu M^2_{\mu\rho}\cJ^\rho - M^{\mu\nu}_1\cJ_\nu
\tM^2_{\mu\rho}\cJ^\rho\), \nonumber \\
F(M_1,M_2) &\equiv& \Tr\(\notp M_1\notp M_2\notp\notgg\notp\notgg 
\gamma_5\) \to 4p^4\Tr\[
\(\tM_1^{\mu\nu}m_2 - m_1\tM^{\mu\nu}_2\)\cJ_\mu\cJ_\nu \] \nonumber \\ & &
+ {4i\over3}p^4\Tr\[\(M^1_{\mu\rho}\tM_2^{\mu\nu} - 
\tM^{\mu\nu}_1 M^2_{\mu\rho}\)\{\cJ^\rho,\cJ_\nu\}\], \nonumber \eea 
\bea F'(M_1,M_2,M_3,M_4) &=& -F'(M_4,M_1,M_2,M_3) \equiv \Tr
\(\notp M_1\notp M_2 \notp M_3 \notp M_4\gamma_5\) \nonumber \\ & &
\to {16i\over3}
p^4\Tr\( \tM_1^{\mu\nu}M_2^{\rho\sigma}M^3_{\mu\nu}M^4_{\rho\sigma} -
M_1^{\mu\nu}M_2^{\rho\sigma}M^3_{\mu\nu}\tM^4_{\rho\sigma}\) \nonumber \\ & &
+ 8ip^4\Tr\(m_1M^2_{\mu\nu}m_3\tM_4^{\mu\nu} - 
m_4M^1_{\mu\nu}m_2\tM_3^{\mu\nu}\),\;\;\;\; \eea
where $M_i=M,\M,M_I,\; \tM_i^{\mu\nu} = 
{1\over 2}\epsilon^{\mu\nu\rho\sigma}(M_i)_{\rho\sigma},$ 
and the traces on the right hand sides are over 
internal indices only.  In evaluating these expressions we used the
fact that since $\Tr\(M_\sigma^1M_\sigma^2M_\sigma^3M_\sigma^4\gamma_5\) =
\Tr\(M_\sigma^4M_\sigma^1M_\sigma^2M_\sigma^3\gamma_5\)$, these terms do not
contribute to 
$${1\over2}\[F'(M_1,M_2,M_3,M_4) - F'(M_4,M_1,M_2,M_3)\]
= F'(M_1,M_2,M_3,M_4).$$ Finally, since the 
expression (3.6) for $\R_5$ is odd in $\gamma_5$: $\[\R_5(\gamma_5)\]^4 =
+ \[\R_5(-\gamma_5)\]^4$, it follows that $\Tr(\R_5)^4$ does not
contribute to $T(\gamma_5) = - T(-\gamma_5).$  The logarithmically divergent 
contributions from the quartic terms in (3.8) are therefore given by:
\bea
& & \Tr\bigg[-\R^3\R_5 + \R_5\R^2\R_5 + \R^2\R^2_5 + (\R\R_5)^2 \nonumber \\
& & \qquad -{4\over 3}\(\R_5^2\R\R_5 + \R_5\R\R_5^2 + \R\R_5^3\)\bigg] 
\to \prr\(T_4 + T'_4\). \eea
For the terms quartic in $M$ we obtain
\bea T'_4 = - {1\over4} F'(M,\M,M,\M) &=& 
- {4i\over3}\Tr\(\tM^{\mu\nu}\M^{\rho\sigma}M_{\mu\nu}\M_{\rho\sigma}
- M^{\mu\nu}\M^{\rho\sigma}M_{\mu\nu}\tbM_{\rho\sigma}\) \nonumber
\\ & & - 2i\Tr\(m\M_{\mu\nu}m\tbM^{\mu\nu} - \m M_{\mu\nu}\m\tM^{\mu\nu}\), 
\;\;\;\;\;\;\;\;\eea
and for the terms quadratic in $M$, we find:
\bea \Tr\R^3\R_5 &\to& 0, \;\;\;\; \Tr(\R\R_5)^2 \to -\prrrr H(M,\M), 
\nonumber \\
\Tr\R^2\R^2_5 &=& \Tr\R_5\R^2\R_5 \to 
\prrrr{1\over2}\[F(M,\M) - F(\M,M)\] =  \prr\(T''_4 + T'''_4\), \nonumber \\
\Tr\R\R_5^3 &=& \Tr\R_5\R\R_5^2 = \Tr\R^2_5\R\R_5 \to \prrrr{1\over 2}\bigg[
H(M,M_I) + H(\M,M_I) \nonumber \\ & &
-F(M,M_I) - F(\M,M_I)
+ F(M_I,\M) + F(M_I,M)\bigg] \nonumber \\
&=& \prr\(T''_4 + T'''_4\) - \prrrr {1\over 2}H(M,\M) = -{1\over2p^4}T_4, 
\nonumber \\
T''_4 &=& \Tr\(\[\{\m,\tM^{\mu\nu}\} - \{ m,\tbM^{\mu\nu}\}\]
[\cJ_\mu,\cJ_\nu]\) \nonumber \\ 
T'''_4 &=& {2i\over3}\Tr\[\(\{M_{\mu\rho},\tbM^{\mu\nu}\}
- \{\M_{\mu\rho},\tM^{\mu\nu}\}\)\{\cJ^\rho,\cJ_\nu\}\].\eea
Then
\bea T_4 &=& -2\(T''_4 + T'''_4\) + \prr H(M,\M) \equiv - 2T_4'' - t_4
\;\;\;\;\;\;\;\;\;\;\;\;\;\;\;\;\;\;\;\;\;\;\;\;\;\;\;\;\;\;\;\;\;\;\;\;
\;\;\;\;\nonumber \\ &=& -2T''_4 - 
{8i\over3}\Tr\(\{\cJ^\rho,M_{\mu\rho}\}\{\cJ_\nu,\tbM^{\mu\nu}\} -
\{\cJ^\rho,\M_{\mu\rho}\}\{\cJ_\nu,\tM^{\mu\nu}\}\).\;\;\;\;\;\;\;\; \eea

To evaluate the cubic and quadratic terms, we use a shorthand notation according
to which the covariant derivatives imply the matrix products:
\beq D_\mu^{\pm}\cJ_\nu \equiv [D_\mu^{\pm},\cJ_\nu], \;\;\;\;
D_\mu^{\pm}M \equiv D_\mu^{\pm}M - MD_\mu^{\mp}, \eeq
where here $M$ is any mass matrix.  
Using the Dirac traces in (A.23), the first identity in 
(B.12), and the additional identities
\bea  \Tr\([A,B]C\) &=& - \Tr\(A[B,C]\),\;\;\;\;
D_\mu(M\M) = [d^+_\mu,M\M], \;\;\;\; D_\mu(\M M) = [d^-_\mu,\M M], 
\nonumber \\ \;[ d_\mu^+, M M_I ] &=& (D_\mu M)M_I + MD^-_\mu M_I , \;\;\;\;
[d^-_\mu,\M M_I] = (D_\mu \M)M_I + \M D^+_\mu M_I, \nonumber \\ 
\Tr\(\{A,B\}CD\) &=& \Tr\(B\{A,CD\}\) = \Tr\(B\{A,C\}D\) - \Tr\(BC[A,D]\),
 \eea
together with the facts [see (A.23)] that $\Tr\(\sigma\cdot 
A\gamma_\mu\sigma\cdot B\gamma_\nu\)$ and $\Tr\(\sigma\cdot 
A\gamma_\mu\sigma\cdot B\gamma_\nu\gamma_5\)$ are symmetric in $\{\mu,\nu\}$,
and that $[L_\mu,\cJ_\nu] = 0$, we obtain 
\bea\Tr\R^2\R_5 &\to& \prr\Tr \bigg\{-
2i\tX_-^{\mu\nu}(M,\M)\tD^+_\mu\cJ_\nu-2i\tX_-^{\mu\nu}(\M,M)\tD^-_\mu\cJ_\nu - 
L(M,\M)\nonumber \\ & & + \[\tX_-^{\mu\nu}(M_I,\M) - 
\tX_-^{\mu\nu}(M,M_I) + \tX_+^{\mu\nu}(M,M_I)\]\tG^+_{\mu\nu} \nonumber \\ & &
+ \[\tX_-^{\mu\nu}(M_I,M) - \tX_-^{\mu\nu}(\M,M_I)
+ \tX_+^{\mu\nu}(\M,M_I)\]\tG^-_{\mu\nu} \nonumber 
\\ & & + {4\over3}\[X^+(M,\M) + X^-(\M,M)\] + 2\(\m M^{\mu\nu} -
m\M^{\mu\nu}\)\hL_{\mu\nu} \nonumber \\ & &
- \hL_{\mu\nu}\[X_-^{\mu\nu}(M_I,M) + X_-^{\mu\nu}(M_I,\M)\]\bigg\}, 
\nonumber \\ \Tr\R\R^2_5 &+& \Tr\R_5\R\R_5 \to \prr\Tr\bigg\{-
4i\[\tX_-^{\mu\nu}(M_I,\M)-\tX_-^{\mu\nu}(M,M_I)\]\tD^+_\mu\cJ_\nu \nonumber 
\\ & & + 4i\[\tX_-^{\mu\nu}(M_I,M) - \tX_-^{\mu\nu}(\M,M_I)\]\tD^-_\mu\cJ_\nu 
- 2L(\M,M_I) + 2L(M_I,M) \nonumber \\ & &
-{8\over3}\[X^+(M,M_I) + X^-(M_I,M)- X^+(M_I,\M) - X^-(\M,M_I)\] \nonumber \\ &
& + \[\tX_+^{\mu\nu}(M_I,M_I) - 
2\tX_-^{\mu\nu}(M_I,M_I)\]\(\tG^+_{\mu\nu}- \tG^-_{\mu\nu}\) \bigg\}, 
\nonumber \\ 
\Tr\R_5^3 &\to& \prr\Tr\bigg\{6i\tX_-^{\mu\nu}(M_I,M_I)
\(\tD^+_\mu\cJ_\nu + \tD^-_\mu\cJ_\nu\) \nonumber \\ & &
- 4\[X^+(M_I,M_I)+X^-(M_I,M_I)\] + 3L(M_I,M_I)\bigg\},
\eea where
\bea X^{\pm}(M_1,M_2) &=& \(\tD^{\pm}_\rho M_1^{\mu\rho}\tM^2_{\mu\nu} - 
\tD^{\pm}_\rho\tM_1^{\mu\rho}M^2_{\mu\nu} + \tM^1_{\mu\nu}\tD^{\mp}_\rho 
M_2^{\mu\rho} - M^1_{\mu\nu}\tD^{\mp}_\rho\tM_2^{\mu\rho}\)\cJ^\nu,\nonumber \\ 
X_{\pm}^{\mu\nu}(M_1,M_2) &=& M_1^{\mu\nu}m_2 \pm m_1M_2^{\mu\nu} \qquad
\tX_{\pm}^{\mu\nu} = {1\over2}\epsilon^{\mu\nu\rho\sigma}\tX^{\pm}_{\rho\sigma},
 \nonumber \\ 
L(M_1,M_2) &=& 2\{L_\mu,m_1\}\{\cJ_\nu,m_2\} + 
{4\over3}\{L_\mu,M_1^{\mu\nu}\}\{\cJ_\nu,M^2_{\mu\nu}\} \nonumber \\ & &
+ {8\over3}\(\{L_\mu,M_1^{\mu\rho}\}\{\cJ^\nu,M^2_{\rho\nu}\} +
\{L^\nu,M_1^{\mu\rho}\}\{\cJ_\mu,M^2_{\rho\nu}\} \).\eea
Again, the traces on the right are over internal indices only.
Here and throughout the remainder of this section, $\tG^{\pm}_{\mu\nu}$ is 
understood as one fourth of the Dirac trace of
$[\tD_\mu^{\pm},\tD_\nu^{\pm}]$, and has no contribution from the spin 
connection, and the derivative operators
$\tD_\mu$ are understood to operate only on the object to their immediate right.
The expressions (3.30) can be simplified further using the relations 
\bea X^{\mu\nu}\(\tD_\mu^+\cJ_\nu + \tD_\mu^-\cJ_\nu\) &=& 
{i\over2}X^{\mu\nu}\(\tG^+_{\mu\nu} -\tG^-_{\mu\nu}\), \nonumber \\ 
X^{\mu\nu}\(\tD_\mu^+\cJ_\nu - \tD_\mu^-\cJ_\nu\) &=&
-2iX^{\mu\nu}[\cJ_\mu,\cJ_\nu], \nonumber \\ 
\{\cJ_\mu,M\} &=& {i\over 2}\(\tD^+_\mu M - \tD^-_\mu M\), \eea
that follow from the definitions (3.10) and (3.27).  Defining
\bea X_1 &=& \Tr\[X^+(M,\M) + X^-(\M,M)\] , \nonumber \\
X_2 &=& \Tr\[X^+(M_I,M_I) + X^-(M_I,M_I)\] =
i\Tr\(\tD^{+\sigma}\tM^I_{\sigma\mu}\tD^-_\rho M_I^{\rho\mu}
- \tD^{+\sigma} M^I_{\sigma\mu}\tD^-_\rho \tM_I^{\rho\mu}\), \nonumber \\
X_3 &=& i\Tr\(\tD^\sigma\tM_{\sigma\mu}\tD^-_\rho M_I^{\rho\mu}
- \tD^\sigma M_{\sigma\mu}\tD^-_\rho \tM_I^{\rho\mu} -
\tD^{+\sigma}\tM^I_{\sigma\mu}\tD_\rho \M^{\rho\mu}
+ \tD^{+\sigma}M^I_{\sigma\mu}\tD_\rho \tbM^{\rho\mu}\),\nonumber \\
X_4 &=& \Tr\[X^+(M,M_I) + X^-(M_I,M) -X^+(M_I,\M) - X^-(\M,M_I)\] 
\nonumber \\ &=& - X_1 +X_3
-i\Tr\(\tD^\sigma M_{\sigma\mu}\tD_\rho \tbM^{\rho\mu}
- \tD^\sigma \M_{\sigma\mu}\tD_\rho \tM^{\rho\mu}\), \eea
where we dropped total derivatives, we obtain
\bea T_3 &=& \Tr\(\R^2\R_5 - \R\R_5^2 - \R_5\R\R_5 +{4\over3}\R_5^3\) 
\to \prr\({4\over3}X_3 - {8\over3}X_2 + t_4 + 2T''_4\) \nonumber \\ & &
- {4i\over3p^4}\Tr\(\tD^\sigma M_{\sigma\mu}\tD_\rho \tbM^{\rho\mu}
- \tD^\sigma \M_{\sigma\mu}\tD_\rho \tM^{\rho\mu}\) \nonumber \\ & &
+ \prr\Tr \bigg\{\tX_-^{\mu\nu}(M,\M)\tG^+_{\mu\nu} 
- \tX_-^{\mu\nu}(\M,M)\tG^-_{\mu\nu} + 2\(\m M^{\mu\nu} - m\M^{\mu\nu}\) 
\nonumber \\ & & - \[X_-^{\mu\nu}(M_I,M) + X_-^{\mu\nu}(M_I,\M)\]\hL_{\mu\nu}
+ \tX_+^{\mu\nu}(M_I,M_I)\(\tG_{\mu\nu}^+ -\tG_{\mu\nu}^-\)
\nonumber \\ & & - \[\tX_+^{\mu\nu}(M,M_I)\tG_{\mu\nu}^+ +
\tX_+^{\mu\nu}(\M,M_I)\tG_{\mu\nu}^-\] - L(M,\M)\bigg\} ,\eea
where $t_4,T''_4$ are defined in (3.25--26), and \beq
t_4 = {4\over3}X_4 - {8\over3}X_2.\eeq
Finally, to obtain the logarithmically divergent parts of Tr$\R\R_5$ and
Tr$\R_5^2$, we use (3.15--17), giving 
\bea 
\Tr\R^2_5 &\to& {8\over3p^4}X_2 - {2\over p^4}L(M_I,M_I)
+ {1\over p^4}\tX_+^{\mu\nu}(M_I,M_I)\(\tG_{\mu\nu}^+ -\tG_{\mu\nu}^-\),
\nonumber \\ \Tr\R\R_5 &\to& {4\over3p^4}X_3
+ {4i\over 3p^4}\Tr\(\{L^\sigma,M_{\sigma\mu}\}\{L_\rho,\tbM^{\rho\mu}\} - 
\{L^\sigma,\M_{\sigma\mu}\}\{L_\rho,\tM^{\rho\mu}\}\) \nonumber \\ & & 
- {1\over p^4}\Tr\[i\(\{L^\rho,m\}\tD_\rho\m - \tD^\rho m\{L_\rho,\m\}\)
+ L(M,\M) + 2L(M_I,M_I)\] \nonumber \\ & & 
- {2i\over 3p^4}\Tr\(\{L^\rho,M_{\sigma\mu}\}\tD_\rho 
\M^{\sigma\mu}- \tD^\rho M_{\sigma\mu}\{L_\rho,\M^{\sigma\mu}\}\)
\nonumber \\ & & 
+ {8i\over 3p^4}\Tr\(\{L^\sigma,M_{\sigma\mu}\}\tD_\rho 
\M^{\rho\mu}- \tD^\sigma M_{\sigma\mu}\{L_\rho,\M^{\rho\mu}\} \)
\nonumber \\ & & - {4i\over 3p^4}\Tr\(\hL_\sigma^{\;\;\rho}\{M^{\sigma\mu},
\M_{\rho\mu}\} \) - {1\over p^4}\Tr\(\hL_{\mu\nu}\[X_-^{\mu\nu}(M,M_I) +
X_-^{\mu\nu}(\M,M_I)\]\)
\nonumber \\ & &  + {1\over p^4}\Tr\[\tX_+^{\mu\nu}(M,M_I)\tG_{\mu\nu}^+ +
\tX_+^{\mu\nu}(\M,M_I)\tG_{\mu\nu}^-\] \nonumber \\ 
& & + {i\over p^4}r^\mu_\nu\Tr\(\tM^{\nu\rho}\M_{\mu\rho} -
M^{\nu\rho}\tbM_{\mu\rho}\) + {\rm total \;derivative}. \eea
Inserting these results in (3.7) gives
\bea -{i\over2}T_- &=& g^{1\over2}\lll\(T'_4 + T_4 + T_3 - \Tr\R\R_5 + 
\Tr\R_5^2\) 
\nonumber \\ &=& g^{1\over2}\lll\Tr\bigg\{T'_4 + 
\[\tX_-^{\mu\nu}(M,\M)\tG^+_{\mu\nu} - \tX_-^{\mu\nu}(\M,M)\tG^-_{\mu\nu}\] 
\;\;\;\;\; \nonumber \\ & &
-{4i\over3}\(\tD^\sigma M_{\sigma\mu}\tD_\rho \tbM^{\rho\mu} - 
\tD^\sigma \M_{\sigma\mu}\tD_\rho \tM^{\rho\mu}\) - 
ir^\mu_\nu\(\tM^{\nu\rho}\M_{\mu\rho} - 
M^{\nu\rho}\tbM_{\mu\rho}\)\qquad \;\;\;\; \nonumber \\ & & 
+ [\hL_{\mu\nu},\m]M^{\mu\nu} - [\hL_{\mu\nu},m]\M^{\mu\nu} 
+ i\(\{L^\rho,m\}\tD_\rho\m - \tD^\rho m\{L_\rho,\m\}\) \nonumber \\ & & 
- {4i\over 3}\(\{L^\sigma,M_{\sigma\mu}\}\{L_\rho,\tbM^{\rho\mu}\} - 
\{L^\sigma,\M_{\sigma\mu}\}\{L_\rho,\tM^{\rho\mu}\}\) \nonumber \\ & &
- {8i\over 3}\(\{L^\sigma,M_{\sigma\mu}\}\tD_\rho 
\M^{\rho\mu}- \tD^\sigma M_{\sigma\mu}\{L_\rho,\M^{\rho\mu}\}\)\nonumber \\ & & 
+ {2i\over 3}\[L^\rho\(\{M_{\sigma\mu},\tD_\rho\M^{\sigma\mu}\} - \{\tD_\rho 
M^{\sigma\mu},\M_{\sigma\mu}\}\) + 2\hL_{\mu\nu}
\{M^{\mu\rho},\M^\nu_{\;\;\rho}\}\]\bigg\}.\;\;\;\;\eea

To evaluate (3.36), we note that the connection is block diagonal in the 
$\chi$-$\lambda$-$\alpha$ sector, and the axial
part is diagonal in the $\lambda$ and $\alpha$ sectors, with
$\cJ_{\lambda\lambda} = - \cJ_{\alpha\alpha}$.  
Using the reality and
symmetry properties of the off-diagonal $\lambda$-$\alpha$ masses:
\beq m_{\lambda\alpha}= -\m_{\lambda\alpha} = m^T_{\lambda\alpha}, \;\;\;\; 
M^{\mu\nu}_{\lambda\alpha} = \M^{\mu\nu}_{\lambda\alpha} = 
- \(M^{\mu\nu}_{\lambda\alpha}\)^T, \eeq
it is easy to see that there is no contribution that involves only these masses.
For the off-diagonal $\lambda$-$\chi$ masses:
\bea & &
m_{\lambda\chi} = m^T_{\lambda\chi},  \;\;\;\;
\tM^{\mu\nu}_{\lambda\chi} = iM^{\mu\nu}_{\lambda\chi}, \;\;\;\;
\tbM^{\mu\nu}_{\lambda\chi} = -i\M^{\mu\nu}_{\lambda\chi},\;\;\;\;
M^{\mu\nu}_{\lambda\chi} = -\(M^{\mu\nu}_{\lambda\chi}\)^T, \nonumber \\ & &
M^{\mu\nu}_{\lambda\chi}\M_{\mu\nu}^{\lambda\chi} =
\tM^{\mu\nu}_{\lambda\chi}\M_{\mu\nu}^{\lambda\chi} = 0, \;\;\;\;
\(M^{\mu\nu}\M_{\rho\nu}\)^a_a = \(M_{\rho\nu}\M^{\mu\nu}\)^a_a. \eea
It follows from these relations that the last line in (3.36) vanishes.

Using the fermion matrix elements given in Appendix C.2, we obtain the 
nonvanishing contributions to $T_-$ listed in Appendices C.3--8.
Note that these expressions are fully covariant, although the
expansion (3.7) of $T_-$ is not.  This noncovariance is necessarily the case 
since $T_-$ contains the chiral anomaly that breaks classical K\"ahler 
invariance.  However, the logarithmically divergent contributions are K\"ahler 
invariant, up to a possible dependence of the effective cut-off on the 
K\"ahler potential~\cite{bg},~\cite{tom},~\cite{mk}.

The ghostino determinant also contains helicity-odd contributions, but since it
has the same form~\cite{us} as that of a four-component
scalar, its evaluation is straightforward; the result is given in Appendix
C.7.

\section{The One-Loop Effective Action}
\setcounter{equation}{0}\indent

The quantum action obtained by the prescriptions defined in I (see section 2 of
that paper) and in Section 2 above takes the form
\begin{equation}
\L_q = -{1\over 2}\Phi^T Z_\Phi\(\hD^2 + H_\Phi\)\Phi + {1\over 2}
\bar{\Theta}Z_\Theta\(i\notD - M_\Theta\)\Theta + \L_{gh} + \L_{Gh}.
\end{equation}
The last two terms are the ghost and ghostino terms, respectively, $\Phi =
(h_{\mu\nu},\hcA^a,\hz^i,\hz^{\m})$ is a $2N+4N_G+10$ component scalar, 
$\Theta = (\psi_\mu,\lambda^a,\chi^I = L\chi^i + R\chi^{\ibar}, \alpha)$ is an 
$N + N_G + 5$ component
Majorana fermion, where $N$ is the number of chiral multiplets, 
$N_G$ is the number of gauge multiplets, and the matrix
valued metrics $Z_\Phi$ and $Z_\Theta$ are defined in Appendix B of I and in 
Appendices C.1 and C.2 below.  As in I we set background fermion fields to 
zero, so $\psi_\mu,\lambda^a,\chi^I$ are the quantum gravitino, gaugino and 
chiral fermions, respectively, and $\alpha$ is the auxiliary field introduced 
to implement the gravitino gauge fixing condition~\cite{us}.
The matrix-valued covariant derivative $D_\mu$ is defined as in 
Appendix A of I, and $\hD_\mu$ includes additional terms in the connections 
that are given in (2.11,17) above.  

The one-loop contribution to the effective action is
\bea 
\L_1 &=& {i\over 2}\Tr\ln(\hD^2 + H_\Phi) -{i\over 2}\Tr\ln(-i\notD + M_\Theta)
\nonumber \\
& & + i\Tr\ln(D^2 + H_{Gh}) - i\Tr\ln(\hD^2 + H_{gh}).
\eea 
The general results obtained 
in~\cite{mkg},~\cite{sigma},~\cite{josh},~\cite{duff} give for the bosonic
determinant:
\bea 
& & {i\over 2}\Tr\ln(\hD^2 + H_\Phi) = 
\sqrt{g}\Bigg\{ {\Lambda^2\over 32\pi^2}\Tr\({1\over 6}r - H_\Phi\) 
\nonumber \\ 
& & \qquad + \lll\Tr\({1\over 2}H_\Phi^2 - {1\over 6}rH_\Phi + 
{1\over 12}\hG^\Phi_{\mu\nu}\hG_\Phi^{\mu\nu} + 
{1\over 120}\[r^2 + 2r^{\mu\nu}r_{\mu\nu}\]\)\Bigg\}, \nonumber \\
 & & \quad \eea 
and for the fermionic determinant we have
\begin{equation}
-{i\over 2}\Tr\ln(-i\notD + M_\Theta) = -{i\over 2}\(T_+ + T_-\) = 
-{i\over 8}\Tr\ln[\hD^2 + H_\Theta] -{i\over 2}T_-,\end{equation}
where in (4.4) $\hD_\mu$ and $H_\Theta$ are the $8\times 8$ matrices defined in 
(2.14--17).  The helicity-averaged part, $T_+$, of the 
fermion trace is $-{1\over 4}$ times (4.3) with the substitutions
$ H_\Phi\to H_\Theta, \;\hG^\Phi_{\mu\nu}\to \hG^\Theta_{\mu\nu}$ 
and the trace includes a trace over Dirac indices, so
$$ {1\over 4}\(\Tr \;{\bf 1}\)_\Theta = \(\Tr \;{\bf 1}\)_\Phi -2N_G 
=  2N+2N_G+10. $$  Similarly, the ghost and
ghostino contributions are equivalent to, respectively, $-2$ times 
the contribution of a $(4+N_G)$-component scalar and $+2$ times the 
contribution of a four-component scalar. For bosons, $H_\Phi$ and $\hD_\mu$ 
are defined in Section 2; the matrix elements of $H$ and of
\begin{equation}
\hG_{\mu\nu} = [\hD_\mu,\hD_\nu],
\end{equation} are given in Appendix C, and the helicity-odd contribution,
$T_-$, of the fermion determinant that was evaluated in Section 3, 
Eq. (3.6) is given in (C.36).  The traces in (4.3--4.4) are given explicitly in
Appendix C below and in Appendix B of I.
Here we list only the contributions involving background Yang-Mills fields 
and/or integration over the quantum Yang-Mills supermultiplet that were 
omitted in I.

If $\L(g,K)$ is the standard
Lagrangian~\cite{cremmer},~\cite{kahler} for $N=1$ supergravity coupled to
matter with space-time metric $g_{\mu\nu}$, K\"ahler potential $K$, 
and gauge kinetic normalization function $f_{ab} = \delta_{ab}(x+iy)$,
then the logarithmically divergent part of the one-loop corrected Lagrangian is
\beq
\L_{eff} = \L\(g_R,K_R\) + \L_0 + \lll\(X^{AB}\L_A\L_B + 
X^A\L_A\) + \sqrt{g}\lll\(L + N_GL_g\), \eeq
where the classical Lagrangian $\L(g,K)$ is given in Appendix C below (see
footnote 1), 
$\L_0$ is the one loop correction found\footnote{The last five lines of (3.6)
in I should read: \bea &-& 4\(\D_\mu\z^{\m}\D^\mu z^iK_{i\m}\)^2 + \({N\over 3}
- 7\)\D_\mu z^j\D^\mu z^i\D_\nu\z^{\m}\D^\nu\z^{\n}K_{i\n}K_{j\m} 
\nonumber \\ &+& 
{2\over 3}\D_\mu\z^{\m}\D^\mu z^i\D_\nu\z^{\n}\D^\nu z^jK_{i\n}K_{j\m} 
- {2\over 3}\D_\rho z^i\D^\rho\z^{\m}K_{i\m}\D^\mu z^j\D_\mu\z^{\n}R_{j\n} 
\nonumber \\ &+& \D_\mu z^j\D^\mu\z^{\m}R^k_{j\m i}
\D_\nu z^{\ell}\D^\nu\z^{\n}R^i_{\ell\n k}
+ \D_\mu z^j\D^\mu z^iR^{k\;\;\ell}_{\;\;j\;\; i}
\D_\nu\z^{\n}\D^\nu\z^{\m}R_{\n k\m\ell} \nonumber \\ &+& 
{1\over3}\D_\mu z^i\D_\nu\z^{\m}K_{i\m}R_{j\n}\(\D^\mu z^j\D^\nu\z^{\n}
- \D^\nu z^j\D^\mu\z^{\n}\)\nonumber \\ &+& 
\D_\mu z^j\D_\nu\z^{\m}R^k_{i\m j}
\D^\mu z^{\ell}\D^\nu\z^{\n}R^i_{k\n \ell} - \D_\mu z^j\D_\nu\z^{\m}R^k_{i\m j}
\D^\nu z^{\ell}\D^\mu\z^{\n}R^i_{k\n \ell} 
+ 4\(\L_i\A^iAe^{-K} +{\rm h.c.}\).\nonumber\eea} in I after renormalization 
of $g,K$ [see Eq.(3.6) of I], and 
\bea L &=& \[\cW^{ab}\(3C_G\delta_{ab} - D_i(T_bz)^jD_j(T_az)^i\) + 
{\rm h.c.}\] - 24e^{-K}a\aa\D
\nonumber \\ & &  + {N + 5\over6}\[\(\cW^{ab} + \cbW^{ab}\)\D_a\D_b -
x\(F^a_{\rho\mu} -i\tF^a_{\rho\mu}\)
\(F_a^{\rho\nu} + i\tF_a^{\rho\nu}\)\D_\nu z^i\D^\mu\z^{\m}K_{i\m}\]
\nonumber \\ & &  + {N+ 5\over3}\[x^2\cW_{ab}\cbW^{ab} + 2\D^2 - 
\D\(K_{i\m}\D_\rho z^i\D^\rho\z^{\m} + 2\hV + 4\D M_\psi^2\)\]\nonumber \\ & & 
+ 14x^2\cW_{ab}\cbW^{ab} + 12\(\cW^{ab} + \cbW^{ab}\)\D_a\D_b + 22\D^2 + 
2\D\(11\hV + 8K_{i\m}\D_\rho z^i\D^\rho\z^{\m}\) \nonumber \\ & & 
+ x\(\cW + \cbW\)\(K_{i\m}\D_\rho z^i\D^\rho\z^{\m} - 2M_\lambda^2 - 2V\) + 
4\D\(27M_\psi^2 + 7M_\lambda^2\) \nonumber \\ & &  
- 26i\D_\mu z^j\D_\nu\z^{\m}K_{i\m}\D^aF_a^{\mu\nu} 
+ {2\over x}\D_\mu z^i\D^\mu\z^{\m} R_{\n i\m j}\D_a D^j(T^a\z)^{\n} 
\nonumber \\ & &  + {2\over x}\D_ae^{-K}R^{k\;j}_{\;n\;i}A_k\A^nD_j(T^az)^i 
+ {e^{-K}\over x}\D_a\[(T^az)^iR_{i\;\;\ell}^{\;\; j\;\; k}\A^{\ell}A_{jk} + 
{\rm h.c.}\] \nonumber \\ & &  
+ 2iF^a_{\mu\nu}D_j(T_az)^iR^j_{i\m k} \D^\mu z^k\D^\nu\z^{\m} + 
{4\over 3}\D e^{-K}R^i_jA_i\A^j + {4\over3}\D\D_\mu z^i\D^\mu\z^{\m}R_{i\m} 
\nonumber \\ & &  
+ {D_i(T_az)^i\over6x}\[4\D_a\(\D_\mu z^j\D^\mu\z^{\m}K_{j\m} + \hV +  
3M_\psi^2 - 2\D\) + 13iF^a_{\mu\nu}K_{\m j}\D^\mu z^j\D^\nu\z^{\m}\]
\nonumber \\ & & 
+ {1\over2}\({19\over x^2} + \rho^i\rho_i\)\D\(\pp_\mu x\pp^\mu x + 
\pp_\mu y\pp^\mu y\) - \({11\over x^2} + 3\rho^i\rho_i\)
\pp_\mu y\pp_\nu x\D^aF_a^{\mu\nu} \nonumber \\ & &  
- {1\over4x}\(1 + 3x^2\rho^i\rho_i\)\(\pp_\rho x\pp^\nu x + \pp_\rho y\pp^\nu 
y\)\(F^a_{\mu\nu} + i\tF^a_{\mu\nu}\)\(F_a^{\mu\rho} - i\tF_a^{\mu\rho}\)
\nonumber \\ & & - 5\lbr\[{i\over x}\(F_a^{\nu\mu} -i
\tF_a^{\nu\mu}\) + {g_{\nu\mu}\over x^2}\D^a\]\(\pp_\nu x + i\pp_\nu y\)
K_{i\m}(T^az)^i\D_\mu\z^{\m} + {\rm h.c.}\rbr \nonumber \\ & &  
- \rho^i\rho_i\lbr\[ix\(F_a^{\nu\mu} -i
\tF_a^{\nu\mu}\) + g_{\nu\mu}\D^a\]\(\pp_\nu x + i\pp_\nu y\)
K_{i\m}(T^az)^i\D_\mu\z^{\m} + {\rm h.c.}\rbr \nonumber \\ & &  
+ 2ix^2\rho^i\rho_i\D_\mu z^j\D_\nu\z^{\m}K_{i\m}\D^aF_a^{\mu\nu} + 
2x^2\rho^i\rho_i\D\[8M_\psi^2 + M_\lambda^2 + 2\hV - 2e^{-K}a\aa\] 
\nonumber \\ & & 
- x^2\rho^i\rho_i\[2x^2\cW_{ab}\cbW^{ab}\(1 - x^2\rho^i\rho_i\) -
4x\(\cW + \cbW\)\D + \(\cW^{ab} + \cbW^{ab}\)\D_a\D_b + 2\D^2\] 
\nonumber \\ & &  + {x^3\over2}\rho_i\rho^i\(F^a_{\rho\mu} -i\tF^a_{\rho\mu}\)
\(F_a^{\rho\nu} + i\tF_a^{\rho\nu}\)\D_\nu z^i\D^\mu\z^{\m}K_{i\m} + 
2x^2\rho_i\rho^i\D\D_\mu z^i\D^\mu\z^{\m}K_{i\m}\nonumber \\ & &  
+ 2x\[4\rho_{ij}(T^az)^i(T^bz)^j\cW_{ab} + i\D_\nu\z^{\m}(T^az)^i
\rho_{\m ij}\(F_a^{\mu\nu} - i\tF_a^{\mu\nu}\)\D_\mu z^j + {\rm h.c.}\]
\nonumber \\ & & + \lbr \rho_{ij}\D^\mu z^j\[{2\over x}\(\pp_\mu x -i\pp_\mu y\)
(T^az)^i\D_a - {\f^i\over2}\cW\(\pp_\mu x + i\pp_\mu y\)\] + {\rm h.c.}\rbr 
\nonumber \\ & & 
+ \lbr\cbW\[2x^3\rho^i\rho_iM_\lambda^2 + \f^ia_i(\aa - \A)e^{-K} - x^2 
\rho^{ij}\(A_{jik}\A^k - A_{ij}\A\)e^{-K}\] + {\rm h.c.}\rbr \nonumber \\ & &
+ \lbr\(e^{-K}\A^j A^{\m} + \D_\mu z^j\D^\mu\z^{\m}\)\[4\rho_{\m ij}
(T^az)^i\D_a - \(\rho_{\m ij} + {\f_{\m}\over x}\rho_{ij}\)\f^i\D\] + 
{\rm h.c.}\rbr \nonumber \\ & & 
- {i\over2}K_{i\m}\[\D^\nu\z^{\m}(T_az)^i - 
\D^\nu z^i(T_a\z)^{\m}\]\[\f^i\rho_{ij}\D^\rho z^j\(F^a_{\rho\nu} - 
i\tF^a_{\rho\nu}\)+ {\rm h.c.}\] \nonumber \\ & &
+ {\D_a\over 2x}\[K_{k\m}(T^az)^k\D^\mu \z^{\m} + {i\over2}\(\pp_\nu x + 
i\pp_\nu y\)\(F^{\nu\mu} -i\tF^{\nu\mu}\) + {\rm h.c.}\]
\(\rho_{ij}\D_\mu z^i\f^j + {\rm h.c.}\) \nonumber \\ & & 
- \[\cW^{ab}\rho_{ij}f^i(T_az)^j\D_b + x^2\D_\rho z^i\D^\rho z^j
\(2\rho_{ij}\cW - R_{\n i\m j}\rho^{\m\n}\cbW\) + {\rm h.c.}\] 
\nonumber \\ & & + 2x^2\rho_{ij}\rho^j_{\m}\D\D_\rho z^i\D^\rho\z^{\m} 
+ x^4\rho_{ij}\rho^{ij}\cW\cbW ,\eea 
\bea L_g &=&  x^6\(\rho^i\rho_i\)^2\cW\cbW - 2M_\lambda^4 + 3M_\psi^4
- 2M_\psi^2M_\lambda^2 + \hV^2 + \D^2
+ 6e^{-K}a\aa M_\psi^2 \nonumber \\ & &  
+ 2\hV\(2M_\psi^2 - M_\lambda^2 + e^{-K}a\aa\) - 
e^{-K}\(\aa^iA_i + {\rm h.c.}\)\(\hV + M_\psi^2 \) \nonumber \\ & &  
+ e^{-2K}a_i\A^i\aa^jA_j - 2e^{-2K}\(\aa^iA_ia\A + {\rm h.c.}\) 
+ x^2\rho_{ij}\D_\mu z^i\D^\mu z^j\rho_{\n\m}\D_\nu\z^{\m}\D^\nu\z^{n} 
\nonumber \\ & &
 + e^{-K}\D_\mu z^i\D^\mu\z^{\m}\[\(a_i - A_i\)\(\aa_{\m} - \A_{\m}\) 
+ x^2\rho_{ik}\A^k\rho_{\m}^jA_j +
{\f_{\m}f_i\over4x^2}a\aa\]
\nonumber \\ & &  + e^{-K}\lbr\D_\mu z^i\D^\mu z^j\[\(a_i - A_i\) 
\({f_j\over2x}\aa - x\rho_{jn}\A^n\) - {f_j\over2x}a_i\A - 
f_i\(a-A\)\rho_{jk}\A^k\] + {\rm h.c.}\rbr \nonumber \\ & &  
+ {e^{-K}\over 2x}\lbr\D_\mu z^i\D^\mu\z^{\m}\f_{\m}\[2\aa a_i 
- x\rho_{ik}(a-A)\A^k\] + {f_if_j\over2x}\D_\mu z^i\D^\mu z^j\aa(2a-A) 
+ {\rm h.c.}\rbr \nonumber \\ & & 
 + x\(\rho_{ij}\D_\mu z^i\D^\mu z^j + {\rm h.c.}\)\(M_\psi^2 - \hV\) 
+ e^{-K}\[x\rho_{ij}\D_\mu z^i\D^\mu z^j\(a_k\A^k 
- 2\A a\) + {\rm h.c.}\] \nonumber \\ & &  
+ {1\over16x^4}\left|\(\pp_\mu x + i\pp_\mu y\)
\(\pp^\mu x + i\pp^\mu y\)\right|^2
- x^3\rho^i\rho_i\(\cW + \cbW\)\(M_\psi^2 + \hV\) \nonumber \\ & & 
+ x^3\rho^k\rho_k\[\cW\(x\rho_{ij}\D_\mu z^i\D^\mu z^j + 
e^{-K}A_i\aa^i - 2e^{-K}\aa A\) + {\rm h.c.}\] \nonumber \\ & & 
+ {1\over6}K_{i\m}K_{j\n}\(4\D_\mu z^i\D^\mu z^j\D_\nu\z^{\m}
\D^\nu\z^{\n} + \D_\mu z^i\D^\mu\z^{\n}\D_\nu\z^{\m}\D^\nu z^j\)
\nonumber \\ & & -{1\over3}\(\D_\mu z^i\D^\mu\z^{\m}K_{i\m}\)^2
+ x^2\cW_{ab}\cbW^{ab} + {1\over2}\(\cW_{ab} + \cbW_{ab}\)\D^a\D^b
\nonumber \\ & & 
- {1\over3}V^2 + {1\over3}M_\lambda^2\(\D_\mu z^i\D^\mu\z^{\m}K_{i\m} -2V\)
- \({\pp_\mu x\pp^\nu x + \pp_\mu y\pp^\nu y\over x^2}\)
K_{i\m}\D_\nu z^i\D^\mu\z^{\m} \nonumber \\ & &
+{1\over3}V\D_\mu z^i\D^\mu\z^{\m}K_{i\m} + \({\pp_\nu x\pp^\nu x\over 6x^2} + 
{\pp_\nu y\pp^\nu y\over6x^2}\)\(2\D_\mu z^i\D^\mu\z^{\m}K_{i\m} - V\)
\nonumber \\ & & + \(F^a_{\rho\mu} + i\tF^a_{\rho\mu}\)
\(F_a^{\rho\nu} - i\tF_a^{\rho\nu}\)\({\pp^\mu x\pp_\nu x + 
\pp^\mu y\pp_\nu y\over4x} - {x\over2}K_{i\m}\D_\nu z^i\D^\mu\z^{\m}\).\eea 
Our notation is defined in Appendix B below.
Here $\cW= \cW^a_a$, where 
\beq \cW^a_b = {1\over4}\(F^a_{\mu\nu}F_b^{\mu\nu} -
iF^a_{\mu\nu}\tF_b^{\mu\nu}\) - {1\over 2x^2}\D^a\D_b \eeq
is the bosonic part of 
the $F$-component of the composite chiral supermultiplet constructed from the
Yang-Mills chiral superfield $W^a(\theta) = \lambda^a_L + O(\theta).$
The renormalized K\"ahler potential is
\beq
K_R = K + \lll\[ e^{-K}\(A_{ij}\A^{ij} -2A_i\A^i - 4A\A\) -4\K_a^a 
- \(12 + 4x^2\rho_i\rho^i\)\D\],\eeq
and the renormalized space-time metric is given by
\bea 
g_{\mu\nu} &=& (1 - \epsilon)g^R_{\mu\nu} + \epsilon_{\mu\nu}, \nonumber \\
\epsilon &=& \epsilon^0 -\lll\[{N_G\over 6}\(r + V\) + {55-N\over6}\D + 
2x^2\rho_i\rho^i\D + {2\over 3x}\D_aD_i(T^az)^i+ 
{N_G\over3}M^2_\lambda\], \nonumber \\ 
\epsilon_{\mu\nu} & = & \epsilon_{\mu\nu}^0 + \lll{N_G\over 2}
\(r_{\mu\nu} - {1\over2}rg_{\mu\nu}\) 
-g_{\mu\nu}{N_G\over6x^2}\(\pp_\rho x\pp^\rho x + \pp_\rho y\pp^\rho y
- \D_\mu z^i\D^\mu\z^{\m}K_{i\m}\)
\nonumber \\ & &   
+ N_G\[2{\nabla_\mu\pp_\nu x\over x} - {\pp_\mu x\pp_\nu x\over x^2} 
+ {\pp_\mu y\pp_\nu y\over x^2}- {1\over2}\(\D_\mu z^i\D_\nu\z^{\m}
+ \D_\nu z^i\D_\mu\z^{\m}\)K_{i\m}\] 
\nonumber \\ & &  -
g_{\mu\nu}xF^a_{\rho\sigma}F_a^{\rho\sigma}\({N+17\over24} + {N_G\over8} 
- {x^2\rho_i\rho^i\over4}\) \nonumber \\ & & 
+ xF^a_{\mu\rho}F_{a\nu}^{\;\;\;\;\rho}\({N+29\over6}
+ {N_G\over2} - x^2\rho_i\rho^i\) \eea 
where the superscript 0 refers to the result of I.  The terms in (4.6) 
proportional to $\L_A$ can be removed by field redefinitions:
\beq \phi^A\to\phi_R^A = \phi^A - \lll\(X^A + {1\over2}X^{AB}\L_B\), \eeq
with 
\bea X_{i\m} &=& {N_G\over 4x^2\sqrt{g}}f_if_{\m}, \;\;\;\;
X_{a\mu,b\nu} = -{1\over x\sqrt{g}}\(7 + x^2\rho_i\rho^i\)
\delta_{ab}g_{\mu\nu}, \nonumber \\
X^i &=& \(X^{\ibar}\)^* = 4e^{-K}\A^iA + {2\over x}\(2 + x^2\rho^j\rho_j\)
\D_a(T^az)^i \nonumber \\ & & 
- 4x\D\rho^i - 2\rho^i_{\m}(T_a\z)^{\m}\D_a - N_G{\pp_\mu x\over x}\D^\mu z^i 
\nonumber \\ & & 
+ N_G{\f^i\over2x}\[x^3\rho^j\rho_j\cW + x\rho_{jk}\D_\mu z^j\D^\mu z^k 
+ e^{-K}\(\aa^jA_j - 2\aa A\) - \hV - M_\psi^2\], \nonumber \\
X_{\mu a} &=& {i\over x}\(16 + 2x^2\rho^i\rho_i\)K_{i\m}\[(T_az)^i
\D_\mu\z^{\m} - (T_a\z)^{\m}\D_\mu z^i\] \nonumber \\ & &
+ x\rho_i\rho^i\(\pp^\rho xF_{a\rho\mu} + \pp^\rho y\tF_{a\rho\mu}\) 
+ 3{\pp^\rho y\over x}\tF_{a\rho\mu} + {\pp^\rho x\over x}\(7 - N_G\)
F_{a\rho\mu} \nonumber \\ & &
+ {1\over2}\[\(F_{a\rho\mu} - i\tF_{a\rho\mu}\)\D^\rho z^i\rho_{ij}\f^j 
+ {\rm h.c.}\] - \(5 + x^2\rho^i\rho_i\){\pp_\mu y\over x^2}\D_a . \eea

The terms in (4.7--8) of the form $g(z,\z)\cW\cbW$ are the bosonic part of the
effective Lagrangian (in the notation of~\cite{kahler})
\beq \L_{|W|^4} = \int d^4\theta Eg(Z,\Z)|WW|^2. \eeq
It should be possible to write the remaining terms in superfield form\footnote{
Note that $F^i = -e^{-K/2}\A^i$ and $M= - 3e^{-K/2}A$ are the bosonic parts of 
auxiliary fields of the chiral superfield $Z^i$ and the gravity superfield,
respectively.  It is easy to show that calculating the one loop
corrections before or after elimination of the auxiliary fields in terms of
their classical solutions gives the same result to the loop order considered.
Our results are expressed in terms of these auxiliary fields in~\cite{us3}.}
[up to total derivatives and field redefinitions of the form (4.11-13)], and
thus to extract the fermionic part of the Lagrangian for these higher dimension
operators.  However, there may be additional fermionic terms, {\it e.g,} those
of the form~\cite{tomw}
\beq \L_{W^{2n}} = \int d^4\theta E g(Z,\Z)(WW)^{n>1} + 
{\rm h.c.}, \eeq
that cannot be obtained in this way, as they have no purely bosonic
components.  The determination of such terms requires retaining fermionic
background fields~\cite{sally},~\cite{sigma},~\cite{casimir}.

Notice that the coefficient of $\ln\Lambda^2F^{\mu\nu}F_{\mu\nu}$ is not a 
holomorphic function, except in the limits of a flat K\"ahler metric 
($D_i\to \pp_i$) and flat space-time ($M_{Pl} \to\infty$, in which case 
operators of dimension greater than four are suppressed).  This 
nonholomorphicity
is distinct from from the holomorphic anomaly~\cite{dixon,russ} that arises 
from the field-dependence of the infrared regulator masses.  In other words, 
when the K\"ahler and/or space-time metric is not flat, there are
corrections that correspond to D-terms as well as the usual F-terms.

The quadratically divergent contributions to the one-loop Lagrangian are given 
by (C.33--C.35). The Pauli-Villars regularization of these terms was given 
in~\cite{mk}; they contribute additional renormalizations of the metric and the
K\"ahler potential that are determined by the field-dependent squared masses
of the Pauli-Villars regulator fields that play the role of effective cut-offs.
The field dependence of the effective cut-offs in the logarithmically divergent
contribution to the renormalized K\"ahler potential will generate additional
terms in the effective Lagrangian proportional to 
$$ D_I\ln \Lambda^2 = 2{D_I\Lambda\over\Lambda},\;\;\;\; I = i,\ibar, $$
that do not grow with the cut-off, and therefore have to be considered together
with the finite terms that we have not evaluated here.

\section{The String Dilaton} 
\setcounter{equation}{0}\indent

In effective supergravity from superstring theory, the classical K\"ahler
potential $K(z,\z)$, superpotential $W(z)$ and Yang-Mills normalization 
function $f_{ab}(z)$ take the forms
\bea K(z,\z) &=& -\ln(s+\s) + G(y^i,\y^{\m}), \;\;\;\; W(z) = W(y^i), 
\nonumber \\ f_{ab}(z) &=& \delta_{ab}k_as, \;\;\;\; y^i,\y^{\m} \ne s.\eea
Although we have restricted our analysis to the case $f_{ab} = \delta_{ab}f$,
it is equally applicable to the case $f_{ab} = \delta_{ab}k_af,\;k_a=$ 
constant, provided we make the substitutions $F^a_{\mu\nu}\to k_a^{1\over2}
F^a_{\mu\nu}, \;A^a_\mu\to k_a^{1\over2}A^a_\mu,\; T^a\to k_a^{-{1\over2}}T^a,
\; c_{abc}\to k_a^{-{1\over2}}c_{abc},\;(c_{abc}\ne 0$ only if $k_a=k_b=k_c$)
in all the relevant equations.  Our results are therefore applicable to all 
known effective tree Lagrangians from superstrings, including those where the 
integers $k_a\ge 1$ correspond to higher affine levels~\cite{gins}.
In this case the operators $a,\rho_{ij},1- x^2\rho_i\rho^i$, and their
covariant derivatives vanish identically.  In particular $M_\lambda^2 =
M_\psi^2 \equiv M^2$, and (4.6) reduces to
\bea 
\L_{eff} &=& \L\(g_R,K_R\) + \L_0 + \lll\(X^{AB}\L_A\L_B + 
X^A\L_A\) + \sqrt{g}\lll\(L + N_GL_g\), \nonumber \\
L &=& \(\cW^{ab} + \cbW^{ab}\)\( 3C_G\delta_{ab} - D_i(T_bz)^jD_j(T_az)^i\) 
+ 2\D\(13\hV + 9K_{i\m}\D_\mu z^i\D^\mu\z^{\m}\) \nonumber \\ & &  
+ {N + 5\over12}\[(s+\s)^2\cW_{ab}\cbW^{ab} + 2\(\cW^{ab} + \cbW^{ab}\)\D_a\D_b 
+ 8\D^2 - 8\(\hV + 2M^2\)\D\] \nonumber \\ & & 
-  {N+5\over12}\[(s+\s)\(F^a_{\rho\mu} - i\tF^a_{\rho\mu}\)\(F_a^{\rho\nu} + 
i\tF_a^{\rho\nu}\) + 4g^\nu_\mu\D\]\D_\nu z^i\D^\mu\z^{\m}K_{i\m}
\nonumber \\ & & + {7\over2}(s+\s)^2\cW_{ab}\cbW^{ab} +
11\(\cW^{ab} + \cbW^{ab}\)\D_a\D_b + 20\D^2 + 154M^2\D \nonumber \\ & & 
+ {(s+\s)\over2}\(\cW + \cbW\)\(K_{i\m}\D_\rho z^i\D^\rho\z^{\m} - 2\hV + 2\D\) 
- 24i\D_\mu z^i\D_\nu\z^{\m}K_{i\m}\D^aF_a^{\mu\nu} \nonumber \\ & & 
 + {(s+\s)\over4}\(F^a_{\rho\mu} - i\tF^a_{\rho\mu}\)\(F_a^{\rho\nu} + 
i\tF_a^{\rho\nu}\)\D_\nu z^i\D^\mu\z^{\m}K_{i\m}\nonumber \\ & & 
+ {D_i(T_az)^i\over3(s+\s)}\[4\D^a\(\hV - 2\D + 3M^2 + K_{j\m}\D_\mu z^j
\D^\mu\z^{\m}\) + 13iF^a_{\mu\nu}K_{\m j}\D^\mu z^j\D^\nu\z^{\m}\]
\nonumber \\ & &
- \({4\over3}\D R_{i\m} + {\D_a\over(s+\s)}R_{\n i\m j}D^j(T^a\z)^{\n}\)
\(e^{-K}\A^iA^{\m} + \D_\mu z^i\D^\mu\z^{\m} \) \nonumber \\ & &
+ 2iF^a_{\mu\nu}D_j(T_az)^iR^j_{i\m k}\D^\mu z^k\D^\nu\z^{\m} + 
{2e^{-K}\over(s+\s)}\D_a\[(T^az)^i R_{i\;\;\ell}^{\;\; j\;\; k}\A^{\ell}
A_{jk} +{\rm h.c.}\] \nonumber \\ & & 
- {12\over s+\s}\lbr\[i\pp_\nu s\(F_a^{\nu\mu} - i\tF_a^{\nu\mu}\) + 
{2\pp^\mu s\over s+\s}\D_a\]\D_\mu\z^{\m}K_{i\m}(T^az)^i + {\rm h.c.}\rbr 
\nonumber \\ & & 
- 2{\pp_\rho s\pp^\nu\s\over(s+\s)}\(F^a_{\mu\nu} + i\tF^a_{\mu\nu}\)
\(F_a^{\mu\rho} - i\tF_a^{\mu\rho}\) + 40{\pp_\mu s\pp^\mu\s\over(s+\s)^2}\D + 
28i{\pp_\mu s\pp_\nu\s\over(s+\s)^2}\D^aF_a^{\mu\nu}, 
\nonumber \\ L_g &=& {(s+\s)^2\over4}\(\cW_{ab}\cbW^{ab} + \cW\cbW\) - 
{s+\s\over2}\(\cW + \cbW\)\(M^2 + \hV\) + \D^2 \nonumber \\ & &
+ {1\over2}\(\cW_{ab} + \cbW_{ab}\)\D^a\D^b  -2\(\D + {1\over3}V\)M^2 + 
{1\over3}\(M^2 + V\)\D_\mu z^i\D^\mu\z^{\m}K_{i\m} \nonumber \\ & &
- {1\over3}V^2 + {1\over6}K_{i\m}K_{j\n}\(4\D_\mu z^i\D^\mu z^j
\D_\nu\z^{\m}\D^\nu\z^{\n} + \D_\mu z^i\D^\mu\z^{\n}\D_\nu\z^{\m}\D^\nu z^j\)
\nonumber \\ & & 
- {1\over3}K_{i\m}K_{j\n}\D_\mu z^i\D^\mu\z^{\m}\D_\nu z^j\D^\nu\z^{\n}  
+ {2\over3}\(2K_{i\m}\D_\mu z^i\D^\mu\z^{\m} - V\) 
{\pp_\nu s\pp^\nu\s\over(s+\s)^2} \nonumber \\ & &
+ {\pp_\mu s\pp^\mu s\pp_\nu\s\pp^\nu\s \over(s+\s)^4}
- {2\pp_\mu s\pp_\nu\s\over(s+\s)^2}K_{i\m}\(\D^\mu z^i
\D^\nu\z^{\m} + \D^\mu\z^{\m}\D^\nu z^i\) \nonumber \\ & & 
+ \(F^a_{\rho\mu} + i\tF^a_{\rho\mu}\)\(F_a^{\rho\nu} - i\tF_a^{\rho\nu}\)
\({\pp_\nu s\pp^\mu\s\over2(s+\s)} - 
{s+\s\over4}K_{i\m}\D_\nu z^i\D^\mu\z^{\m}\), \eea 
with, instead of (4.10),
\beq K_R = K + \lll\( e^{-K}\[A_{ij}\A^{ij} -2A_i\A^i + (N_G - 4)A\A\]
-4\K_a^a - 16\D\).\eeq

Here we have considered only the standard chiral multiplet formulation 
of supergravity. Their is reason to 
believe~\cite{anomalies},~\cite{tom},~\cite{warren} that the dilaton in the
effective field theory from superstrings should be described, in fact, by a
linear multiplet, which is dual to the chiral multiple used here. 
It has been shown~\cite{frad} that a variety of classically dual theories remain
equivalent at the quantum level.  In~\cite{mk} it was observed that
once the ambiguous matrix elements (3.9) have been fixed in a supersymmetric way
that admits Pauli-Villars regularization, the axion $y$ of the dilaton
supermultiplet appears only through its dual $h^{\nu\rho\sigma}= 
\epsilon^{\nu\rho\sigma\mu}\pp_\mu y/4x^2$.  This suggests that the properly
regulated chiral supergravity theory also remains equivalent to the linear
multiplet version for the dilaton at the quantum level.  Some loop 
corrections using the
linear multiple formulation have been carried out in~\cite{quevedo}.

As shown in I, further simplifications occur\footnote{ 
The four-derivative terms of (4.4) of I should read: \bea & &
- 4\(\D_\mu\z^{\m}\D^\mu z^iK_{i\m}\)^2 + \({N\over 3}
- 7\)\D_\mu z^j\D^\mu z^i\D_\nu\z^{\m}\D^\nu\z^{\n}K_{i\n}K_{j\m} 
\nonumber \\ & & 
+ {2\over 3}\D_\mu\z^{\m}\D^\mu z^i\D_\nu\z^{\n}\D^\nu z^jK_{i\n}K_{j\m} 
- {2\over 3}\D_\rho z^i\D^\rho\z^{\m}K_{i\m}\sum_\alpha\(N_\alpha+1\)\D^\mu 
z^j\D_\mu\z^{\n}K^\alpha_{j\n} \nonumber \\ & &
+ {1\over3}\D_\mu z^i\D_\nu\z^{\m}K_{i\m}\sum_\alpha\(N_\alpha+1\)K^\alpha_{j\n}
\(\D^\mu z^j\D^\nu\z^{\n} - \D^\nu z^j\D^\mu\z^{\n}\)\nonumber \\ & & 
+ \sum_\alpha\bigg[\(N_\alpha + 1\)\(\D_\mu z^i\D^\mu\z^{\m}K^\alpha_{i\m}\)^2 
+ \(N_\alpha + 7\)\D_\mu z^j\D^\mu z^i\D_\nu\z^{\n}\D^\nu\z^{\m}
K^\alpha_{i\m}K^\alpha_{j\n} \nonumber \\ & & 
- \(N_\alpha + 1\)\D_\mu\z^{\m}\D^\mu z^i\D_\nu z^j\D^\nu\z^{\n}
K^\alpha_{j\m}K^\alpha_{i\n}\bigg].\nonumber \eea}
in specific models, such as the untwisted sectors
from orbifold compactifications where the scalar Riemann tensor is covariantly
constant and the Ricci tensor is proportional to the K\"ahler metric for each
untwisted sector.

\section{Conclusions} 
\setcounter{equation}{0}\indent

In this paper we have completed the results of I by including the gauge sector.
The complete divergent part of the 
one-loop Lagrangian, obtained from the results of this paper and from
I, will be presented elsewhere in a short communication~\cite{us3}.

Some comments on the implications and applications of our results are in order.
It has already been shown~\cite{mk} that, using the gauge fixing and expansion
procedures defined here, the one-loop quadratic divergences, as well as the 
logarithmic divergences in the flat space limit and in the absence of a dilaton,
can be regulated \`a la Pauli-Villars.  Regularization of the full supergravity 
divergences without a dilaton are under study~\cite{rey}.  An objective of
this study is to determine the extent to which, in the string theory context, a
modular invariant regularization procedure can be achieved that preserves the
continuous $SL(2,R)$ symmetry of the classical effective Lagrangian.  To obtain
the full one-loop Lagrangian, including all finite contributions, requires a
resummation of the derivative expansion.  A procedure for resummation will be
described elsewhere~\cite{rey}.

We have presented our results for one-loop corrections to the classical general
supergravity Lagrangian~\cite{cremmer,kahler} with at most two-derivative terms.
As seen in Section 5, the result simplifies considerably for the classical
effective Lagrangian derived from string theory, due to the the absence of a
potential for the dilaton and the special form of its K\"ahler potential.  These
features are modified when the effective Lagrangian includes a nonperturbatively
induced~\cite{nilles} superpotential for the dilaton and/or the 
Green-Schwarz counterterm~\cite{anomalies} that is necessary to restore modular
invariance.  The latter term destroys the no-scale nature of Lagrangians from
torus compactification and the untwisted sector of orbifold compactification,
and generally destabilizes the effective scalar potential.
However this term is of one-loop order and therefore should be
considered together with the full one-loop corrections.  An interesting
question, that will be addressed elsewhere, 
is whether these corrections can restabilize the potential.

An important unresolved issue in the construction of effective supergravity
Lagrangians for gaugino condensation is the correct form of the kinetic term
for the composite chiral multiplet that represents the lightest bound state of
the confined Yang-Mills sector.  It has recently been shown~\cite{kin}, in the 
context of both the linear and chiral multiplet formulations for the dilaton,
that such terms can be generated by higher dimension operators.  The
contribution (4.14) to the effective Lagrangian determines the leading one-loop
contribution to these operators; similar terms occur in string
theory~\cite{tomd}.  This is one example of how the determination of
loop corrections can serve as guide to the construction of such an
effective theory.

\vskip .3in
\noindent{\bf Acknowledgements.} We thank Josh Burton for his collaboration at
the initial stage of this work.  This work was supported in part by the
Director, Office of Energy Research, Office of High Energy and Nuclear Physics,
Division of High Energy Physics of the U.S. Department of Energy under Contract
DE-AC03-76SF00098 and in part by the National Science Foundation under grants
PHY-90-21139 and PHY--93--09888. 
\vskip .3in

\appendix
\noindent{\large \bf Appendix}
\def\ksubsection{\Alph{subsection}}
\def\theequation{\ksubsection.\arabic{equation}} 

     
\catcode`\@=11

\def\thesubsection{\Alph{subsection}.}
\def\thesubsubsection{\arabic{subsubsection}.}

\subsection{Dirac algebra}
\setcounter{equation}{0}\indent

We work in the Weyl representation for the Dirac matrices; for a flat metric: 
\bea \gamma_0 &=& \gamma^0 = \pmatrix{0&-1\cr -1&0\cr}, \;\;\;\; \gamma^i = 
-\gamma_i = \pmatrix{0&\sigma^i\cr -\sigma^i&0\cr},\nonumber \\ 
\gamma_5 &=& i\gamma^0\gamma^1\gamma^2\gamma^3 = 
\pmatrix{1&0\cr 0& -1\cr}, \;\;\;\; \sigma^{\mu\nu} = 
{i\over 2}[\gamma^\mu,\gamma^\nu]. \eea
To evaluate the fermion determinant, we note that an arbitrary $4\times 4$
Dirac matrix $\fM$ can be written as 
\beq \fM = RA R + LB L + RCL + LDR, \eeq
where $A,B$ contain an even number of Dirac matrices $\gamma_\nu$, $C,D$
contain an odd number, $A,B,C,D$ have no explicit $\gamma_5$-dependence, and
$L= {1\over 2}(1 -\gamma_5)$ and $R= {1\over 2}(1 +\gamma_5)$ are
the helicity projection operators.
Then $\Tr\fM = \Tr RA + \Tr LB = \Tr\eM$, where $\eM$ is the $8\times 8$ matrix
\beq \eM = \pmatrix{RA R &RCL\cr LDR&LBL\cr}, \eeq and
$ \Tr f(\fM) = \Tr f(\eM)$, where $f$ is any function that can be expanded in a
Taylor series.  Writing $\fM\equiv \fM(\gamma_5)$, we have
\bea \fM(-\gamma_5) &=& RB R + LA L + RDL + LCR,\;\;\;\;\;\;\;\;\;\;\; 
\nonumber \\
{1\over 2}\[\Tr\fM(\gamma_5)+\Tr\fM(-\gamma_5)\] &=& 
{1\over 2}\(\Tr A + \Tr B\) = {1\over 2}\Tr\pmatrix{A&C\cr D&B\cr}. \eea
Similarly, if $f$ is an arbitrary function of $\fM$,
\beq{1\over 2}\lbr\Tr f\[\fM(\gamma_5)\]+\Tr f\[\fM(-\gamma_5)\]\rbr =
{1\over 2}\Tr f(P), \;\;\;\; P = \pmatrix{A&C\cr D&B\cr}. \eeq
Setting $\displaystyle{\fM = -i\notD + M_\Theta ,\; f(\fM)=\ln\fM}$, 
(A.5) gives the trace
$T_+$ that has been evaluated previously\footnote{The contributions from the
terms $M_{\mu\nu}\sigma^{\mu\nu}$ were not fully included 
in~\cite{josh}.}{~\cite{us}--\cite{sigma}. 
To evaluate the determinant $T_-$ we define 
\beq \fM = \gamma_0\(-i\notD + M_\Theta\), \eeq 
which is a $4\times 4$ matrix in Dirac space that we write~\cite{josh} in 
terms of the $2\times 2$ Pauli $\sigma$-matrices as
\bea \fM &=& -\pmatrix{i\stA& C\cr {\tilde D}&i\sB}, 
\;\;\;\; \sigma^\mu_{\pm} = (1,
\pm \vec\sigma) , \;\;\;\; \sigma_{\pm}^{\mu\nu} = {i\over2}\(
\sigma_{\pm}^\mu\sigma_{\mp}^\nu - \sigma_{\pm}^\nu\sigma_{\mp}^\mu\),
\nonumber \\ 
\stA &=& \sigma^\mu_+d^+_\mu = \sigma^\mu_+\[\tD^+_\mu - \tL_\mu\(\sigma_-,
\sigma_+\)\], \nonumber \\ 
\sB &=& \sigma^\mu_-d^-_\mu = \sigma^\mu_-\[\tD^-_\mu - \tL_\mu\(\sigma_+,
\sigma_-\)\], \nonumber \\
C &=& m + M_{\mu\nu}\sigma_+^{\mu\nu}\equiv M(\sigma_+^{\mu\nu}), 
\;\;\;\; {\tilde D} = \m + \M_{\mu\nu}\sigma_-^{\mu\nu}\equiv 
\M(\sigma_-^{\mu\nu}) \nonumber \eea
\beq \tL_\mu\(\sigma_-,\sigma_+\) =
{L_\mu\over24}\epsilon_{\lambda\nu\rho\sigma}
\sigma_-^\lambda\sigma_+^\nu\sigma_-^\rho\sigma_+^\sigma.\eeq
The matrix elements in $\fM$ are defined, up to the $\gamma_5$ ambiguity noted 
in~\cite{mk}, in terms of those appearing in the 
fermionic part of the action (4.1) by:
\beq D_\mu = \tD_\mu + i\gamma_5L_\mu 
= iD^+_\mu R + iD^-_\mu L, \;\;\;\; M_\Theta = \M(\sigma^{\mu\nu})R + 
M(\sigma^{\mu\nu})L.\eeq
The matrix-valued derivative operator $\tD_\mu$ is defined in (A.12)
of I, the additional gaugino connection $L_\mu$ is given in (C.19) below,
 and the elements of the mass matrix $M_\Theta = \M R + M L$ are given in 
(2.16), (2.17), (A.11) and (B.10) of I, together with (C.15) below.  The tilde 
operation on $\sA,\sB,C,D$
amounts to the interchange $\sigma_+\leftrightarrow\sigma_-$.  Thus 
\bea \notA L &=& \pmatrix{0 & -\sA\cr 0 & 0\cr}, \;\;\;\;
\notA R = \pmatrix{0 & 0\cr -\stA & 0\cr}, \nonumber \\
\sA\stA &=& R\(\nottD^+ - {\notL\over24}\epsilon_{\lambda\nu\rho\sigma}
\gamma^\lambda\gamma^\nu\gamma^\rho\gamma^\sigma\)^2R = R\notD^2R, \nonumber \\
\stB\sB &=& L\(\nottD^- - {\notL\over24}\epsilon_{\lambda\nu\rho\sigma}
\gamma^\lambda\gamma^\nu\gamma^\rho\gamma^\sigma\)^2L = L\notD^2L,\eea
where the appropriate zero's in the transition from $2\times 2$ to $4\times 4$
matrices is implicit in the last two lines.  More, generally, products of
$\sigma_{\pm}^\mu$ can be converted into products of $\gamma^\mu$ by
\bea \(\sigma_+\sigma_-\)^n\sigma_+ &\to& - L\gamma^{2n+1}R, \;\;\;\;
\(\sigma_-\sigma_+\)^n\sigma_- \to - R\gamma^{2n+1}L, \nonumber \\
\(\sigma_+\sigma_-\)^n &\to& L\gamma^{2n}L, \;\;\;\;
\(\sigma_-\sigma_+\)^n \to R\gamma^{2n}R.\eea
Then defining
\bea S_{\pm} &=& {1\over 2}\[\Tr\ln\fM(M,\vec\sigma) \pm 
\Tr\ln\fM(-M,-\vec\sigma) \],\nonumber \\ 
\fM(-M,-\vec\sigma) &=& -\pmatrix{i\sA&-{\tilde C}\cr - D&i\stB} = \fM(-M,
-\gamma_5)\gamma_0,\eea
(A.9--10) immediately gives:
\bea S_+ &=& {1\over 2}\Tr\ln\[\fM(-M,-\vec\sigma)\fM(M,\vec\sigma)\] 
\nonumber \\
&=& {1\over 2}\Tr\ln\pmatrix{-R[\notD_+^2 + M\M]R & -R[i\notD^+ M - Mi\notD^-]L
\cr -L[i\notD^-\M - \M i\notD^+]R & -L[\notD_-^2 + \M M]L \cr}\nonumber \\ 
&=& {1\over 2}\Tr\ln\(-\notD^2 - M_\Theta^2 + i[\notD, M_\Theta]\)
= {1\over 2}\Tr\ln\(-\hD^2 - H_\Theta^2\).\eea
where  $\hD=\hD_\Theta$ and $H_\Theta$ are defined in (2.17)
Although the matrix in (A.12) is $8\times 8$, the helicity projection operators
$L,R$ project out half the elements, so the counting of states is unchanged
when we take the Dirac trace.  Since $\Tr\ln\cM(M) = 
\Tr\ln\cM(-M)$, we have\footnote{In~\cite{josh} it was 
incorrectly stated that $S_- =0$.} $S_{\pm} = T_{\pm}$, and (A.12) is 
equivalent to (A.5), up to the ambiguity described in~\cite{mk}: terms even and
odd in $\gamma_5$ can be interchanged using 
$\gamma_5 = (i/24)\epsilon^{\mu\nu\rho\sigma}\gamma_\mu
\gamma_\nu\gamma_\rho\gamma_\sigma$.

The next step is to cast $S_- = T_-$ in the form of (3.3) and to
take its Fourier transform to obtain an expression of the form (3.4), but 
before performing the $p$-integration we write
\bea &&\cM^{-1}[\cM(\gamma_5) - \cM(-\gamma_5)] = \cM^{-1}
\cM_0^{-1}\cM_0[\cM(\gamma_5) - \cM(-\gamma_5)] \nonumber \\ &&\qquad
= 2\(D^2 - {i\over2}\sigma_{\mu\nu}G^{\mu\nu} + iD_\mu M^\mu\)^{-1}
iD_\nu N^\nu,\eea
where $\cM_0$ is\footnote{It might seem more efficient to take instead 
$\cM_0 = \fM(-M,-\vec\sigma)$ but this form turns out to introduce a
spurious quadratic divergent term involving $M_{\mu\nu}$. To explicitly
regulate ultraviolet (or infrared) divergences, one should introduce a
regulator mass matrix $\mu_0$ and set $\cM_0 \to\cM_0 + \mu_0$; see the 
discussion in Section 3.} the matrix (A.11) with $$C = D = 0,\;\;\;\;
A_\mu = \tD^+_\mu + \tL_\mu\(\sigma_+,\sigma_-\),\;\;\;\;
 B_\mu = \tD^-_\mu + \tL_\mu\(\sigma_-,\sigma_+\),$$
\bea {1\over 2}\[\cM(\gamma_5)-\cM(-\gamma_5)\] &=& -\pmatrix{\tj &
M_I(\sigma^{\mu\nu}_+)\cr -M_I(\sigma^{\mu\nu}_-) & -\sj\cr}, \nonumber \\
{\cal J}_\mu = {i\over 2}(D_\mu^+ - D_\mu^-), \;\;\;\; & &
M_I = {1\over 2}(M - \M),\eea
and 
\beq N_\mu = \pmatrix{-R\gamma_\mu\notgg R& R\gamma_\mu M_IL\cr 
- L\gamma_\mu M_IR & L\gamma_\mu\notgg L\cr}, \;\;\;\;
M_\mu = \pmatrix{ 0 & R\gamma_\mu ML\cr 
 L\gamma_\mu\M R & 0 \cr}. \eeq
We then redefine the integrand by~\cite{mkg} 
\beq T(p,x)\to UT(p,x)U^{-1}, \;\;\;\; U = \exp\(-id\cdot{\pp\over\pp p}\)
\exp\(i\pp\cdot{\pp\over\pp p}\),\eeq
which leaves the (properly regulated) integral unchanged.  In the absence of
background space-time curvature, the $8\times 8$ matrix valued operator 
$d_\mu$ is simply
\beq d_\mu = D_\mu = {\pp\over\pp x^\mu} + a_\mu(x) .\eeq
In the presence of space-time curvature, one  has to expand~\cite{sigma}
the action at $x' = x + y$ in terms of normal coordinates, $\xi^\mu =
y^\mu + {1\over2}\gamma^\mu_{\rho\nu}(x)y^\rho y^\nu + O(\xi^3)$:
\beq d_\mu = {\pp\over\pp \xi^\mu} + a_\mu(x,\xi).\eeq
where $\gamma^\mu_{\rho\nu}$ is the affine connection, and the full connection
$a_\mu(x,\xi)$ includes terms that depend on the affine connection and its
derivatives.  The expansion of (A.13) for this case is determined 
in~\cite{sigma}.  We then obtain the expression (3.4) with 
\bea T(p,x) &=& -{1\over2}\Tr\ln\[1 + 2\Delta(x,p)p^2\R_5(x,p)\],
\nonumber \\ 
\Delta^{-1} &=& - T^{\mu\nu}\Delta_\mu\Delta_\nu + \hh + X
+ \(p^\nu + G^\nu\)P_{\mu\nu}\hM^\mu , \nonumber \\ 
\Delta_\mu &=& p_\mu + G_\mu + \delta_\mu, \qquad 
- p^2\R_5 = \(p^\nu + G^\nu\)P_{\mu\nu}\hN^\mu, 
\qquad h = -{i\over2}\sigma_{\mu\nu}G^{\mu\nu} \nonumber \\
G_{\mu} &=& \sum_{m=0}{m+1\over (m+2)!}\(-iD\cdot{\partial\over\partial p}\)^m
G_{\nu\mu}{\partial\over \partial p_\nu}, \;\;\;\; 
G_{\mu\nu} = [D_\mu,D_\nu], \nonumber \\ 
\hF &=& \sum_0^{\infty}{(-i)^n\over n!}
\left(D\cdot{\partial\over \partial p}\right)^nF, \;\;\;\; 
F=h,M^\mu,N^\mu,\quad D\cdot{\partial\over \partial p}X \equiv 
[D_\mu,X]{\partial\over \partial p_\mu}, \nonumber \\ 
P^{\mu\nu}\gamma_\nu &=& P^\mu = \gamma^\mu - 
{1\over6}r^{\mu\rho\sigma\nu}\gamma_\nu
{\partial^2\over \partial p^\rho\partial p^\sigma}
+ O\(\pp^3\over\pp p^3\),\nonumber \\ 
T^{\mu\nu} &=& g^{\mu\nu} - 
{1\over3}r^{\mu\rho\sigma\nu}{\partial^2\over \partial p^\rho\partial p^\sigma} 
- {i\over6}\nabla^\lambda r^{\mu\rho\sigma\nu}{\partial^3\over \partial p^\rho
\partial p^\sigma \partial p^\lambda} + O\(\pp^4\over\pp p^4\),\nonumber \\ 
X &=& - {r\over3} - {i\over3}\nabla^\mu r{\partial\over\partial p^\mu} + 
O\(\pp^2\over\pp p\),\nonumber \\
\delta_\mu &=& {i\over9}\(\nabla_\mu r_{\rho\nu} - \nabla_\nu r_{\rho\mu}\)
{\partial^2\over \partial p_\nu\partial p_\rho}
+ O\(\pp^3\over\pp p^3\),\eea 
Finally we write $ \Delta^{-1} = -p^2(1 + \R)$ and expand 
\beq \Delta = (1 + \R)^{-1}(-p^{-2}) = 
\sum_{n=0}(-\R)^n(-p^{-2}) \eeq
to obtain the expression (3.8), where we have set $\mu_0=0$.

Once all these manipulations have been performed we can simplify the
expression for the fermion connection by using simply
\beq D^{\pm}_\mu = \tD^{\pm}_\mu + i\gamma_5L_\mu. \eeq 
The point is that the part
of the gaugino connection arising from the dilaton has been included in the 
``vector'' ($\cJ^V_\mu$):
\beq\pp_\mu + \cJ^V_\mu = {1\over2}\(D^+_\mu + D^-_\mu\) = {1\over2}\(\tD^+_\mu 
+ \tD^-_\mu\) + i\gamma_5L_\mu,\eeq
rather than in the ``axial vector'' ($\cJ_\mu$) part of the connection.

We conclude this appendix by listing some Dirac traces that are useful in 
the evaluation of $T_-$ and of the ghostino and fermion determinants:
$$ \Tr\gamma_5\gamma^\mu\gamma^\nu\gamma^\rho\gamma^\sigma =
-4i\epsilon^{\mu\nu\rho\sigma}, \;\;\;\; 
\Tr\gamma_5\sigma^{\alpha\beta}\sigma^{\mu\nu} = 
4i\epsilon^{\alpha\beta\mu\nu}, \;\;\;\; \epsilon^{0123} = -g^{-1}
\epsilon_{0123} = g^{-{1\over2}}, $$
$$ \leqno{\Tr(\gamma_5\gamma^\alpha\gamma^\beta\gamma^\gamma\gamma^\delta
\gamma^\epsilon\gamma^\zeta) = -4i[\epsilon^{\gamma\delta\epsilon\zeta}
g^{\alpha\beta} + \epsilon^{\alpha\beta\gamma\delta}g^{\epsilon\zeta}} $$
$$+ \epsilon^{\alpha\beta\epsilon\gamma}g^{\delta\zeta} + 
\epsilon^{\alpha\beta\gamma\zeta}g^{\delta\epsilon} + 
\epsilon^{\alpha\beta\delta\epsilon}g^{\gamma\zeta}
+ \epsilon^{\alpha\beta\zeta\delta}g^{\gamma\epsilon}
 + \epsilon^{\alpha\beta\epsilon\zeta}g^{\gamma\delta}], $$
$$ \Tr(\gamma_5\sigma^{\alpha\beta}\gamma^\gamma
\sigma^{\delta\epsilon}\gamma^\zeta) = 
4i[\epsilon^{\alpha\beta\gamma\delta}g^{\epsilon\zeta} 
+ \epsilon^{\alpha\beta\epsilon\gamma}g^{\delta\zeta} + 
\epsilon^{\alpha\beta\delta\epsilon}g^{\gamma\zeta}
+ \epsilon^{\alpha\beta\zeta\delta}g^{\gamma\epsilon}
 + \epsilon^{\alpha\beta\epsilon\zeta}g^{\gamma\delta}], $$
$$ \Tr(\gamma_5\sigma^{\alpha\beta}\sigma^{\gamma\delta}
\gamma^\epsilon\gamma^\zeta) = 
4i[\epsilon^{\alpha\beta\gamma\delta}g^{\epsilon\zeta} 
+ \epsilon^{\alpha\beta\epsilon\gamma}g^{\delta\zeta} 
 + \epsilon^{\alpha\beta\gamma\zeta}g^{\delta\epsilon} + 
\epsilon^{\alpha\beta\delta\epsilon}g^{\gamma\zeta}
+ \epsilon^{\alpha\beta\zeta\delta}g^{\gamma\epsilon}], $$
$$ \Tr\sigma_{\rho\sigma}\sigma^{\mu\nu}F_a^{\rho\sigma}F^b_{\mu\nu} = 
8F_a^{\mu\nu}F^b_{\mu\nu}, \;\;\;\; \Tr\sigma_{\rho\sigma}\sigma_{\mu\nu}
\sigma_{\lambda\tau}F_a^{\rho\sigma}F_b^{\mu\nu}F_c^{\lambda\tau} = 32i
F_a^{\mu\nu}F_{b\mu\rho}F_{c\;\;\;\nu}^\rho, $$
\bea & & \Tr\(\sigma\cdot A\sigma\cdot B\sigma\cdot C\sigma\cdot D\)
= 16\[A^{\mu\nu}B^{\rho\sigma}C_{\mu\nu}D_{\rho\sigma} + 
(A\cdot B)(C\cdot D) + A^{\mu\nu}(B\cdot C)D_{\mu\nu}\] \nonumber \\ & & \qquad
+ 64\(A^{\mu\nu}B_{\mu\rho}C^{\rho\sigma}D_{\nu\sigma} -
A^{\mu\nu}B_{\mu\rho}C_{\nu\sigma}D^{\rho\sigma}
- A^{\mu\nu}B^{\rho\sigma}C_{\mu\rho}D_{\nu\sigma}\), \nonumber \eea \bea
\Tr\(\gamma^\mu\sigma\cdot A\gamma^\nu\sigma\cdot B\) &=& 8\(g^{\mu\nu}
A_{\rho\sigma}B^{\rho\sigma} + 2A^{\mu\rho}B_\rho^{\;\;\;\nu}
+ 2A_\rho^{\;\;\;\nu}B^{\mu\rho}\),\;\;\;\;\;\;\nonumber \\
\Tr\(\gamma^\mu\gamma^\nu\sigma\cdot A\sigma\cdot B\) &=& 8\(g^{\mu\nu}
A_{\rho\sigma}B^{\rho\sigma} + 2A^{\mu\rho}B_\rho^{\;\;\;\nu}
- 2A_\rho^{\;\;\;\nu}B^{\mu\rho}\),\;\;\;\;\;\; \nonumber \\
\Tr\(Z_{\mu\nu}\gamma^\mu\sigma\cdot A\gamma^\nu\sigma\cdot B\gamma_5\) &=&
8ir^\mu_\nu\(\tA^{\nu\rho}B_{\mu\rho} - A^{\nu\rho}\tB_{\mu\rho}\),\eea
where $\sigma\cdot A = \sigma_{\mu\nu}A^{\mu\nu},\;etc.$, and
$Z_{\mu\nu} = {1\over4}\gamma^\rho\gamma^\sigma r_{\rho\sigma\nu\mu}$
is the field strength arising from the spin connection (note that
$\gamma_\mu\gamma_\nu Z^{\mu\nu} = {1\over2}r$).  To evaluate the last
trace in (A.23) we used the relations (B.14) and (C.25).

\subsection{Relations among operators}
\setcounter{equation}{0}\indent

In this appendix we derive relations among the various operators that appear 
in the traces needed to evaluate the one-loop effective action.
We adopt the gauge sign conventions of~\cite{cremmer},~\cite{fz}:
\bea \D_\mu &=& \nabla_\mu + iA_\mu, \;\;\;\; A_\mu = T_aA^a_\mu, \;\;\;\; 
T^{\ibar}_{a\bj} = \(T^i_{aj}\)^*, \nonumber \\
\D_\mu z^i &=& \pp_\mu z^i + iA^a_\mu(T_az)^i, \;\;\;\;
\D_\mu\z^{\m} = \pp_\mu\z^{\m} - iA^a_\mu(T_a\z)^{\m} ,\nonumber \\
F_{\mu\nu} &=&  {1\over i}[\D_\mu,\D_\nu] = \nabla_\mu A_\nu - \nabla_\nu A_\mu
+ i[A_\mu,A_\nu], \nonumber \\
F^a_{\mu\nu} &=& \nabla_\mu A^a_\nu - \nabla_\nu A^a_\mu
- c^a_{\;\;bc}A^b_\mu A^c_\nu. \eea
Our other conventions and notations are given in Appendix A of I.  

We first consider constraints on covariant scalar derivatives that follow from
gauge invariance.  We define 
\bea
\K_{ab} &=& {1\over x}K_{\m j}(T_a\z)^{\m}(T_bz)^j, \;\;\;\;
\D_a = K_i(T_az)^i, \;\;\;\; \D = {1\over 2x}\D_a\D^a, \nonumber \\
f_{ab}(z) &=& \delta_{ab}f(z), \;\;\;\; f = x + iy.
\eea
The classical scalar potential is $\hV + \D$, where $\hV$ has been defined 
in I. It follows from the gauge invariance of the K\"ahler potential $K$ that:
\bea
\delta_a K &=& K_i(T_az)^i - K_{\m}(T_a\z)^{\m}=0, \;\;\;\; D_iD_j\D_a = 
D_{\m}D_{\n}\D_a = 0, \nonumber \\
K_{i\m}D_{\n}(T_a\z)^{\m} &=& K_{j\n}D_i(T_az)^j, \;\;\;\; 
D^i(T_a\z)^{\m} = D^{\m}(T_az)^i, \nonumber \eea
\beq K_{ij}(T_az)^j + K_j(T_a)^j_i = K_{i\m}(T_a\z)^{\m}, \;\;\;\;
D_kD_j(T_az)^i = - R_{j\m k}^{\;\;\;\;\;\; i}(T_a\z)^{\m},
\end{equation}
where $K_{ij} = \pp_i\pp_j K = \pp_i K_j$, and
the second and third lines follow from the first by taking successive 
scalar derivatives.  Here $\pp_I = \pp/\pp z^I,\; I = i,\ibar,\; D_I$ is the
reparameterization covariant scalar derivative, and $R_{i\m j\n}$ is the 
K\"ahler curvature tensor.  Indices are lowered and raised, respectively, 
with the K\"ahler metric $K_{i\m}$ and its inverse $K^{i\m}$.
Similarly, it follows from the gauge invariance of $f$ that
\bea
\delta_a f &=& f_i(T_az)^i =0, \nonumber \\
f_{ij}(T_az)^j\f^i &=& -f_iD_j(T_az)^i\f^j, \;\;\;\; 
f_{ij}(T_az)^j(T_bz)^i = -f_i(T_az)^jD_j(T_bz)^i, \nonumber \\
f_i\f_{\m}D^i(T_a\z)^{\m} &=& - f^{\n}\f_{\n\m}(T_a\z)^{\m} 
= - \f^if_{ij}(T_az)^j,\;\;\;\; f_{ij} = D_iD_j f,
\eea
and from the gauge invariance of the superpotential $W$ that
\bea
A_i(T_az)^i &=& A_{\m}(T_a\z)^{\m} = \D_a A, \nonumber \\
A_{ij}(T_az)^i + A_iD_j(T_az)^i &=& \D_a A_j + K_{j\m}(T_a\z)^{\m}A, 
\nonumber \eea
\beq A_{ijk}(T_az)^i + A_{ij}D_k(T_az)^i + A_{ik}D_j(T_az)^i + 
A_iD_kD_j(T_az)^i$$
$$= \D_aA_{jk} + K_{j\m}(T_a\z)^{\m}A_k + K_{k\m}(T_a\z)^{\m}A_j.
\end{equation}
The tensors $A_{i_1\cdots i_n}$ are reparameterization invariant
covariant derivatives~\cite{us} of $A = e^KW$. 
Using (B.3) and the definitions (B.2) we obtain
\begin{equation}
\K_{ab} - \K_{ba} = {i\over x}c_{abc}\D^c, \;\;\;\; 
\K^{ab}(\K_{ab} - \K_{ba}) = -{1\over 2x^2}C_{G}^{(a)}\D_a\D^a,
\end{equation}
where $C_{G}^{(a)}$ is the Casimir in the adjoint representation, $c_{abc}$
are the structure constants of the gauge group, and
\begin{equation}
(T_bz)^iD_i(T_az)^j = (T_az)^iD_i(T_bz)^j +ic_{abc}(T^cz)^j,$$
$$\D_b K_{\m j}(T_a\z)^{\m}(T^az)^iD_i(T^bz)^j = 
\D_b K_{\m j}(T_a\z)^{\m}(T^bz)^iD_i(T^az)^j - {1\over 2}C_{G}^{(a)}\D_a\D^a.
\end{equation}
Combining (B.3) and (B.5) we obtain
\bea
A_iD^i(T_a\z)^{\m} &=& A_iD^{\m}(T_az)^i = -A^{\m}_i(T_az)^i + A^{\m}\D_a 
+ A(T_a\z)^{\m}, \nonumber \\
\A^iD_i(T_az)^k &=& \A_{\n}D^{\n}(T_az)^k = \A_{\n}D^k(T_a\z)^{\n} 
= -\A^k_{\n}(T_a\z)^{\n} + \A^k\D_a + \A(T_az)^k, \nonumber \eea
\bea &&\D^a\A^{jk}A_{ij}D_k(T_az)^i = -{1\over 2}\A^{jk}A_{ijk}(T_az)^i\D^a 
\nonumber \\ &&\qquad + {1\over 2}R_{j\m k}^{\;\;\;\;\;\; i}
\A^{jk}A_i\D^a(T_a\z)^{\m} + x\D A_{ij}\A^{ij} + \D^a(T_a\z)^{\m}\A^i_{\m}A_i.
\eea 

To evaluate the one-loop effective action, we find it convenient to
introduce the scalar field reparameterization covariant derivatives
of the variable $\rho$, defined as the squared gauge coupling:
\bea \rho = {1\over x} &=& g^2, \;\;\;\; \rho_i = D_i\rho =
-{f_i\over2x^2}, \;\;\;\; \rho^i = K^{i\m}D_{\m}\rho = K^{i\m}\rho_{\m},  
\nonumber \\ 
\rho_{ij} &=& D_iD_j\rho = -{1\over 2x^2}\(f_{ij} - {f_if_j\over x}\), 
\nonumber \\
D_{\m}\rho_i = \rho_{\m i} &=& -{1\over x}\f_{\m}\rho_i =
2x\rho_{\m}\rho_i , \nonumber \\
D_j\(x^2\rho_i\rho^i\) &=& x^2\rho^i\rho_{ij}, \;\;\;\; 
D_{\m}\(x^2\rho_i\rho^i\) = x^2\rho^i_{\m}\rho_{i}, \nonumber \\
D_jD_k\(x^2\rho_i\rho^i\) &=& x^2\rho^i\rho_{ijk}, \;\;\;\;  etc.,
\nonumber \\
f_{\m ij} = R^k_{i\m j}f_k &=& -2x^2\rho_{\m ij} -2x\f_{\m}\rho_{ij}
- {f_if_j\f_{\m}\over2x^2}.  \eea
It follows from $[D_{\m},D_i](x^2\rho_i\rho^i) = 0$ that
\beq \f^k\rho^j_{\;\;ki} +{1\over x}\f^k\f^j\rho_{ki} =
f_k\rho_i^{\;\;kj} +{1\over x}f_kf_i\rho^{kj}. \eeq
In addition we introduce the variable
\beq a = A + {\f^i\over 2x}A_i = e^{K/2}\(\m_\psi - \m_\lambda\), \;\;\;\;
a_{i_1\cdots i_n} = D_{i_1}\cdots D_{i_n}a. \eeq
The variables $a,\rho_{ij}$ and $1-x^2\rho^i\rho_i$, and all covariant
derivatives thereof, vanish for effective supergravity theories obtained 
from superstrings in the classical limit: $f(z) = s, \; K = -\ln(s+\s) +
G(z,\z\ne s,\s),\; W_s = 0.$

We will also need the following identities involving the Yang-Mills field
strength and the space-time curvature.  It follows from manipulating products 
of the antisymmetric tensor $\epsilon_{\mu\nu\rho\sigma}$ that 
\bea \tM_1^{\mu\nu}M^2_{\mu\rho} &=& {1\over 2}g^\nu_\rho M_1^{\mu\sigma}
\tM^2_{\mu\sigma} - M^1_{\mu\rho}\tM_2^{\mu\nu}, \;\;\;\;  \tM^i_{\mu\nu} = 
{1\over 2}\epsilon_{\mu\nu\rho\sigma}M_i^{\rho\sigma},
\nonumber \\
\tF^a_{\mu\nu}F_b^{\mu\nu}\tF^b_{\rho\sigma}F_a^{\rho\sigma} &=& - 
(F^a_{\mu\nu}F_b^{\mu\nu})^2 - \(F^a_{\mu\nu}F_a^{\mu\nu}\)^2 + 
4F^a_{\rho\mu}F_a^{\rho\nu}F_b^{\sigma\mu}F^b_{\sigma\nu},\nonumber \\
(\tF^a_{\mu\nu}F_a^{\mu\nu})^2 &=& -2(F^a_{\mu\nu}F_b^{\mu\nu})^2
+ 4 F^a_{\mu\nu}F_{a\rho\sigma}F^{b\mu\rho}F_b^{\nu\sigma}, \eea
where $M^i_{\mu\nu}$ is any antisymmetric tensor-valued operator. 
Using the first of these gives 
\bea 
\Tr A^{\mu\nu}B_{\mu\rho}\tC_{\nu\sigma}D^{\sigma\rho} &=& {1\over4}\Tr
\[(\tD\cdot A)(B\cdot C) - (A\cdot B)(\tC\cdot D)
- \tA^{\mu\nu}B_{\rho\sigma}C_{\mu\nu}D^{\rho\sigma}\],
\nonumber \\
\Tr A^{\mu\nu}B_{\mu\rho}\tC^{\sigma\rho}D_{\sigma\nu} &=& {1\over4}\Tr
\[(A\cdot B)(\tC\cdot D) + (D\cdot A)(\tB\cdot C) 
- A^{\mu\nu}\tB_{\rho\sigma}C_{\mu\nu}D^{\rho\sigma}\],
\nonumber \\
\Tr A^{\nu\sigma}B^{\rho\mu}\tC_{\mu\nu}D_{\rho\sigma} &=& {1\over4}\Tr
\[(\tA\cdot B)(C\cdot D) - (D\cdot A)(\tB\cdot C)  
- \tA^{\mu\nu}B_{\rho\sigma}C_{\mu\nu}D^{\rho\sigma}\], \nonumber \\
\tF^a_{\mu\nu}F_a^{\mu\rho} &=& {1\over 4}g^\rho_\nu\tF^a_{\mu\sigma}
F_a^{\mu\sigma}.\eea
It follows from the the symmetry properties of the space-time Riemann tensor
that
\beq r_{\rho\sigma\mu\nu}F_a^{\nu\sigma}F^{a\mu\rho} =
{1\over2}r_{\mu\nu}^{\;\;\;\;\;\rho\sigma}F_a^{\mu\nu}F^a_{\rho\sigma},\eeq
and, using (B.12) with $M_1 = F, \;M_2 = \tF, \;\tM_2 = -F$,
\bea r_{\rho\sigma\mu\nu}\tF_a^{\nu\sigma}\tF^{a\mu\rho} &=& 
{1\over2}r_{\mu\nu}^{\;\;\;\;\;\rho\sigma}\tF_a^{\mu\nu}\tF^a_{\rho\sigma} 
\nonumber \\ &=& 2r^\mu_\nu F^a_{\mu\rho}F_a^{\nu\rho} - 
{1\over 2}rF^a_{\mu\rho}F_a^{\mu\rho} - {1\over 2}
r_{\mu\nu}^{\;\;\;\;\;\rho\sigma}F_a^{\mu\nu}F^a_{\rho\sigma}.\eea
In addition: \bea
F^a_{\mu\nu}[\D^\mu,\D_\rho]F_a^{\rho\nu} &=& c_{abc}F^a_{\mu\nu}F^{b\mu\rho}
F^{c\nu}_{\;\;\;\;\rho} \nonumber \\ & &
+ r^\mu_\nu F^a_{\mu\rho}F_a^{\nu\rho} - {1\over2}
r_{\mu\nu}^{\;\;\;\;\;\rho\sigma}F_a^{\mu\nu}F^a_{\rho\sigma}.\eea

It is convenient to isolate terms that do not contribute to the S-matrix,
using the classical equations of motion:  
\bea g^{-{1\over2}}\L_I &=& -K_{IJ}D_\mu\D^\mu z^J - \hV_I - 
{1\over x}\D_a(T^az)^JK_{IJ} - {1\over 2}f_I\cases{\cW\cr\cbW\cr}, 
\;\;\;\; I,J=\cases{i,\bj\cr\ibar,j\cr}, \nonumber \\ 
(xg)^{-{1\over2}}\L_{a\mu} &=& (xg)^{-{1\over2}}g_{\mu\nu}
{\pp\L\over\pp A^a_\nu} = \D''^\nu\F_{a\nu\mu} + 
\tcF_{a\nu\mu}{\pp^\nu y\over x} \nonumber \\ & & \quad + 
{i\over\sqrt{x}}K_{i\m}\(\D_\mu\z^{\m}(T_az)^i - \D_\mu z^i(T_a\z)^{\m}\).
\eea 
The first of these gives, in particular $(M_\psi^2 = m_\psi\m_\psi,\;
M_\lambda^2 = m_\lambda\m_\lambda)$:
\bea
{f_i\over\sqrt{g}}\L^i = \({\f^i\over\sqrt{g}}\L_i\)^* &=& -\nabla^2 x - 
i\nabla^2 y - 2x^4\rho^i\rho_i\cbW 
+{1\over x}\(\pp_\nu x+i\pp_\nu y\)\(\pp^\nu x+i\pp^\nu y\) 
\nonumber \\ & & - 2x^2\rho_{ij}\D_\mu z^j\D^\mu z^i + 2xe^{-K}\(2\aa A - 
\aa^iA_i\) + 2x\(\hV + M_\psi^2 - M_\lambda^2\), \nonumber \eea \bea 
-{\pp_\mu x\over\sqrt{g}x}\D^\mu z^i\L_i + {\rm h.c.} &=&
\({\pp_\mu x\over x}\D^\mu z^iK_{i\m}D^\nu\D_\nu\z^{\m} + {\rm h.c.}\) 
- {\nabla^2x\over x}V \nonumber \\ & & + {\pp_\mu x\pp^\mu x\over x^2}\(V +
{x\over4}F^2\) + {\pp_\mu y\pp^\mu x\over 4x}F\tF + {\rm total\;
derivative}, \nonumber \\ 
{a + bx^2\rho^i\rho_i\over x\sqrt{g}}\D^a(T_az)^I\L_I &=& 
{a + bx^2\rho^i\rho_i\over x}\Bigg(
2K_{j\m}\D^\mu z^i\D_\mu\z^{\m}\D^aD_i(T_az)^j + 8x\D M_\psi^2
\nonumber \\ & & \quad
- 2\D^a\D^b\K_{ab} -e^{-K}\[\D^a(T_az)^iA_{ij}\A^j + {\rm h.c.}\]
\nonumber \\ & & \quad + \lbr K_{i\n}K_{j\m}\D^\mu z^j
(T_a\z)^{\m}\[(T_az)^i\D_\mu\z^{\n} + (T_a\z)^{\n}\D_\mu z^i\] + {\rm h.c.}\rbr
\nonumber \\ & & \quad
- {\pp_\mu x\over x}\D^aK_{j\m}\[\D^\mu z^j(T_a\z)^{\m} + 
(T_az)^j\D^\mu\z^{\m}\]\Bigg) \nonumber \\ & & 
+ bx\D^a\[\D_\mu z^k\rho^j\rho_{kj}K_{i\m}\(\D_\mu\z^{\m}(T_az)^i + 
\D_\mu z^i(T_a\z)^{\m}\) + {\rm h.c.}\] \nonumber \\ & & 
+ {\rm total\;derivative}. \eea
We absorb a part of the one loop correction into the K\"ahler potential;
a shift $\delta K$ in the K\"ahler potential gives a shift $\Delta_{\delta K}\L
$ in the Lagrangian:
\bea {1\over\sqrt{g}}\Delta_{\delta K}\L &=& - \delta K\hV + 
\delta K_{i\m}\(e^{-K}\A^iA^{\m} + \D_\mu z^i\D^\mu\z^{\m}\) \nonumber \\ & &
- \lbr\delta K_i\[e^{-K}\A^i A + {1\over2x}\D_a(T^az)^i\] + {\rm h.c.}\rbr.\eea
Taking $\delta K = \D$, the last equation in (B.18) can be written as 
\bea {a + bx^2\rho^i\rho_i\over x\sqrt{g}}\D^a(T_az)^I\L_I &=& 
\(a + bx^2\rho^i\rho_i\)\Bigg({2\over\sqrt{g}}\Delta_{\D}\L
+ 2\D\[e^{-K}a\aa - 3M_\psi^2 -3M_\lambda^2 - \hV\]
\nonumber \\ & & \qquad + \[K_{i\n}K_{j\m}\D_\mu z^i\D^\mu z^j(T_a\z)^{\m}
(T_a\z)^{\n} + {\rm h.c.}\] \nonumber \\ & & \qquad 
+ i{\pp_\mu y\over x^2}\D^a\[K_{i\m}(T_az)^i\D_\mu\z^{\m} - {\rm h.c.}\]
\nonumber \\ & & \qquad - {1\over x^2}\D\[\pp_\mu x\pp^\mu x + 
\pp_\mu y\pp^\mu y\]\Bigg) \nonumber \\ & & 
+ bx\D^a\[\D_\mu z^k\rho^j\rho_{kj}K_{i\m}\(\D_\mu\z^{\m}(T_az)^i + 
\D_\mu z^i(T_a\z)^{\m}\) + {\rm h.c.}\] \nonumber \\ & & 
+ {\rm total\;derivative}. \eea

\subsection{Matrix Elements and Supertraces}
\setcounter{equation}{0}\indent

In this Appendix we list matrix elements of operators appearing in Eqs.
(4.2--4.5) and traces needed to evaluate the divergent contributions to the
one-loop effective action (4.6). Notation and conventions are defined in 
Appendix A of I, and
the relevant part of the tree Lagrangian~\cite{cremmer},~\cite{kahler} 
is\footnote{In I we defined $\epsilon^{0123} = 1$; here we denote by 
$\epsilon^{\mu\nu\rho\sigma}$ the covariantly constant tensor--see (A.23).
With this definition there is no factor $g^{-{1\over2}}$ multiplying the $F\tF$
term in the Lagrangian. See also footnote 1.} 
\bea 
{1\over\sqrt{g}}\L(g,K,f) &=& {1\over 2}r + K_{i\m}\D^\mu z^i\D_\mu\z^{\m}
- {x\over 4}F_{\mu\nu}F^{\mu\nu} - 
{y\over 4}{\tilde F}_{\mu\nu}F^{\mu\nu} - V \nonumber \\
& & + {ix\over 2}\bl\notD\lambda + iK_{i\m}\(\bc_L^{\m}\notD\chi_L^i +
\bc_R^i\notD\chi_R^{\m}\) \nonumber \\ & & 
+ e^{-K/2}\({1\over 4}f_i\A^i\bl_R\lambda_L - A_{ij}\bc^i_R\chi^j_L
+ {\rm h.c.}\) \nonumber \\ & & 
+ \(i\bl^a_R\[2K_{i\m}(T_a\z)^{\m} - {1\over 2x}f_i\D_a -
{1\over 4}\sigma_{\mu\nu}F^{\mu\nu}_af_i\]\chi^i_L + {\rm h.c.}\) 
\nonumber \\ & & 
+ \L_\psi + {\rm four-fermion \; terms}, \nonumber \\
{1\over\sqrt{g}}\L_\psi &=& 
{1\over 4}\bps_\mu\gamma^\nu(i\notD + M)\gamma^\mu\psi_\nu -
{1\over 4}\bps_\mu\gamma^\mu(i\notD + M)\gamma^\nu\psi_\nu 
-\bigg[{x\over 8}\bps_\mu\sigma^{\nu\rho}\gamma^\mu\lambda_aF^a_{\nu\rho}
\nonumber \\ 
 & & + \bps_\mu\notcD \z^{\m}K_{i\m}\gamma^\mu L\chi^i -
{1\over 4}\bps_\mu\gamma^\mu\gamma_5\lambda^a\D_a 
+ i\bps_\mu\gamma^\mu L\chi^im_i + {\rm h.c.}\bigg], \nonumber \\ 
\M &=& (M)^{\dag} = e^{K/2}\(WR + \W L\), \;\;\;\; m_i = e^{-K/2}A_i.\eea 

If we define
\begin{equation}
\STr F = \Tr F_\Phi - {1\over 4}\btr F_\Theta - 2\Tr F_{gh} + 2\Tr F_{Gh},
\;\;\;\; -{i\over2}T_- = \sqrt{g}\lll T, \end{equation}
where $\btr F_\Theta$ is defined below [see (C.24)],
the effective Lagrangian (4.2) is \begin{equation}
{1\over\sqrt{g}}\L_1 = -{\Lambda^2\over 32\pi^2}\STr H 
 + \lll\[\STr\({1\over 2}H^2 - {1\over 6}rH + {1\over 12}\hG_{\mu\nu}
\hG^{\mu\nu}\) + T\],\end{equation}
In the following subsections we list the matrix elements that were not included
in I; the subscript 
$0$ refers to the contributions without the Yang-Mills sector that are given in
Appendix B of I, except that ordinary derivatives are replaced by gauge 
covariant derivatives.\footnote{In (B.21) of I ${1\over2}\STr H^2$
should be modified as follows: the last term in the first line should be
multiplied by $e^K$, the term $-{1\over2}re^{-K}A_{ij}\A^{ij}$ should be added, 
and the third and forth lines from the bottom should read: 
\bea & &  + {N -47\over 4}\D_\mu z^j\D^\mu z^i\D_\nu\z^{\m}\D^\nu\z^{\n}
K_{i\n}K_{j\m} 
- {N + 17\over 4}\D_\mu\z^{\m}\D^\mu z^i\D_\nu\z^{\n}\D^\nu z^jK_{i\n}K_{j\m} 
\nonumber \\ & & 
+ {1\over2}\D_\mu z^i\D_\nu\z^{\m}K_{i\m}R_{j\n}\(\D^\mu z^j\D^\nu\z^{\n}
- \D^\nu z^j\D^\mu\z^{\n}\) \nonumber. \eea  In addition, the term 
$ - {1\over6}\D_\mu z^i\D_\nu\z^{\m}K_{i\m}R_{j\n}\(\D^\mu z^j\D^\nu\z^{\n}
- \D^\nu z^j\D^\mu\z^{\n}\) $ should be added to the right hand side of 
${1\over 12}\STr G_{\mu\nu}G^{\mu\nu} $ in the same equation. }

The contributions to STr$H$ from each supermultiplet have been given
in~\cite{mk}; below we list the analogous contributions to STr$H^2$ and
STr$G^2$; we drop all total derivatives.
\subsubsection{Boson matrix elements}

As in~\cite{noncan} we rescale the quantum gauge fields: 
$A_\mu = \sqrt{x}\cA_\mu.$  Then the operator $H_\Phi$ can be expressed as
\bea Z_\Phi H_\Phi &=& H + X + Y - N - S - K, \nonumber \\
\Phi^T Z_\Phi H_\Phi\Phi &=& z^IH_{IJ}z^J + h^{\mu\nu}X_{\mu\nu,\rho\sigma}
h^{\rho\sigma} + 2h^{\mu\nu}Y_{\mu\nu I}z^I - \hcA^\mu N_{\mu\nu}\hcA^\nu 
\nonumber \\ & & - 2\hcA^\mu S_{\mu I}z^I - 2h^{\mu\nu}K_{\mu\nu,\rho}\hcA^\rho,
\eea 
with, in addition to the matrix elements of $Z_\Phi$ given in I,
\begin{equation}
Z_{i,a\mu} = Z_{\mu\nu,a\rho} = 0, \;\;\;\; Z_{a\mu,b\nu} = -g_{\mu\nu}
\delta_{ab}.
\end{equation}
Using the results of~\cite{noncan} and Section 2 above, the elements of 
$H,X,Y$ are modified with respect to those given in (B.3) of I by\footnote{The
Lorentz indices in $U_{IJ}$ and ${\cal{R}}_{IJ}$ in Eq.(B.3) of I 
should be contracted.}
\bea H_{IJ} &=& (H_0)_{IJ} +\D_{IJ} + q^a_Iq_{aJ} + v_{IJ} - 
\(V_\mu V^\mu\)_{IJ}, \;\;\;\; 
q_a^i = -{i\over \sqrt{x}}(T_az)^i, \nonumber \\ 
q^a_i &=& {i\over \sqrt{x}}(T^a\z)^{\m}K_{i\m}, \;\;\;\;
v_{i\m} = v_{\m i} = \(V_\mu V^\mu\)_{i\m} = \(V_\mu V^\mu\)_{\m i} = 0, 
\nonumber \\ 
\(V_\mu V^\mu\)_{IJ} &=& {f_If_J\over xf_{IJ}}v_{IJ}
= {1\over 8x^2}f_If_J\(\F^{\mu\nu}_a\F^a_{\mu\nu}\mp 
i\tcF^{\mu\nu}_a\F^a_{\mu\nu}\), \nonumber \eea
\bea v_{IJ} - \(V_\mu V^\mu\)_{IJ} = -{x\over 4}\rho_{IJ}\(\F^{\mu\nu}_a
\F^a_{\mu\nu}\mp i\tcF^{\mu\nu}_a\F^a_{\mu\nu}\),
\;\;\;\;I,J= \cases{i,j\cr\ibar,\bj\cr}, \nonumber \eea
\bea Y_{\mu\nu I} &=& -{1\over2}\(D_\mu\D_\nu + D_\nu\D_\mu\)K_{IJ}z^J 
-{1\over 8}f_IF^a_{\mu\rho}F_{a\nu}^{\;\;\;\rho} \pm {i\over 32}g_{\mu\nu}
f_IF_a^{\sigma\rho}\tF^a_{\sigma\rho},\;\;\;\; I,J =\cases{i,\bj\cr\ibar,j\cr},
\nonumber \\ 
X_{\mu\nu,\rho\sigma} &=& (X_0)_{\mu\nu,\rho\sigma} -2P_{\mu\nu,\rho\sigma}\D
+ {1\over 2}P_{\mu\nu,\rho\sigma}\F^a_{\lambda\tau}
\F_a^{\lambda\tau} + {1\over 4}\(\F^a_{\mu\rho}\F_{a\nu\sigma} +
\F^a_{\nu\rho}\F_{a\mu\sigma}\) \nonumber \\ & &
- {1\over 16}\(\F^a_{\mu\lambda}\F_{a\rho}^{\;\;\;\;\lambda}g_{\nu\sigma} +
\F^a_{\nu\lambda}\F_{a\rho}^{\;\;\;\;\lambda}g_{\mu\sigma}
+ \F^a_{\mu\lambda}\F_{a\sigma}^{\;\;\;\;\lambda}g_{\nu\rho} +
\F^a_{\nu\lambda}\F_{a\sigma}^{\;\;\;\;\lambda}g_{\mu\rho}\), \eea
where $f_I\equiv f_i(\f_{\ibar})$ for $I = i(\ibar)$, {\it etc.}.
The potential $V=\hV+\D$ now includes the D-term $\D$ defined in (B.2) above:
\bea 
\D_i &=& -{1\over 2x}f_i\D + {1\over x}\D_a K_{i\m}(T^a\z)^{\m}, \nonumber \\
\D_i^j &=& {1\over 2x^2}f_i\f^j\D -{1\over 2x^2}f_i\D_a(T^az)^j
-{1\over 2x^2}\f^j\D_a K_{i\n}(T^a\z)^{\n} \nonumber \\ & &
+ {1\over x} (T_az)^j K_{i\n}(T^a\z)^{\n}
+ {1\over x}\D_a D_i(T^az)^j, \nonumber \\
\D_{ij} &=& x\rho_{ij}\D 
-{1\over 2x^2}\D_a(f_i K_{j\m} + f_j K_{i\m})(T^a\z)^{\m} \nonumber \\ & &
+ {1\over x}K_{j\m}(T^a\z)^{\m} K_{i\n}(T_a\z)^{\n}.\eea
The additional nonvanishing elements of $Z_\Phi H_\Phi$:
are $-N_{a\mu,b\nu},\;S_{a\mu,I}$, and $K_{\mu\nu,a\rho}$, with\footnote{
In~\cite{josh},~\cite{noncan}, there is an additional graviton-gauge mass term
$Q_{\mu\nu,a\rho}$; this term drops out when the prescription (2.10) is 
adopted.}
\bea N_{a\mu,b\nu} &=& g_{\mu\nu}\(\K_{ab} + \K_{ba}
-{1\over2}\F_{a\rho\sigma}\F_b^{\rho\sigma}\) +2c_{abc}F^c_{\mu\nu}
+{1\over2}\(5\F_{a\mu\rho}\F_{b\nu}^{\;\;\;\;\rho}
-\F_{a\nu\rho}\F_{b\mu}^{\;\;\;\;\rho}\) \nonumber \\ & &
- {x^2\over2}\rho_i\rho^i\(\F_{a\mu\rho}\F_{b\nu}^{\;\;\;\;\rho} +
\F_{a\nu\rho}\F_{b\mu}^{\;\;\;\;\rho} -{1\over2}g_{\mu\nu}
\F_{a\rho\sigma}\F_b^{\rho\sigma}\) \nonumber \\ & &
- g_{\mu\nu}\delta_{ab}\({\nabla^2 x\over2x}-{\pp_\rho x\pp^\rho x\over4x^2}
+ {\pp_\rho y\pp^\rho y\over2x^2}\) + \delta_{ab}\({\nabla_\mu\pp_\nu x\over x} 
- {\pp_\mu x\pp_\nu x\over x^2}+{\pp_\mu y\pp_\nu y\over2x^2} + r_{\mu\nu}\),
\nonumber \\ 
S_{a\mu,I} &=& \pm i{2\over\sqrt{x}}K_{IK}\[D_\mu(T_az)^K - {\pp_\mu x\over 
2x}(T_az)^K\]- {x\over2}\rho_{IJ}\(\F_{a\nu\mu}\mp i\tcF_{a\nu\mu}\)\D^\nu z^J 
\nonumber \\ & &
+ {1\over4x}f_I\[\D'^\nu\F_{a\nu\mu} + {3\pp^\nu x\over2x}\(\F_{a\nu\mu} \mp 
i\tcF_{a\nu\mu}\)\] \nonumber \\ & &
+ 2\D^\nu z^KK_{IK}\F_{a\mu\nu}, \;\;\;\; I,J,K = 
\cases{i,j,\bar{k}\cr\ibar,\bj,k\cr},\nonumber \\ 
K^a_{\mu\nu,\rho} &=& -{1\over 2}\(\D''_\mu\F^a_{\nu\rho} 
+ \D''_\nu \F^a_{\mu\rho}\) 
+ {1\over 4}\(g_{\mu\rho}\D''^\sigma\F^a_{\sigma\nu}  + g_{\nu\rho}\D''^\sigma
\F^a_{\sigma\mu}\) \nonumber \\ & &
- {\pp^\sigma x\over8x}
\(g_{\mu\rho}\F^a_{\sigma\nu} + g_{\nu\rho}\F^a_{\sigma\mu}\)
+ {3\pp^\sigma y\over8x} 
\(g_{\mu\rho}\tcF^a_{\sigma\nu} + g_{\nu\rho}\tcF^a_{\sigma\mu}\)
\nonumber \\ & &
- {1\over 8x}\(\pp_\mu y\tcF^a_{\nu\rho} + \pp_\nu y\tcF^a_{\mu\rho}\) 
- g_{\mu\nu}\tcF^a_{\sigma\rho}{\pp^\sigma y\over4x} . \eea
In writing the above expressions we used the notation in (2.2--3) and 
the first identity in (B.12) with $M_1=\F_a,\;M_2=\tcF_b,\;\tM_2=-\F_b$.
The inverse metric $Z^{-1}$ must be included in evaluating the traces of these
operators, which are defined such that
\bea \Tr H_\Phi &=& \Tr H + \Tr X + \Tr N, \nonumber \\ 
\Tr H_\Phi^2 &=& 
\Tr H^2 + \Tr X^2 + \Tr N^2 + 2\Tr Y^2 - 2\Tr K^2 - 2\Tr S^2.\eea
In the expressions for the traces\footnote{There is a term missing from Tr$Y^2$
in I, namely: $$ - 4\D_\mu\z^{\m}\D^\mu\z^{\n}\D_\nu z^j\D^\nu z^iR_{\n j\m i}
+ 4\D_\mu\z^{\m}\D^\mu z^i\D_\nu z^j\D^\nu\z^{\n}R_{\m j\n i}.$$} to be given
below, space-time indices are raised with $g^{\mu\nu}$ and
scalar indices are raised with $K^{i\m}$.

Finally we need \footnote{In (B8) of I the expression for
Tr$R_{\mu\nu}R^{\mu\nu}$ should be multiplied by 2 and 
the fourth line of (B8) of I should read: $
 \(G^G_{\mu\nu}\)_{\gamma\delta,\alpha\beta} 
  = \delta_{\alpha\beta,\rho\sigma}\(r^\rho_{\gamma\mu\nu}
g^\sigma_\delta + r^\rho_{\delta\mu\nu}g^\sigma_\gamma\).$}
\bea 
\hG_{\mu\nu} &=& \(G_z + G_G + G_g + G_{gz} + G_{Gz} + G_{gG}\)_{\mu\nu}, 
\nonumber \\ 
\(G^z_{\mu\nu}\)^I_J &=& \(G^z_{0\mu\nu}\)^I_J \pm iF^a_{\mu\nu}D_J(T_az)^I
, \;\;\;\; I,J = \cases{i,j\cr\ibar,\bj\cr},
\nonumber \\ 
\(G^z_{\mu\nu}\)^I_J &=& \(G^z_{0\mu\nu}\)^I_J ,\;\;\;\; I,J = \cases{i,\bj
\cr\ibar,j\cr}, \nonumber \\ 
\(G^G_{\mu\nu}\)_{\alpha\beta,\gamma\delta} &=& 
\(G^G_{0\mu\nu}\)_{\alpha\beta,\gamma\delta} + {1\over 4}
\bigg[\F^a_{\alpha\mu}\F_{a\gamma\nu}g_{\beta\delta}  \nonumber \\ & &
+ \F^a_{\alpha\mu}\F_{a\delta\nu}g_{\beta\gamma} + 
\F^a_{\beta\mu}\F_{a\gamma\nu}g_{\alpha\delta} + 
\F^a_{\beta\mu}\F_{a\delta\nu}g_{\alpha\gamma} - (\mu\leftrightarrow\nu)\bigg], 
\nonumber \\ 
\(G^g_{\mu\nu}\)_{a\rho,b\sigma} &=& g_{\rho\sigma}\(c_{abc}F^c_{\mu\nu}
+ {1\over 2}\[\F_{a\lambda\mu}\F^\lambda_{b\;\nu} - \F_{a\lambda\nu}
\F^\lambda_{b\;\mu}\]\) + \delta_{ab}r_{\sigma\rho\mu\nu} \nonumber \\ & &
- \delta_{ab}\(\epsilon_{\rho\nu\sigma\lambda}\[{\nabla_\mu\pp^\lambda y\over
2x} - {\pp^\lambda y\pp_\mu x\over 2x^2}\] - (\mu\leftrightarrow\nu)\) 
\nonumber \\ & &
- \delta_{ab}{1\over4x^2}\(\pp_\lambda y\pp^\lambda yg_{\rho\nu}g_{\mu\sigma} +
\pp_\sigma y\pp_\nu yg_{\rho\mu} + \pp_\rho y\pp_\mu yg_{\nu\sigma}
- (\mu\leftrightarrow\nu)\) \nonumber \\ & &
+ {1\over 2}\[\F_{a\sigma\mu}\F_{b\rho\nu} - \F_{a\rho\mu}\F_{b\sigma\nu} 
+ x^2\rho_i\rho^i\(\F_{a\mu\rho}\F_{b\nu\sigma} + 
\tcF_{a\mu\rho}\tcF_{b\nu\sigma}\) - (\mu\leftrightarrow\nu)\] \nonumber \\ 
\(G^{gz}_{\mu\nu}\)_{a\rho,I} &=& \(G^{gz}_{\mu\nu}\)_{I,a\rho } =
-\D_\mu\[{x\over2}\rho_I\(\F_{a\nu\rho} \mp i\tcF_{a\nu\rho}\)\]\nonumber \\ & &
- \epsilon_{\rho\mu\sigma\lambda}{\pp^\lambda y\over8x^2}f_I
\(\F_{a\nu}^{\;\;\;\;\sigma}\mp i\tcF_{a\nu}^{\;\;\;\;\sigma}\)
-(\mu\leftrightarrow\nu),\;\;\;\;I = \cases{i\cr\ibar\cr},\nonumber \\
\(G^{Gz}_{\mu\nu}\)_{\alpha\beta,I} &=& -4\(G^{Gz}_{\mu\nu}\)_{I,\alpha\beta} =
\pm{ix\rho_I\over2}\[\tcF^a_{\alpha\mu}\F_{a\beta\nu}
+ \tcF^a_{\beta\mu}\F_{a\alpha\nu}- (\mu\leftrightarrow\nu)\],\;\;\;\;I = 
\cases{i\cr\ibar\cr},\nonumber \\ 
\(G^{gG}_{\mu\nu}\)_{a\rho,\alpha\beta} &=& {1\over4}\[\(g_{\beta\rho}\D_\mu -
{\pp^\lambda y\over2x}\epsilon_{\rho\mu\beta\lambda}\)\F_{a\alpha\nu} + \(
g_{\alpha\rho}\D_\mu - {\pp^\lambda y\over2x}\epsilon_{\rho\mu\alpha\lambda}\)
\F_{a\beta\nu} - (\mu\leftrightarrow\nu)\], \nonumber \\
\(G^{gG}_{\mu\nu}\)_{\alpha\beta,a\rho} &=& \(g_{\beta\rho}\D_\mu -
{\pp^\lambda y\over2x}\epsilon_{\rho\mu\beta\lambda}\)\F_{a\alpha\nu} 
+ \(g_{\alpha\rho}\D_\mu - {\pp^\lambda y\over2x}\epsilon_{\rho\mu\alpha\lambda}
\)\F_{a\beta\nu} \nonumber \\ & &
- g_{\alpha\beta}\(\D_\mu\F_{a\rho\nu} - {\pp^\lambda y\over2x}
\epsilon_{\rho\mu\sigma\lambda}\F^\sigma_{a\;\;\nu}\) - 
(\mu\leftrightarrow\nu). \eea 

\subsubsection{Fermion matrix elements}

As described in I, we take the Landau gauge condition $G=0$, where
\bea G &=& -\gamma^\nu(i\notD - \M)\psi_\nu 
- 2(\notcD z^iK_{i\m}R\chi^{\m} + \notcD\z^{\m}K_{i\m}L\chi^i) \nonumber \\ & &
+ {x\over 2}\sigma^{\nu\rho}\lambda_aF^a_{\nu\rho} + 2im_I\chi^I
- \gamma_5\D_a\lambda^a, \eea 
which we implement by introducing an auxiliary field $\alpha$.  After an
appropriate shift in the gravitino field $\psi_\mu$, we obtain for the 
bilinear fermion couplings of the gravity sector:
\bea {1\over\sqrt{g}}\L_{\psi+\alpha} &=& 
-{1\over 2}\bps^\mu(i\notD - \M)\psi_\mu -\ba(i\notD + 2M)\alpha \nonumber \\
& & +ix\bps_\mu\notF^\mu_a\lambda^a - 2\bps_\mu(\D^\mu\z^{\m}K_{i\m}L\chi^i + 
\D^\mu z^iK_{i\m}R\chi^{\m})  \nonumber \\ & & 
- \ba\({x\over 2}\sigma^{\nu\rho}\lambda_aF^a_{\nu\rho} 
- 2im_I\chi^I + \gamma_5\D_a\lambda^a\).\eea 
To obtain the ghostino determinant we use the supersymmetry
transformations~\cite{cremmer} 
\bea i\delta\chi^i &=& {1\over 2}(\notcD z^iR - i\m^iL)\epsilon, \;\;\;\; 
i\delta\chi^{\m} = \[{1\over 2}(\notcD\z^{\m}L - im^{\m}R)\]\epsilon, 
\nonumber \\
i\delta\psi_\mu &=& (iD_\mu - {1\over 2}\gamma_\mu M)\epsilon, \;\;\;\; 
i\delta\lambda^a = \[{i\over 4}\gamma^\mu\gamma^\nu F^a_{\mu\nu}
- {1\over 2x}\gamma_5\D^a\]\epsilon, \eea 
yielding
\bea D^2 + H_{Gh} = {\pp\delta G\over\pp\epsilon} &=& D^\mu D_\mu - 
{1\over 2}\gamma^\mu\gamma^\nu[D_\mu,D_\nu] - i[\notD,M]- 2M\M + \m^im_i + \D
\nonumber \\ 
& & +2i\m_{\m}\notcD\z^{\m}L + 2im_i\notcD z^iR 
+{x\over 2}\sigma_{\sigma\rho}F_a^{\sigma\rho}
[{1\over 4}\sigma^{\mu\nu}F^a_{\mu\nu} - {1\over x}\gamma_5\D^a] \nonumber\\
& & - \D_\mu z^iK_{i\m}\D^\mu\z^{\m} 
+ {1\over 2}\gamma_5[\gamma^\mu,\gamma^\nu]\D_\mu\z^{\m}K_{i\m}\D_\nu z^i.\eea 

The metric for the gaugino field, as obtained from the classical supergravity
Lagrangian given in (A.9) of I, is $Z_{ab} = \delta_{ab}x.$  
Following~\cite{josh} we rescale the gaugino field $\lambda = \sqrt{x}
\lambda'$, so for the rescaled field $\lambda'$, $Z_{ab} = \delta_{ab}.$ The
matrix elements of $M_\Theta$ are given by (2.17), (A.11) and (B.9-10)
of I and by\footnote{(B.10) of I should read $M^\mu_I = -2Z_{IJ}\D^\mu z^J, 
\;\;\;\; M^I_\mu = \D_\mu z^I.$ The equation before (2.16) should read
$A = e^KW = e^{K/2}\M$.}
\bea M^a_b &=& \(\M^a_b\)^* = \delta^a_b m_\lambda, \;\;\;\; 
m_\lambda = -{e^{-K/2}\over 2x}f_k\A^k, \nonumber \\ 
M^a_I &=&  \delta^{ab}\(m_{bI} + M^{\mu\nu}_{bI}\sigma_{\mu\nu}\), \;\;\;\;
M_a^I = {1\over2}K^{IJ}\(m_{Ja} + M^{\mu\nu}_{Ja}\sigma_{\mu\nu}\), 
\nonumber \\  m_{ai} &=&  m_{ia} = {i\over\sqrt{x}}\({1\over 2x}f_i\D_a - 
2K_{i\m}(T_a\z)^{\m}\) = m^*_{a\ibar}, \nonumber \\ 
M^{\mu\nu}_{aI} &=&  -M^{\mu\nu}_{Ia} = -{ix\over4}\rho_I\(\F^{\mu\nu}_a
\mp i\tcF^{\mu\nu}_a\),\;\;\;\;I =\cases{i\cr\ibar\cr}, \nonumber \\ 
2M^\alpha_a &=&  -\M^a_\alpha = m_{\alpha a} + M^{\mu\nu}_{\alpha
a}\sigma_{\mu\nu},\;\;\;\; 2\M^\alpha_a = -M^a_\alpha =  \m_{\alpha a} + 
\M^{\mu\nu}_{\alpha a}\sigma_{\mu\nu}, \nonumber \\ 
m_{\alpha a} &=&  -\m_{\alpha a} = {1\over\sqrt{x}}\D_a, \;\;\;\;
M^{\mu\nu}_{\alpha a} = \M^{\mu\nu}_{\alpha a}= -{1\over2}\F_a^{\mu\nu}, \eea 
with covariant derivatives as defined in (A.21) [see also (B.11) of I]
\bea D_\mu m^\lambda &=& -e^{-K/2}\(\D_\mu\z^{\m}\[\aa_{\m} - \A_{\m}\]
+ \D_\mu z^i\[{f_i\over 2x}\aa - x\rho_{ik}\A^k\]\), \nonumber \\ 
D_\rho M_{aA} &=& \tD_\rho M_{aA} - i{\pp_\rho y\over2x}M_{aA}\gamma_5,\;\;\;\;
D_\rho M_{Aa} = \tD_\rho M_{Aa} + i{\pp_\rho y\over2x}M_{Aa}\gamma_5,\;\;\;\;
A = i,\m,\alpha, \nonumber \\
\tD_\rho M^{\mu\nu}_{ai} &=& -\tD_\rho M^{\mu\nu}_{ia} = 
-\(\tD_\rho\M^{\mu\nu}_{a\ibar}\)^* = \(\tD_\rho\M^{\mu\nu}_{\ibar a}\)^* 
\nonumber \\ &=& -{ix\over4}
\[\rho_i\(\D_\rho + i{\pp_\rho y\over x}\) + \D_\rho z^j\rho_{ij} 
\]\(\F_{a\mu\nu} -i\tcF_{a\mu\nu}\), \nonumber \\ 
\tD_\rho m_{ai} &=& \tD_\rho m_{ia} = \(\tD_\rho m_{a\ibar}\)^* 
\nonumber \\ &=& {i\over\sqrt{x}}\Bigg[\D_a\({f_i\over4x^2}\[2i\pp_\rho y -
\pp_\rho x\]  - x\rho_{ij}\D_\rho z^j\) +
{\pp_\rho x\over x}K_{i\m}(T_a\z)^{\m} \nonumber \\ & &
+ {1\over 2x}f_i(K_{j\m}(T_a\z)^{\m}\D_\rho z^j + {\rm h.c.}) 
- 2K_{i\m}D_{\n}(T_a\z)^{\m}\D_\rho \z^{\n}\Bigg], \nonumber \\ 
\tD_\mu m_{\alpha a} &=& -\(\tD_\mu\m_{\alpha a}\)^* = {1\over \sqrt{x}}
\(K_{i\m}\[\D_\mu z^i(T_a\z)^{\m} + \D_\mu\z^{\m}(T_az)^i\] 
- {\pp_\mu x\over2x}\D_a\), \nonumber \\ 
\tD_\rho M^{\mu\nu}_{\alpha a} &=& \(\tD_\rho\M^{\mu\nu}_{\alpha a}\)^* = -  
{1\over2}\D_\rho\F_a^{\mu\nu}.\eea
Here $\alpha$ is the auxiliary field introduced in I to implement the gravitino
gauge fixing; its couplings to chiral and Yang-Mills matter are given in (3.10)
of I.  In addition, there is a $\lambda$-$\psi$ 
connection~\cite{josh}, $(D_\mu)_{a\nu} = (D_\mu)_{\nu a} =  -\F_{a\nu\mu}$,
that contributes 
as follows to the covariant derivatives of the fermion mass matrix:
\bea \(D_\rho M\)_{a\mu} &=& -\(D_\rho M\)_{\mu a} = -e^{-K/2}\aa\F_{a\mu\rho}, 
\nonumber \\ 
\(D^\rho M\)^\mu_I &=& - 2K_{IJ}D^\rho\D^\mu z^J- M^a_I\F_a^{\mu\rho}, \;\;\;\;
\(D_\rho M\)^I_\mu = D_\rho\D_\mu z^I + M^I_a\F^a_{\mu\rho}, \nonumber \\ 
\(D_\rho M\)^a_I &=& D_\rho M^a_I + 2K_{IJ}\D^\mu z^J\F^a_{\mu\rho},\;\;\;\; 
\(D_\rho M\)^I_a = D_\rho M^I_a + \D^\mu z^I\F^a_{\mu\rho}, \nonumber \\ 
\(D^\rho M\)^\mu_\alpha &=& - M^a_\alpha\F_a^{\mu\rho}, \;\;\;\;
\(D_\rho M\)^\alpha_\mu = M^\alpha_a\F^a_{\mu\rho}, \eea
The nonvanishing matrix elements of $G_{\mu\nu}$ involving the gaugino field are
\bea \(G^{\pm}_{\mu\nu}\)_{ab} &=& c_{abc}F^c_{\mu\nu} +
\delta_{ab}\(\pm\Gamma_{\mu\nu} + i\gamma_5 L_{\mu\nu} + Z_{\mu\nu}\)
+ \(\F_{a\rho\mu}\F_{b\;\;\nu}^\rho - \mu\leftrightarrow\nu\), \nonumber \\ 
\(G^{\pm}_{\mu\nu}\)_{a\rho} &=& -\[(\D_\mu + i\gamma_5L_\mu)\F_{a\rho\nu} -
(\mu\leftrightarrow\nu)\], \nonumber \\ 
\(G^{\pm}_{\mu\nu}\)_{\rho a} &=& -\[(\D_\mu - i\gamma_5L_\mu)\F_{a\rho\nu} -
(\mu\leftrightarrow\nu)\]. \eea 
As in I, $\D_\mu$ is the gauge and general coordinate covariant derivative,
$\Gamma_{\mu\nu}$ and $Z_{\mu\nu}$ are given in (B.13) of I, and\footnote{We use
the notation $L_\mu,L_{\mu\nu},$ to denote the field operators defined in 
(C.19), and also the matrices defined by these fields multiplying the unit
projection operator in the space of gauginos, as in (3.9--11), (A.22--23), 
(C.22), {\it etc}.}
\begin{equation}
\F_{\mu\nu} = \sqrt{x}F_{\mu\nu}, \;\;\;\; L_\mu = -{\pp_\mu y\over 2x},\;\;\;
\; L_{\mu\nu} = {1\over 2x^2}\(\pp_\mu x \pp_\nu y
-\pp_\nu x \pp_\mu y\).\eeq 
The other matrix elements of $G_{\mu\nu}$ are as given in Appendix (B.12) of I,
except that now the chiral matter connection includes the gauge field:
\beq \(G_{\mu\nu}\)^I_J = \(R_{\mu\nu}\)^I_J \pm iF^a_{\mu\nu}D_J(T_az)^I
+ \delta^I_J\(Z_{\mu\nu} \pm \Gamma_{\mu\nu}\),
\;\;\;\;I,J=\cases{i,j\cr\ibar,\bj}, \eeq
where $(R_{\mu\nu})^I_J$ is defined in (B.8) of I, and the $\psi$-$\lambda$
connection gives an additional contribution to the gravitino matrix
element\footnote{The last line of Eq. (B12) of I should read 
$(G_{\mu\nu})^\rho_\sigma = \delta^\rho_\sigma(\gamma_5\Gamma_{\mu\nu}+
Z_{\mu\nu}) - r^\rho_{\sigma\nu\mu}$.} of $G_{\mu\nu}$:
\beq \(G^{\pm}_{\mu\nu}\)_{\rho\sigma} = g_{\rho\sigma}\(\pm\Gamma_{\mu\nu}+
Z_{\mu\nu}\) - r_{\rho\sigma\mu\nu}
+ \(\F^a_{\rho\mu}\F_{a\sigma\nu} - \mu\leftrightarrow\nu\). \eeq
Finally, in the $8\times8$ matrix notation of (2.14--17), setting $G_{\mu\nu} = 
\tG_{\mu\nu} + i\gamma_5L_{\mu\nu}$,
\bea H_\Theta &=& M_\Theta M_\Theta + 
{1\over4}[\gamma^\mu,\gamma^\nu]\tG_{\mu\nu}
- i\notD M_\Theta - 2D^\mu M^\Theta_{\mu\nu}\gamma^\nu 
- 4\gamma^\rho\gamma_\sigma M_{\mu\rho}^\Theta M^{\mu\sigma}_\Theta 
\nonumber \\ & & \qquad - 2L_\mu L^\mu + i\tD^\mu L_\mu\gamma_5  +
2i\gamma_\mu\gamma_\rho\gamma_\nu\gamma_5[L^\rho,M_\Theta^{\mu\nu}], 
\nonumber \\ \hD^\Theta_\mu &=& \tD_\mu + 2\gamma^\nu M_{\mu\nu}^\Theta + 
\sigma_{\mu\nu}\gamma_5L^\nu, \nonumber \\ 
\hG^\Theta_{\mu\nu} &=& \tG_{\mu\nu} + 
2\gamma^\rho\(\tD_\mu M^\Theta_{\nu\rho} - \tD_\nu M^\Theta_{\mu\rho}\)
+ 4\gamma^\rho\gamma^\sigma\(M_{\mu\rho}^\Theta M_{\nu\sigma}^\Theta -
M_{\nu\rho}^\Theta M_{\mu\sigma}^\Theta\) \nonumber \\ & & 
+ \[\sigma_{\rho\mu}\(\gamma_5\tD_\nu L^\rho - 2iL_\nu L^\rho\)
- \(\mu\leftrightarrow\nu\)\] - 2iL_\rho L^\rho\sigma_{\mu\nu} -
4i[\notL,M_{\mu\nu}^\Theta]\gamma_5 \nonumber \\ & & - 2\[\gamma_\mu
\(\{L^\rho,\tM^\Theta_{\rho\nu}\} - i[L^\rho,M^\Theta_{\rho\nu}]\gamma_5\) + 
\{L_\mu,\tM^\Theta_{\nu\rho}\}\gamma^\rho - \(\mu\leftrightarrow\nu\)\], 
\nonumber \\
M_\Theta &=& m_\Theta + M_\Theta^{\mu\nu}\sigma_{\mu\nu} =
m_\Theta + M_\sigma. \eea
Then, defining $H_\Theta = H_1 + H_2 + H_3$, with
\bea H_1 &=& M_\Theta M_\Theta 
- 4\gamma^\rho\gamma_\sigma M_{\mu\rho}^\Theta M^{\mu\sigma}_\Theta, 
\nonumber \\ H_2 &=& - i\notD M_\Theta - 2\gamma^\nu D^\mu M^\Theta_{\mu\nu} 
+ 2i\gamma_\mu\gamma_\rho\gamma_\nu\gamma_5[L^\rho,M_\Theta^{\mu\nu}], 
\nonumber \\ H_3 &=& {1\over4}[\gamma^\mu,\gamma^\nu]\tG_{\mu\nu}
- 2L_\mu L^\mu + i\tD^\mu L_\mu\gamma_5, \nonumber \\ 
G'_{\mu\nu} &=& \tG_{\mu\nu} - Z_{\mu\nu}, \eea
we find the following traces ($\btr$ includes the Dirac trace): 
$\btr 1 \equiv 8\Tr 1$, where Tr is over internal symmetry indices only):
\bea {1\over 8}\btr H_1 &=& {1\over 8}\btr\[m_\Theta m_\Theta - 
2M^\Theta_{\mu\nu}M_\Theta^{\mu\nu}\] = \Tr\[\m m - 
2\M_{\mu\nu}M^{\mu\nu}\] \nonumber \\ & &
\nonumber \\ &=& {1\over 8}\btr\(M_\Theta^2\)_0 + 4\K_a^a 
- 2\D\(1- x^2\rho^i\rho_i\)  + N_G M_\lambda^2
+ {x\over2}F^a_{\mu\nu}F_a^{\mu\nu}, \nonumber \\ 
{1\over 8}\btr H_1^2 &=& {1\over 8}\btr\Bigg[\(m_\Theta m_\Theta\)^2 +
\(\sigma_{\mu\nu}\sigma_{\rho\sigma}M_\Theta^{\mu\nu}M_\Theta^{\rho\sigma}\)^2
 + 4m_\Theta M_\Theta^{\mu\nu}m_\Theta M^\Theta_{\mu\nu} 
\nonumber \\ & & + 16M_\Theta^{\mu\nu}M^\Theta_{\rho\nu}\(
M^\Theta_{\mu\sigma}M_\Theta^{\rho\sigma} 
- M_\Theta^{\rho\sigma}M^\Theta_{\mu\sigma}\)\Bigg]\nonumber \\ &=&
\Tr\Bigg[\(\m m\)^2 + 2\m M^{\mu\nu}\m M_{\mu\nu} + 2\M^{\mu\nu}m\M_{\mu\nu}m
+ 4\M_{\mu\nu}M_{\rho\sigma}\M^{\mu\nu}M^{\rho\sigma} \nonumber \\ & & 
+ 8\(\M^{\mu\nu} M_{\mu\nu}\)^2 - 16\M^{\mu\nu}M^{\rho\sigma}
\M_{\mu\rho}M_{\nu\sigma} \Bigg], \eea
and using (B.12), partial integration and the relation
\bea \notD\gamma^\mu\gamma^\nu M_{\mu\nu} &=& 2\gamma^\nu D^\mu M_{\mu\nu} +
2i\gamma^\nu D^\mu\tM_{\mu\nu}\gamma_5, \eea
we obtain \bea
-{1\over 8}\btr H_2^2 &=& -{1\over 8}\btr\lbr -i\notD m_\Theta 
+ 2\gamma_\nu [L_\mu,\tM^{\mu\nu}_\Theta] + 2i \gamma_\nu\gamma_5\tD_\mu
\tM^{\mu\nu}_\Theta\}\rbr^2 
\nonumber \\ &=& \Tr\Big\{\tD_\mu\m \tD^\mu m - 4\tD_\mu\tbM^{\mu\nu} 
\tD^\rho\tM_{\rho\nu} - 4[L_\mu\tbM^{\mu\rho}][L^\nu,\tM_{\nu\rho}]
\nonumber \\ & & + [L_\mu,\m] [L^\mu, m] - i\([\hL_{\mu\nu},\m]\tM^{\mu\nu} +
[\hL_{\mu\nu},m]\tbM^{\mu\nu}\)\Big\},\qquad\qquad \eea
where $\hL_{\mu\nu}$ is defined in (3.11).  
The remaining traces needed to evaluate $\btr H_\Theta,\btr H_\Theta^2$ are:
\bea {1\over2}\btr H_3 &=& (N + N_G + 5)r - 2N_G{\pp_\mu y\pp^\mu y\over x^2}, 
\nonumber \\
{1\over2}\btr H_3^2 &=& N_G\btr h_3^2 + (N + 5){r^2\over 4} - 
\btr\([\Gamma'_\mu,L^\mu]\)^2 - {1\over4}\btr\(G'_{\mu\nu}G'^{\mu\nu}\) 
\nonumber \\ {1\over2}\btr\(H_1H_3\) &=& {1\over 2}\btr\[\({r\over 4} - 
2L_\mu L^\mu\)H_1 - 2M_{\mu\nu}^\Theta\tM^{\mu\nu}_\Theta\tD_\rho L^\rho - 
iG'_{\mu\nu}\{M_\Theta^{\mu\nu},m_\Theta\}\] \nonumber \\ 
{1\over2}\btr\hG^\Theta_{\mu\nu}\hG_\Theta^{\mu\nu} &=& 
{1\over 2}\btr\Bigg\{\tG_{\mu\nu}\tG^{\mu\nu} + 16\tG_{\mu\nu}
M^{\mu\rho}_\Theta M^{\nu\sigma}_\Theta\gamma_\rho\gamma_\sigma 
+ 8\tD_\mu M_{\nu\rho}^\Theta\(\tD^\mu M^{\nu\rho}_\Theta -
\tD^\nu M^{\mu\rho}_\Theta\) \nonumber \\ & & 
+ 16\(\tD_\mu M^{\mu\nu}_\Theta\{L^\rho,\tM_{\rho\nu}^\Theta\} 
- \tD_\mu\tM^{\mu\nu}_\Theta\{L^\rho,M_{\rho\nu}^\Theta\}\) 
+ 4[\Gamma'_\mu,L_\nu][\Gamma'^\mu,L^\nu]\nonumber \\ & & 
+ 2\([\Gamma'_\mu,L^\mu]\)^2 - 32\(M_{\mu\nu}^\Theta M^{\mu\nu}_\Theta 
M_{\rho\sigma}^\Theta M^{\rho\sigma}_\Theta + M_{\nu\rho}^\Theta 
M^{\mu\rho}_\Theta M_{\mu\sigma}^\Theta M^{\nu\sigma}_\Theta\) \nonumber \\ & & 
- 32M_{\mu\sigma}^\Theta M_{\nu\rho}^\Theta\(
M^{\mu\sigma}_\Theta M^{\nu\rho}_\Theta - 3M^{\nu\sigma}_\Theta
M^{\mu\rho}_\Theta\) + 80L_\mu L^\nu M^{\mu\rho}_\Theta M_{\nu\rho}^\Theta 
\nonumber \\ & & - 80L_\rho L^\rho M^{\mu\nu}_\Theta M_{\mu\nu}^\Theta 
+ 24\tD_\mu L^\mu\tM_{\nu\rho}^\Theta M^{\nu\rho}_\Theta\Bigg\} + N_G
\btr\(\hgg^2 - \tgg^2\),\eea
where $\btr h_3^2,\btr\hgg^2,\btr\tgg^2$ are given in (C.66), and 
$\Gamma'_\mu$ is the gaugino-gravitino connection.

\subsubsection{Ghost matrix elements}

For the gravitino ghost, $H_{Gh}$ is defined by (C.14).
For the bosonic ghosts we have
\bea 
H_{gh}^{\mu\nu} &=& \(H_{gh}^{\mu\nu}\)_0 + 
{3\over 2}\F^a_{\mu\rho}\F^{\;\;\;\;\rho}_{a\nu}, \nonumber \\ 
H_{gh}^{ab} &=& \K_{ab} + \K_{ba} - {1\over 2}\F^a_{\mu\nu}\F_b^{\mu\nu} 
- \delta_{ab}\({\nabla^2 x\over2x} - {\pp_\mu x\pp^\mu x\over4x^2}\), 
\nonumber \\ 
H^{gh}_{\mu a} &=& {1\over\sqrt{2}}\D^\nu\F_{a\mu\nu} + \F_{a\mu\nu}{\pp^\nu
x\over\sqrt{2}x} + \sqrt{2}q_{aI}\D_\mu z^I, \nonumber \\ 
(H^{gh})^a_{\;\;\nu} &=& -{1\over\sqrt{2}}\D^\mu\F^a_{\nu\mu} -
\F^a_{\nu\mu}{\pp^\mu x\over\sqrt{2}x} - \sqrt{2}q^a_I\D_\nu z^I ,\eea
\bea 
\(\hG^{gh}_{\mu\nu}\)_{\rho\sigma} &=& - r_{\rho\sigma\mu\nu} + 
{1\over 2}\(\F^a_{\rho\mu}\F_{a\sigma\nu} - (\mu\leftrightarrow\nu)\), 
\nonumber \\ 
\(\hG^{gh}_{\mu\nu}\)_{ab} &=& c_{abc}F^c_{\mu\nu} +
{1\over 2}\(\F_{a\rho\mu}\F^{\rho}_{b\;\;\nu} - (a\leftrightarrow b)\), 
\nonumber \\ 
(\hG^{gh}_{\mu\nu})^a_{\;\;\rho} &=& (\hG^{gh}_{\mu\nu})_\rho^{\;\;a}
= -{1\over\sqrt{2}}\(\D_\mu\F^a_{\rho\nu} -\D_\nu\F^a_{\rho\mu}\).\eea

\subsubsection{Chiral multiplet supertraces}

Defining 
\beq {1\over 2}\STr H_\chi^2 = H^i_jH^j_i + H_{ij}H^{ij} - {1\over8}\btr
\(H_\Theta^{IJ}H^\Theta_{IJ}\),\;\;\;\; h^\chi_{\m i} = (\m m)_{\m i},\eeq
we have \bea
{1\over8}\btr\(H_1^\chi\)^2 &=& \Tr \;h_\chi^2 + {x^4(\rho^i\rho_i)^2\over8}
\D_a\D^bF^a_{\mu\nu}F_b^{\mu\nu},
\nonumber \eea
\bea (h^\chi)_i^j &=& e^{-K}\(A_{ki}\A^{jk} - A_i\A^j\) - 
2\D_\mu z^j\D^\mu\z^{\m}K_{i\m} + {1\over 4x^2}f_i\f^j\D \nonumber \\ & & 
-{1\over 2x^2}f_i\D_a(T^az)^j -{1\over 2x^2}\f^j\D_a K_{i\n}(T^a\z)^{\n} 
+ {2\over x} (T_az)^j K_{i\n}(T^a\z)^{\n}, \nonumber \\
H_i^j &=& (h^\chi)_i^j + \delta_i^j\(\hV + M_\psi^2\) + 
R^j_{k\m i}\(e^{-K}\A^kA^{\m} + \D_\mu z^k\D^\mu\z^{\m}\) \nonumber \\ & & 
+ {1\over 4x^2}f_i\f^j\D + {1\over x}\D_a D_i(T^az)^j, \nonumber \\ 
H_{ij} &=& e^{-K}\(A_{jik}\A^k - A_{ij}\A\) - \D_\mu\z^{n}\D^\mu\z^{\m}\( 
2K_{j\n}K_{i\m}  + R_{i\m j\n}\) \nonumber \\ & &
-{1\over 2x^2}\D_a(f_i K_{j\m} + f_j K_{i\m})(T^a\z)^{\m} -x^2\rho_{ij}\cW,
\eea 
where \beq \cW = \cW^a_a, \;\;\;\;\cW^a_b = {1\over 4}\(F^a_{\mu\nu}F_b^{\mu\nu}
-iF^a_{\mu\nu}\tF_b^{\mu\nu}\) - {1\over2x^2}\D^a\D_b \eeq
is the bosonic part of the F-component of the chiral superfield $W^a_\alpha
W_b^\alpha$, and $W^a_\alpha = \lambda^a_\alpha + O(\theta)$ is the Yang-Mills
field strength supermultiplet.  Thus: \bea
{1\over8}\btr\(H_1^\chi\)^2 &=& \Tr \;h_\chi^2 + {x^4(\rho^i\rho_i)^2\over4}
\[\(\cW^{ab} + \cbW^{ab}\)\D_a\D_b + 4\D^2\], 
\nonumber \\ \btr\(H_2^\chi\)^2 &=& \btr\(H_2^\chi\)^2_0, \;\;\;\;
{1\over8}\btr H_3^\chi = {1\over8}\btr\(H_3^\chi\)_0 \nonumber \\
{1\over8}\btr\(H_3^\chi\)^2 &=& {1\over8}\btr\(H_3^\chi\)^2_0 
+ {1\over2}D_i(T_az)^jD_j(T^bz)^iF^a_{\mu\nu}F_b^{\mu\nu} \nonumber \\ & & 
- 2iF^a_{\mu\nu}\(D_j(T_az)^iR^j_{i\m k} + 
D_i(T_az)^iK_{\m k}\)\D^\mu z^k\D^\nu\z^{\m} \nonumber \\ 
{1\over4}\btr H_3^\chi H_1^\chi &=& {1\over4}\btr\(H_3^\chi H_1^\chi\)_0 
- t^\chi + {r\over2} \Tr \;h^\chi , \eea
where
\bea \Tr \;h^\chi &=& e^{-K}\A^{ij}A_{ij} - \hV - 3M_\psi^2 - 
2\D_\mu z^i\D^\mu\z^{\m}K_{i\m} + x^2\rho^i\rho_i\D + 2\K^a_a, \nonumber \\ 
t^\chi &=& \[\(x\cW^{ab} + {1\over2x}\D^a\D^b\)\rho_{ij}(T_az)^i\(2(T_bz)^j +
x\rho^j\D_b\) + {\rm h.c.}\] \nonumber \\ & & + {i\over2}
x^2\rho^i\rho_i\D_\mu z^j\D_\nu\z^{\m}K_{i\m}\D^aF_a^{\mu\nu},\eea
and the chiral fermion contributions to the helicity-odd operator $T$ are
\bea
T^\chi &=& T^\chi_3 + T^\chi_4, \nonumber \\
T^\chi_3 &=& \[\tX_-^{\mu\nu}(M,\M)\]^{\m}_{\n}\(G'^+_{\mu\nu}\)^{\n}_{\m}
- \[\tX_-^{\mu\nu}(\M,M)\]^i_j\(G'^-_{\mu\nu}\)_i^j 
+ r^\mu_\nu\Tr\(\tM^{\nu\rho}\M_{\mu\rho} - M^{\nu\rho}\tbM_{\mu\rho}\)_i^j
\nonumber \\ &=& t^\chi + {1\over8}x^3\rho^i\rho_i
\(r^\mu_\nu F_a^{\nu\rho}F^a_{\mu\rho} - {1\over4}rF_a^{\mu\nu}F^a_{\mu\nu}\), 
\nonumber \\ T_4^\chi &=& {8\over3}\(\M^{\mu\nu}M^{\rho\sigma}\)^i_j
\(\M_{\mu\nu}M_{\rho\sigma}\)_i^j 
-2\[\(\m M^{\mu\nu}\)^i_j\(\m M_{\mu\nu}\)_i^j + {\rm h.c.}\] \;\;\;\;\;\;\;\;
\nonumber \\ &=& {x^6(\rho^i\rho_i)^2\over96}
\[(F^a_{\mu\nu}F_b^{\mu\nu})^2 + (F^a_{\mu\nu}\tF_b^{\mu\nu})^2\] -
{x^4(\rho^i\rho_i)^2\over8}\D_a\D^bF^a_{\mu\nu}F_b^{\mu\nu}. \;\;\;\;\;\;\;\;
\eea
Then we obtain \bea
\STr H_\chi &=& \STr \(H_\chi\)_0 + 
2x^{-1}\D_aD_i(T^az)^i + 2x^2\rho_i\rho^i\D, \nonumber \\ 
{1\over 2}\STr H_\chi^2 &=& {1\over 2}\STr \(H_\chi^2 \)_0 - T_3^\chi +
2x\rho_i\rho^i\D_a\D_b\K^{ab} - {1\over x}\D^a(T_az)^ik_i
- {2\over x^2}C_G^a\D_a\D^a\nonumber \\ & &
- \(\cW_{ab} + \cbW_{ab}\)D_i(T^bz)^jD_j(T^az)^i - 
{x^4(\rho^i\rho_i)^2\over4}\(\cW^{ab} + \cbW^{ab}\)\D_a\D_b  \nonumber \\ & &
+ {1\over8}x^3\rho^i\rho_i\(r^\mu_\nu F_a^{\nu\rho}F^a_{\mu\rho} 
- {1\over4}rF_a^{\mu\nu}F^a_{\mu\nu}\) 
- r\(\K^a_a + {x^2\rho^i\rho_i\over2}\D\) \nonumber \\ & & 
+ 2x^4(\rho_i\rho^i)^2\D^2 + 2\D\(6M_\psi^2 - 3M_\lambda^2 -\hV\)
- 6x^2\rho^i\rho_iM_\lambda^2\D \nonumber \\ & &
+ 4\(\hV +M_\psi^2\)\(\K^a_a + x^2\rho_i\rho^i\D\)
- 2e^{-K}\D\f^i\A^j A_k\(\rho^k_{\;\;ij} + 
{\f^k\over x}\rho_{ij}\)\nonumber \\ & &
+ {4e^{-K}\over x}(T_az)^i(T^a\z)^{\m}R^k_{\;n\m i}A_k\A^n + 
2e^{-K}\D\(a_i\aa^i + 2a\aa + A_{ij}\A^{ij}\) \nonumber \\ & &
+ {2\over x}e^{-K}\lbr\[a_j\(2\A-\aa\) - a_{ij}\A^i + 
A_k\A^i\(x\rho^k_{\;\;ij} + \rho_{ij}\f^k\)\]\D_a(T^az)^j +{\rm h.c.}\rbr 
\nonumber \\ & & + {2\over x}\D_a\[\(\hV + M_\psi^2\)D_i(T^az)^i + 
e^{-K}R^{k\;j}_{\;n\;i}A_k\A^nD_j(T^az)^i\] \nonumber \\ & &
+ {e^{-K}\over x}\lbr\D_a(T^az)^i\[4A_{ij}\A^j 
+  R_{i\;\;\ell}^{\;\; j\;\; k}\A^{\ell}A_{jk}\]
- \D\(2a_i\A^i - {\f^if_j\over 2x^2}a_i\A^j\) +{\rm h.c.}\rbr \nonumber \\ & &
+ x^2\[2\D_\rho z^i\D^\rho z^j\rho_{ij}\cW - e^{-K}\rho^{ij}\(A_{jik}\A^k - 
A_{ij}\A\)\cbW + {\rm h.c.}\] + x^4\rho_{ij}\rho^{ij}\cW\cbW \nonumber \\ & &
- {4\over x}\D^\mu z^i\D_\mu\z^{\m}K_{j\m}
\D_aD_i(T^az)^j - {1\over x^2}\(\pp_\mu x\pp^\mu x + 
\pp_\mu y\pp^\mu y\)\(1 + 2x^2\rho^i\rho_i\)\D \nonumber \\ & &
+ \D^a\[{2\over x^2}\f_{\m}\D_\mu\z^{\m}K_{i\n}(T_az)^i\D^\mu\z^{\n} + 
\(\cW(T_az)^j - \cW_{ab}(T^bz)^j\)\rho_{ij}\f^i + {\rm h.c.}\] \nonumber \\ & &
+ 2\D_\mu z^i\D^\mu\z^{\m} R_{\n i\m j}\[{2\over x}(T_az)^j(T^a\z)^{\n} 
+ {1\over x}\D_a D^j(T^a\z)^{\n}\] \nonumber \\ & &
+ \lbr\[2(T^az)^j\D_a - \f^j\D\]\[\rho_{\m ij} + 
{1\over x}\f_{\m}\rho_{ij}\]\D^\mu\z^{\m}\D_\mu z^i + {\rm h.c.}\rbr 
\nonumber \\ & & + \lbr\D_\mu z^i\D^\mu z^j\[x^2R_{\n i\m j}\rho^{\m\n}\cbW - 
2\rho_{\m ij}\D_a(T^a\z)^{\m}\] + {\rm h.c.}\rbr \nonumber \\ & &
+ 2iF^a_{\mu\nu}\(D_j(T_az)^iR^j_{i\m k} + 
D_i(T_az)^iK_{\m k}\)\D^\mu z^k\D^\nu\z^{\m} \nonumber \\ & &
+ ix^2\rho^i\rho_i\D_\mu z^j\D_\nu\z^{\m}K_{j\m}\D^aF_a^{\mu\nu} +
4x\(\cW^{ab}\rho_{ij}(T_az)^i(T_bz)^j + {\rm h.c.}\), \eea
where
\beq k_i = D_ik,\;\;\;\; k = e^{-K}A_{ij}\A^{ij} -2\hV -10M_\psi^2 - 4\K^a_a.
\eeq
Finally we have\footnote{The term $ + 4\D_\mu 
z^i\D_\nu\z^{\m}K_{i\m}R_{j\n}\(\D^\mu z^j\D^\nu\z^{\n} - \D^\nu 
z^j\D^\mu\z^{\n}\)$ should be included in the right hand side of (B.14) of I.}
\bea {1\over12}\STr\hG_{\mu\nu}^\chi\hG^{\mu\nu}_\chi &=& {1\over12}
\STr\(\hG_{\mu\nu}^\chi\hG^{\mu\nu}_\chi\)_0 - {x^3\rho^i\rho_i\over12}\(
r^\mu_\nu F^a_{\mu\rho} F_a^{\nu\rho} - {r\over4}F^a_{\mu\nu} F_a^{\mu\nu}\) 
\nonumber \\ & & + {x^6\(\rho^i\rho_i\)^2\over96}\[
\(F^a_{\mu\nu} F_b^{\mu\nu}\)^2 + \(F^a_{\mu\nu}\tF_b^{\mu\nu}\)^2\]
\nonumber \\ & & + {i\over6}F^a_{\mu\nu}K_{i\m}\D^\mu
z^j\D^\nu\z^{\m}D_i(T_az)^i. \eea

\subsubsection
{Mixed chiral-gauge supertraces}
For the bose sector we have $H_\Phi^{\chi g} = -S$, and
\bea \Tr S^2 &=& {8\over x}K_{i\m}\[D_\mu(T_az)^i][D^\mu(T_a\z)^{\m}\] 
- 4{\pp^\mu x\over x^2}\[(T_a z)^iK_{i\m}D_\mu(T^a\z)^{\m} + {\rm h.c.}\] 
 \nonumber \\ & &
+ 2\K^a_a{\pp_\mu x\pp^\mu x\over x^2} - x\[\rho_{ij}(T^az)^i(T_bz)^j
F_a^{\nu\mu}\(F^b_{\nu\mu} - i\tF^b_{\nu\mu}\) + {\rm h.c.}\] \nonumber \\ & &
- 2i\[x\rho_{\m ij}\D_\nu\z^{\m}(T^az)^i\D_\mu z^j
\(F_a^{\mu\nu} - i\tF_a^{\mu\nu}\) - {\rm h.c.}\] \nonumber \\ & &
+ {x^2\over 2}\rho_i\rho^i\(\D''_\nu\F_a^{\nu\rho} + {\pp_\nu
y\over x}\tcF_a^{\nu\rho}\)^2 
+ x\rho_i\rho^i F_a^{\nu\mu}F^a_{\rho\mu}\({\pp_\nu y\pp^\rho y\over2} 
+ {5\pp_\nu x\pp^\rho x\over4}\) \nonumber \\ & &
- x\rho_i\rho^i\[F_a^{\nu\mu}F^a_{\nu\mu}\({9\pp_\rho x\pp^\rho x\over16} +
{\pp_\rho y\pp^\rho y\over4}\)
+ \tF_a^{\nu\mu}F^a_{\nu\mu}{\pp_\rho y\pp^\rho x\over8}\] \nonumber \\ & &
- {\sqrt{x}\over 4}\(\D''_\nu\F_a^{\nu\rho} + {\pp_\nu y\over x}
\tcF_a^{\nu\rho}\)\[\f^i\rho_{ij}\D^\mu z^j\(F^a_{\mu\rho} - i\tF^a_{\mu\rho}\)
+ {\rm h.c.}\] \nonumber \\ & &
+ {\sqrt{x}\over 2}x\rho^i\rho_i\(\D''_\nu\F_a^{\nu\rho} + {\pp_\nu y\over x}
\tcF_a^{\nu\rho}\)\(\pp^\mu xF_{\mu\rho}^a - 2\pp^\mu y\tF_{\mu\rho}^a\)
\nonumber \\ & &
- F_a^{\nu\mu}F^a_{\rho\mu}\lbr\[\({\pp_\nu x\over2} + i{\pp_\nu y\over4}\)
\D^\rho z^j\rho_{ij}\f^i + {\rm h.c.}\] + x^3\D_\nu z^i
\D^\rho\z^{\m}\rho_{ij}\rho^j_{\m}\rbr \nonumber \\ & &
+ F_a^{\nu\mu}F^a_{\nu\mu}\lbr\[\({3\pp_\rho x\over16}+ i{\pp_\rho y\over8}\)
\D^\rho z^j\rho_{ij}\f^i + {\rm h.c.}\] 
+ {x^3\over 4}\D_\rho z^i\D^\rho\z^{\m}\rho_{ij}\rho^j_{\m}\rbr
\nonumber \\ & & 
- {1\over16}\tF_a^{\nu\mu}F^a_{\nu\mu}\[i\(\pp_\rho x+ i\pp_\rho y\)
\D^\rho z^j\rho_{ij}\f^i + {\rm h.c.}\] \nonumber \\ & & 
+ 8xF_a^{\nu\mu}F^b_{\nu\mu}\K^a_b + 4i\({2\over\sqrt{x}}\D''_\mu\F^{\mu\nu} -
{\pp_\mu x\over x}F_a^{\mu\nu}\)\[(T^a z)^iK_{i\m}\D_\nu\z^{\m} - {\rm h.c.}\] 
\nonumber \\ & &
- 2{\pp^\mu x\over\sqrt{x}}F^a_{\mu\rho}\D''_\nu\F_a^{\nu\rho} - {3\pp_\nu
y\pp^\nu x\over4x}F_a^{\mu\rho}\tF^a_{\mu\rho} 
- F_a^{\nu\mu}F^a_{\rho\mu}{\pp_\nu x\pp^\rho x\over x} \nonumber \\ & &
+ 2xF_a^{\nu\mu}F^a_{\rho\mu}\[4\D^\rho z^i\D_\nu\z^{\m}K_{i\m} +x\(\D^\rho z^i
\D^\nu z^j\rho_{ij} + {\rm h.c.}\)\] \nonumber \\ & &
- {i\over2}x^2\tF_a^{\nu\mu}F^a_{\nu\mu}\(\D_\rho z^j\D^\rho z^i\rho_{ij} - 
{\rm h.c.}\) . \eea 
In writing this expression we dropped total derivatives and used (B.10)
and (B.12--B.14), as well as the Yang-Mills Bianchi identity.
In addition we used (B.3--5) and (B.8) and
\bea & &\sqrt{x}\rho_{ij}\(\F_a^{\nu\mu} - i\tcF_a^{\nu\mu}\)\D_\nu z^iD_\mu
(T^az)^j = - \D_\nu z^i\rho_{ij}(T^az)^j\(\sqrt{x}\D_\mu''\F_a^{\nu\mu} - 
i\tF_a^{\nu\mu}\pp_\mu x\) \nonumber \\ & & \qquad
- x(T^az)^j\(F_a^{\nu\mu} - i\tF_a^{\nu\mu}\)\(\rho_{ij}D_\mu\D_\nu z^i
+ \rho_{\m ij}\D_\mu\z^{\m}\D_\nu z^i\) + {\rm total\;derivative},
\nonumber \\ & &
-iF^a_{\mu\nu}\[\D^\mu z^i\D^\nu \z^{\m}K_{j\m}D_i(T_az)^j 
- {\rm h.c.}\] = {\rm total \; deriv.} \nonumber \\ & &
\qquad + i\D^\mu F^a_{\mu\nu}K_{i\m}\[\D^\nu\z^{\m}(T_az)^i -
\D^\nu z^i(T_a\z)^{\m}\] + xF^a_{\mu\nu}F_b^{\mu\nu}\K^b_a, \nonumber \\ & &
\qquad F^a_{\mu\nu}D^\mu\D^\nu z^I = \pm {i\over2}F^a_{\mu\nu}F_b^{\mu\nu}
(T^b z)^I, \;\;\;\; I = \cases{i\cr\ibar\cr}. \eea

To evaluate the fermion matrix elements we use (3.34); we have
\bea {1\over 8}\btr\(H_1^{\chi g}\)^2 &=& 
\Tr h_{\chi g}^2 + 2\[\(\m M^{\mu\nu}\)^i_a\(\m M_{\mu\nu}\)^a_i + 
\(\M^{\mu\nu}m\)^i_a\(\M_{\mu\nu}m\)^a_i + {\rm h.c.}\]
\nonumber \\ &=& - T_4^{\chi g} + e^{-K}\D\(2a_i\aa^i + 8 a\aa \)
+ 2\D\(\hV - M_\psi^2\) \nonumber \\ & &
+ {e^{-K}\over x}\[ 4(T_az)^iA_{ij}(T^az)^j\(\aa - \A\) - 2\((T_az)^i
\D^aA_{ij}\aa^j + {\rm h.c.}\)\] 
\nonumber \\ & &
+ 4{e^{-K}\over x}(T_az)^i(T^a\z)^{\n}A_{ki}\A_{\n}^k + 
2M_\lambda^2\[2\K^a_a + \(3x^2\rho^i\rho_i - 4\)\D\]\nonumber \\ & & 
+ 2{e^{-K}\over x}\lbr a_i\(\aa - \A\)\[\f^i\D - (T^a z)^i\D_a\]
+ {\rm h.c.} \rbr, \nonumber \eea \bea
-{1\over 8}\btr\(H^{\chi g}_2\)^2 &=& 2\(\tD_\mu\m\)^i_a\(\tD^\mu m\)^a_i 
- 8\(\tD_\mu\M^{\mu\nu}\)^i_a\(\tD^\rho M_{\rho\nu}\)^a_i 
\nonumber \\ & & + \lbr [\hL_{\mu\nu},\m]^i_a\(M^{\mu\nu}\)^a_i +
[\hL_{\mu\nu},\m]_{\m}^a\(M^{\mu\nu}\)_a^{\m} + {\rm h.c.}\rbr\nonumber \\ & & 
+ 2[L_\mu,\m]^i_a[L^\mu, m]^a_i 
+ 8L^\rho L_\mu\(\M_{\rho\nu}\)^i_a\(M^{\mu\nu}\)^a_i  \eea  
with \bea & & - 8\(\tD_\mu\M^{\mu\nu}\)^i_a\(\tD^\rho M_{\rho\nu}\)^a_i = 
- {x^2\rho_i\rho^i\over 4}\(\D''_\nu\F^{\nu\mu} + {\pp_\nu y\over\sqrt{x}}
\tF^{\nu\mu}\)^2 \nonumber \\ & & \qquad
+ {x\rho_i\rho^i\over 4}\(\sqrt{x}\D''_\nu\F^{\nu\mu} + \pp_\nu y 
\tF^{\nu\mu} - {1\over2}\pp_\nu xF^{\nu\mu}\)\pp^\rho xF_{\rho\mu} 
\nonumber \\ & & \qquad
+ {x\rho_i\rho^i\over 4}\({1\over8}\pp_\rho x\pp^\rho xF^{\nu\mu}F_{\nu\mu} + 
{1\over4}\pp_\rho x\pp^\rho yF^{\nu\mu}\tF_{\nu\mu} - \pp_\nu y\pp^\rho y 
F^{\nu\mu}F_{\rho\mu}\) \nonumber \\ & & \qquad 
+ {1\over8}\lbr\D^\rho z^i\rho_{ij}\f^j\(F_{\rho\mu} - i\tF_{\rho\mu}\)\[
\sqrt{x}\D''_\nu\F^{\nu\mu} - i\pp_\nu y\(F^{\nu\mu} + i\tF^{\nu\mu}\)
\] + {\rm h.c.}\rbr \nonumber \\ & & \qquad 
- {1\over32}\[\D^\rho z^i\rho_{ij}\f^j\pp_\rho x\(F^{\nu\mu}F_{\nu\mu} -i
F^{\nu\mu}\tF_{\nu\mu}\) + {\rm h.c.}\] \nonumber \\ & & \qquad 
- {x^3\over2}\D^\rho z^i\D_\nu\z^{\m}\rho_{ij}\rho^j_{\m}F^{\nu\mu}F_{\rho\mu} 
+ {x^3\over8}\D^\rho z^i\D_\rho\z^{\m}\rho_{ij}\rho^j_{\m}F^{\nu\mu}F_{\nu\mu} 
,\nonumber \\ & & 8L^\rho L_\mu\(\M_{\rho\nu}\)^i_a\(M^{\mu\nu}\)^a_i = 
{x\rho^i\rho_i\over32}\(4\pp_\nu y\pp^\rho yF_{\rho\mu}F^{\nu\mu} - 
\pp_\rho y\pp^\rho yF_{\mu\nu}F^{\mu\nu}\),
\nonumber \\ & & \lbr [\hL_{\mu\nu},\m]^i_a\(M^{\mu\nu}\)^a_i +
[\hL_{\mu\nu},\m]_{\m}^a\(M^{\mu\nu}\)_a^{\m} + {\rm h.c.} \rbr = 
\tau^{\chi g}_3, \eea
and, using (C.40),
\bea & & 2\(\tD_\mu\m\)^i_a\(\tD^\mu m\)^a_i + 2[L_\mu,\m]^i_a[L^\mu, m]^a_i = 
-2{\pp^\mu x\over x^2}\[(T_a z)^iK_{i\m}D_\mu(T^a\z)^{\m} + {\rm h.c.}\] 
\nonumber \\ & & \qquad +
{4\over x}K_{i\m}D_{\n}(T_a\z)^{\m}\D_\mu \z^{\n}D_j(T^az)^i\D^\mu z^j 
+ 4x\K^b_aF^a_{\mu\nu}F_b^{\mu\nu} \nonumber \\ & & \qquad
+ x\rho_i\rho^i\lbr K_{i\n}K_{j\m}\D^\mu z^j
(T_a\z)^{\m}\[(T^az)^i\D_\mu\z^{\n} + (T^a\z)^{\n}\D_\mu z^i\] + {\rm h.c.}\rbr
\nonumber \\ & & \qquad
- \pp^\mu x\rho_i\rho^i\D^a K_{j\m}\[
(T_az)^j\D_\mu \z^{\m} + (T_a\z)^{\m}\D_\mu z^j\] \nonumber \\ & & \qquad
- 2\lbr K_{j\m}(T_a\z)^{\m}\D_\mu z^j\[\rho_{ik}(T^az)^i\D^\mu z^k 
+ \rho_{\m\n}(T^a\z)^{\m}\D^\mu\z^{\n}\] + {\rm h.c.}\rbr \nonumber \\ & & 
\qquad
- {1\over 2x}\D_a\lbr \rho_{ij}\D_\mu z^i\f^jK_{k\m}\[(T^az)^k\D^\mu \z^{\m} 
+ (T^a\z)^{\m}\D^\mu z^k\] + {\rm h.c.}\rbr \nonumber \\ & & \qquad
+ 2\D_a\[\rho_{ij}\D_\mu z^i\D^\mu z^k\D_k(T^az)^j + {\rm h.c.}\]
- {2i\pp_\mu y\over x}\D_a\[\rho_{ik}(T^az)^i\D^\mu z^k - {\rm h.c.}\]
\nonumber \\ & & \qquad
+ \(\pp_\mu x\pp^\mu x + 3\pp_\mu y\pp^\mu y\){\rho_i\rho^i\over2}\D 
+ 2x^2\D\rho_{ij}\D_\mu z^i\D^\mu \z^{\m}\rho^j_{\m} \nonumber \\ & & \qquad
+ \[{2\over x}F_a^{\mu\nu}\pp_\nu yK_{i\m}\D_\mu z^i(T^a\z)^{\m} - 
x\D\rho^i\rho_{ij}\D_\mu z^i\(\pp_\mu x + 2i\pp_\mu y\) + {\rm h.c.}\]
\nonumber \\ & & \qquad
+ 4xF^a_{\mu\nu}F_a^{\mu\rho}\D_\rho z^i\D^\nu\z^{\m}K_{i\m}
+ \K^a_a\( {\pp_\mu x\pp^\mu x\over x^2} 
- {\pp_\mu y\pp^\mu y\over x^2}\) \nonumber \\ & & \qquad
+ 2i\({2\over\sqrt{x}}\D''_\mu\F_a^{\mu\nu} -{\pp_\mu x\over x}F_a^{\mu\nu}\)
K_{i\m}\[\D_\nu\z^{\m}(T_az)^i - \D_\nu z^i(T_a\z)^{\m}\]. \eea
We write the $\chi$-$\lambda$ contribution to $T$ as 
\bea T^{\chi g} &=& T_4^{\chi g} + T_3^{\chi g} + \tau^{\chi g}_3 + t_3^{\chi g}
= T'_{\chi g} + t_3^{\chi g}, \nonumber \\
T_4^{\chi g} &=& -4\(\m M^{\mu\nu}\)^i_a\(\m M_{\mu\nu}\)^a_i + {\rm h.c.}
\nonumber \\ &=& \(x\cW + \D\)\(x^2\rho^i\rho_iM_\lambda^2 + (a-A)
\aa^i{f_i\over2x}\) + {\rm h.c.}, \nonumber \\
t_3^{\chi g} &=& - {16\over3}\[\(\tD^\sigma\M_{\sigma\mu}\)^i_a\(\tD_\rho 
M^{\rho\mu}\)^a_i + L^\sigma L_\rho\(\M_{\sigma\mu}\)^i_a\(M^{\rho\mu}\)^a_i
\] \nonumber \\ & &
+ {16i\over 3}L^\sigma\[\(\M_{\sigma\mu}\)^i_a\(\tD_\rho M^{\rho\mu}\)^a_i
- {\rm h.c.} \],\nonumber \\ 
\tau^{\chi g}_3 &=& - {\rho^i\rho_i\over2}\pp_\mu y\pp_\nu xF_a^{\mu\nu}\D^a 
+ {\pp_\mu y\over 2x}F^a_{\rho\nu}\(\pp^\rho x\tF^{\mu\nu}_a - 
\pp^\rho yF^{\mu\nu}_a\)\nonumber \\ 
T_3^{\chi g} &=&  2iL^\rho m^a_i\tD_\rho\m^i_a + {\rm h.c.} \nonumber \\ 
&=& - 2\pp_\mu y\pp^\mu y \rho_i\rho^i\D + 2{\pp_\mu y\over x}F^{\mu\nu}_a
\[K_{i\m}\D_\nu z^i(T^a\z)^{\m} + {\rm h.c.}\] + {\pp_\mu x\pp_\nu y\over x^2}
F^{\mu\nu}_a\D^a \nonumber \\ & & 
- {i\pp_\mu y\over 2x}\lbr\D^\mu z^i\[\D\f^j\rho_{ij} + 
{4\over x}K_{j\m}(T^a\z)^{\m}D_i(T^az)^j\] - {\rm h.c.}\rbr \nonumber \\ & & 
+ {2i\over x}\D^a\pp^\mu y\[\rho_{ij}(T_az)^i\D_\mu z^j - {\rm h.c.}\], \eea
where
\bea & & 8iL^\sigma\(\M_{\sigma\mu}\)^i_a\(\tD_\rho M^{\rho\mu}\)^a_i
+ {\rm h.c.} = {i\over32}\[\D^\rho z^i\rho_{ij}\f^j\(4\pp_\nu yF_{\mu\rho}^a
F_a^{\mu\nu} - \pp_\rho yF_a^{\mu\nu}F^a_{\mu\nu}\) - {\rm h.c.}\]
\nonumber \\ & & \qquad + {x\rho_i\rho^i\over4}\pp^\rho y\[\tF^a_{\rho\nu}
\(\sqrt{x}\D''_\mu\F_a^{\mu\nu} + \pp_\mu y\tF_a^{\mu\nu} - \pp_\mu x
F_a^{\mu\nu}\) + \pp_\nu y F_{\mu\rho}^aF_a^{\mu\nu}\].\eea
In addition we have
\bea \Tr\(\hG_\Phi^{\chi g}\)^2 &=& 4\(G^{gz}_{\mu\nu}\)_{a\rho,i}
\(G^{gz}_{\mu\nu}\)^{i,a\rho } = \btr\(\hG_\Theta^{\chi g}\)^2 
\nonumber \\ &=& 64\(\tD_\mu\M_{\nu\rho}\)_a^i\(\tD^\mu M^{\nu\rho} - 
\tD^\nu M^{\mu\rho}\)_i^a - 128i\[L^\nu\(\M_{\nu\rho}\)_a^i
\(\tD_\mu M^{\mu\rho}\)_i^a - {\rm h.c.}\],  \nonumber \eea\bea 
& & 64\(\tD_\mu\M_{\nu\rho}\)_a^i\(
\tD^\mu M^{\nu\rho} - \tD^\nu M^{\mu\rho}\)_i^a = - 2x^2 
\rho_i\rho^i\(\D''_\mu\F_a^{\mu\nu} + {\pp_\mu y\over\sqrt{x}}
\tF_a^{\mu\nu}\)^2 \nonumber \\ & & \qquad
- {1\over4}\lbr\D^\rho z^i\rho_{ij}\f^j\[F_a^{\mu\nu}F^a_{\mu\nu}\pp_\rho x -
i\tF_a^{\mu\nu}F^a_{\mu\nu}\(\pp_\rho x + i\pp_\rho y\)
+ 4i\pp_\nu y F_{\mu\rho}^aF_a^{\mu\nu}\] + {\rm h.c.}\rbr 
\nonumber \\ & & \qquad + \(\sqrt{x}\D''_\mu\F_a^{\mu\nu} + \pp_\mu y 
\tF_a^{\mu\nu}\)\lbr 2x\rho^i\rho_i\pp^\rho xF^a_{\rho\nu} + \[\(F^a_{\rho\nu} 
- i\tF^a_{\rho\nu}\)\D^\rho z^i\rho_{ij}\f^j + {\rm h.c.}\]\rbr 
\nonumber \\ & & \qquad 
+ F_a^{\mu\nu}F^a_{\mu\nu}\[{x\over4}\rho_i\rho^i\pp_\rho x\pp^\rho x 
+ x^3\rho_{ij}\rho^j_{\m}\D_\rho z^i\D^\rho\z^{\m}\]
+ {x\rho_i\rho^i\over2}\tF_a^{\mu\nu}F^a_{\mu\nu}\pp^\rho y\pp_\rho x
\nonumber \\ & &
\qquad - F_a^{\mu\rho}F^a_{\nu\rho}\[x\rho_i\rho^i\(\pp_\mu x\pp^\nu x + 2
\pp_\mu y\pp^\nu y\) + 4x^3\rho_{ij}\rho^j_{\m}\D_\mu z^i\D^\nu\z^{m}\],\eea
Using the classical equations of motion (B.17--20), we obtain, with
$k^1 = -4\K^a_a$,
\bea {1\over2}\STr H^2_{\chi g} &=& - T'_{\chi g} + \(\D_\mu z^i\D^\mu\z^{\m} + 
\A^iA^{\m}e^{-K}\)\(k^1_{i\m} - {4\over x}(T^az)^j(T_a\z)^{\n}R_{i\m j\n}\)
\nonumber \\ & & - e^{-K}\(k^1_i\A^iA + {\rm h.c.}\) 
- {4x^2\rho^i\rho_i\over\sqrt{g}}\Delta_{\D}\L - {x^2\rho^i\rho_i\over gx}
\L_{a\mu}\L^{a\mu} \nonumber \\ & & 
+ {2x\rho^i\rho_i\over\sqrt{g}}\[i\L^{a\mu}\(
K_{i\m}\D_\mu\z^{\m}(T_az)^i - {\rm h.c.}\) + \D^a(T_az)^I\L_I\] 
\nonumber \\ & & 
+ {1\over\sqrt{g}}\L_a^\rho\lbr x\rho^i\rho_i
\pp^\mu y\tF_{\mu\rho}^a + \[{\f^i\over2}\rho_{ij}\D^\nu z^j\(F^a_{\nu\rho} 
- i\tF^a_{\nu\rho}\)+ {\rm h.c.}\]\rbr \nonumber \\ & &
- {1\over12}\STr\hG_{\chi g}^2 - t_3^{\chi g} + 4x^2\rho^i\rho_i\D\[3M_\psi^2 
+ \hV - e^{-K}a\aa \] + 10x^2\rho^i\rho_i\D M_\lambda^2 
\nonumber \\ & & + 4x\[\rho_{ij}(T^az)^i(T^bz)^j 
\(\cW_{ab} + {1\over2x^2}\D_a\D_b\) + {\rm h.c.}\] - 4M_\psi^2\K^a_a 
\nonumber \\ & & + 2\[ix\rho_{\m ij}\D_\nu\z^{\m}(T^az)^i\D_\mu z^j
\(F_a^{\mu\nu} - i\tF_a^{\mu\nu}\) + {\rm h.c.}\] \nonumber \\ & &
- {3x\rho_i\rho^i\over4}\(F^{\nu\mu} - i\tF^{\nu\mu}\)\(F_{\rho\mu} + 
i\tF_{\rho\mu}\)\(\pp_\nu x\pp^\rho x + \pp_\nu y\pp^\rho y\) \nonumber \\ & & 
- \({1\over x^2} + \rho^i\rho_i\)\pp_\mu y\pp_\nu x\D^aF_a^{\mu\nu} 
+ {\rho_i\rho^i\over2}\D\(5\pp_\mu x\pp^\mu x + 3\pp_\mu y\pp^\mu y\) 
\nonumber \\ & &
 -ix\rho^i\rho_iK_{i\m}\[\D^\rho\z^{\m}(T_az)^i - \D^\rho z^i(T_a\z)^{\m}\]
\pp^\mu y\tF_{\mu\rho}^a \nonumber \\ & &
+ \lbr\cW\[2x^3\rho^i\rho_i M_\lambda^2 + (a-A)f_i\aa^ie^{-K} - \(\pp_\rho x + 
i\pp_\rho y\)\D^\rho z^j\rho_{ij}{\f^i\over2}\] + {\rm h.c.}\rbr 
\nonumber \\ & & + 2x^2\rho_{ij}\rho^j_{\m}\D\D_\rho z^i\D^\rho\z^{\m} 
- 2x^2\[\(F_a^{\nu\mu}F^a_{\rho\mu} -iF_a^{\nu\mu}\tF^a_{\rho\mu}\)
\D^\rho z^i\D^\nu z^j\rho_{ij} + {\rm h.c.}\] \nonumber \\ & &
+ 2{\pp^\mu x\over\sqrt{x}}F^a_{\mu\rho}\D''_\nu\F_a^{\nu\rho} +
{\pp_\nu y\pp^\nu x\over x}F_a^{\mu\rho}\tF^a_{\mu\rho} 
+ F_a^{\nu\mu}F^a_{\rho\mu}\({\pp_\nu x\pp^\rho x\over x} -
{\pp_\nu y\pp^\rho y\over x}\) \nonumber \\ & &
- e^{-K}\D\(2a_i\aa^i + 16a\aa \)
+ 2\D\(3\hV + 17M_\psi^2\) + 4M_\lambda^2\(\D + \K^a_a\)\nonumber \\ & &
+ {e^{-K}\over x}\lbr 2\D^a(T_az)^iA_{ij}\(\aa^j - 2\A^j\)- 
a_i\(\aa - \A\)\[\f^i\D - 2(T^az)^i\D_a\] + {\rm h.c.}\rbr \nonumber \\ & &
- 4x\K^b_aF^a_{\mu\nu}F_b^{\mu\nu} - \rho_i\rho^i\[\(\pp^\mu x + 2i\pp^\mu y\) 
K_{j\m}(T_az)^j\D_\mu \z^{\m} + {\rm h.c.}\]\D^a \nonumber \\ & & 
- {i\over2}K_{i\m}\(\D^\nu\z^{\m}(T_az)^i - \D^\nu z^i(T_a\z)^{\m}\)
\[\f^i\rho_{ij}\D^\rho z^j\(F^a_{\rho\nu} 
- i\tF^a_{\rho\nu}\)+ {\rm h.c.}\] \nonumber \\ & &
- 2\lbr K_{j\m}(T_a\z)^{\m}\D_\mu z^j\[\rho_{ik}(T^az)^i\D^\mu z^k 
+ \rho_{\bell\n}(T^a\z)^{\bell}\D^\mu\z^{\n}\] + {\rm h.c.}\rbr 
\nonumber \\ & & 
+ {1\over 2x}\D_a\lbr \rho_{ij}\D_\mu z^i\f^jK_{k\m}\[(T^az)^k\D^\mu \z^{\m} 
+ (T^a\z)^{\m}\D^\mu z^k\] + {\rm h.c.}\rbr \nonumber \\ & & 
+ \lbr \rho_{ij}\D^\nu z^i\[2\D_a\D_k(T^az)^j\D_\nu z^k 
+ \f^jF_a^{\mu\rho}\({\pp_\mu x\over2}F^a_{\nu\rho} - 
{\pp_\nu x\over8}F^a_{\mu\rho}\)\] + {\rm h.c.}\rbr
\nonumber \\ & & - 4xF^a_{\mu\nu}F_a^{\mu\rho}\D_\rho z^i\D^\nu\z^{\m}K_{i\m}
+ \K^a_a\({\pp_\mu x\pp^\mu x\over x^2} 
+ {\pp_\mu y\pp^\mu y\over x^2}\) \nonumber \\ & & 
- 2i\(2{\D''_\mu\F_a^{\mu\nu}\over\sqrt{x}} - {\pp_\mu x\over x}F_a^{\mu\nu}\)
K_{i\m}\[\D_\nu\z^{\m}(T_az)^i - \D_\nu z^i(T_a\z)^{\m}\] \nonumber \\ & &
+ {x\rho_i\rho^i\over4}\(F_a^{\mu\nu}F^a_{\mu\nu}\pp_\rho x\pp^\rho x +
F_a^{\mu\nu}\tF^a_{\mu\nu}\pp_\rho x\pp^\rho y\) \nonumber \\ & & 
- {13x\rho_i\rho^i\over96}\(F_a^{\mu\nu}F^a_{\mu\nu}\pp_\rho y\pp^\rho y - 
4F_a^{\mu\rho}F^a_{\nu\rho}\pp_\mu y\pp^\nu y\), 
\nonumber \\ {1\over12}\STr\hG_{\chi g}^2 &=& - t_3^{\chi g}
- {x^2\rho_i\rho^i\over4}\(\D''_\mu\F_a^{\mu\nu} +
{\pp_\mu y\over\sqrt{x}}\tF_a^{\mu\nu}\)^2 \nonumber \\ & & 
+  x\rho_i\rho^i\[{\pp^\rho x\over4}F^a_{\rho\nu} 
\(\sqrt{x}\D''_\mu\F_a^{\mu\nu} + \pp_\mu y\tF_a^{\mu\nu}\)
+ {\pp^\rho y\pp_\rho x\over16}\tF_a^{\mu\nu}F^a_{\mu\nu}\]\nonumber \\ & & 
+ {1\over8}\[\(F^a_{\rho\nu} - i\tF^a_{\rho\nu}\)
\D^\rho z^i\rho_{ij}\f^j + {\rm h.c.}\]\(\sqrt{x}\D''_\mu\F_a^{\mu\nu} +
\pp_\mu y\tF_a^{\mu\nu}\) \nonumber \\ & &
+ {1\over32}F^{\mu\nu}_a\tF^a_{\mu\nu}\[i\(\pp_\rho x + i\pp_\rho y\)
\D^\rho z^i\rho_{ij}\f^j + {\rm h.c.}\] \nonumber \\ & &
- {x^3\over2}\rho_{ij}\rho^j_{\m}\(F_a^{\mu\rho}F^a_{\nu\rho}
\D_\mu z^i\D^\nu\z^{m} - {1\over 4}F_a^{\mu\nu}F^a_{\mu\nu}
\D_\rho z^i\D^\rho\z^{m}\) \nonumber \\ & &
+ {x\rho_i\rho^i\over32}\[F_a^{\mu\nu}F^a_{\mu\nu}\pp_\rho x\pp^\rho x
- 4F_a^{\mu\rho}F^a_{\nu\rho}\(\pp_\mu x\pp^\nu x + 2\pp_\mu y\pp^\nu y\)\] 
\nonumber \\ & &
- {1\over32}\[\(F^a_{\mu\nu}\pp_\rho x + 4iF^a_{\mu\rho}\pp_\nu y\)F_a^{\mu\nu}
\rho_{ij}\f^i\D^\rho z^j + {\rm h.c.}\] \nonumber \\ & &
+ {x\rho_i\rho^i\over48}\(F_a^{\mu\nu}F^a_{\mu\nu}\pp_\rho y\pp^\rho y
- 4F_a^{\mu\rho}F^a_{\nu\rho}\pp_\mu y\pp^\nu y\). \eea

\subsubsection {Mixed chiral-gravity supertraces}

For the bosonic sector $H_\Phi^{\chi G} = S$; using (C.39) we obtain
\bea \Tr Y^2 &=& \Tr Y^2_0 + 2x\K^a_bF^b_{\mu\nu}F_a^{\mu\nu} \nonumber \\ & &
+ {x^4\rho_i\rho^i\over8}\[\(F^a_{\mu\nu}F_b^{\mu\nu}\)^2
+ \(F^a_{\mu\nu}\tF_b^{\mu\nu}\)^2 + \(F^a_{\mu\nu}F_a^{\mu\nu}\)^2 + 
\(F^a_{\mu\nu}\tF_a^{\mu\nu}\)^2\] \nonumber \\ & &
+ 2i\({\D''_\mu\F_a^{\mu\rho}\over\sqrt{x}} -{\pp_\mu x\over x}F_a^{\mu\rho} \)
K_{i\m}\[\D_\rho\z^{\m}(T^az)^i - \D_\rho z^i(T^a\z)^{\m}\]
\nonumber \\ & &
- x^2\[\rho_{ij}\D_\mu z^i\D^\mu z^jF^a_{\nu\rho}\(F_a^{\nu\rho} - {i\over2}
\tF_a^{\nu\rho}\) - 2\rho_{ij}\D_\nu z^i\D^\mu z^j F^a_{\mu\rho} F_a^{\nu\rho} 
+ {\rm h.c.} \]
\nonumber \\ & &
- F^a_{\mu\nu}F_a^{\mu\nu}\({\nabla^2x\over2} - {\pp_\rho x\pp^\rho x 
- \pp_\rho y\pp^\rho y\over x}\) + 
{2\pp_\mu y\pp^\nu y\over x}F_a^{\mu\rho}F^a_{\nu\rho}
\nonumber \\ & &
-2F^a_{\mu\rho}\D''_\nu\F_a^{\nu\rho}{\pp^\mu x\over\sqrt{x}} 
-F^a_{\mu\nu}\tF_a^{\mu\nu}\({\nabla^2y\over2} - 
{\pp_\rho x\pp^\rho y\over x}\),\eea  
For the fermions, we have
\bea {1\over 8}\btr\(H_1^{\chi G}\)^2 &=& 
\Tr h_{\chi G}^2 + 4\[\(\m M^{\mu\nu}\)^i_\alpha\(\m M_{\mu\nu}\)^\alpha_i + 
{\rm h.c.}\] + 8\(\M_{\mu\nu}M_{\rho\sigma}\)^i_\alpha
\(\M^{\mu\nu}M^{\rho\sigma}\)^\alpha_i \nonumber \\ & & + 
16\(\M^{\mu\nu}M_{\mu\nu}\)^i_\alpha\(\M^{\rho\sigma}M_{\rho\sigma}\)^\alpha_i 
- 32\(\M^{\mu\nu}M^{\rho\sigma}\)^i_\alpha\(\M_{\mu\rho}M_{\nu\sigma}\)^\alpha_i
\;\;\;\;\;\; \nonumber \\
&=& {1\over 8}\btr\(H_1^{\chi G}\)^2_0 + 2e^{-K}\D\(A_i\aa^i + {\rm h.c.}\) 
+ 4\D\(2M_\lambda^2 - \hV - e^{-K}a\aa \) \nonumber \\ & &
- 2x^2\rho^i\rho_i\D^2 - {2\over x}\D_a\D_b\K^{ab} - 2
{e^{-K}\over x}\D^a\[(T_az)^iA_{ij}\A^j + {\rm h.c.}\] \nonumber \\ & &
- {x^2\over4}\rho^i\rho_i\D_a\D^bF^a_{\mu\nu}F_b^{\mu\nu}
- {x^4\rho_i\rho^i\over32}\[\(F^a_{\mu\nu}F_a^{\mu\nu}\)^2 + 
\(F^a_{\mu\nu}\tF_a^{\mu\nu}\)^2\], \nonumber \eea \bea
&& - {1\over 8}\btr\(H_2^{\chi G}\)^2 = 2\sum_{A=\alpha,\nu}\(\tD_\mu\m\)^i_A
\(\tD^\mu m\)^A_i - 8\(\tD_\sigma\M^{\sigma\nu}\)^i_\mu
\(\tD^\rho M_{\rho\nu}\)^\mu_i = \nonumber \\ &&\quad - 
{1\over 8}\btr\(H_2^{\chi G}\)^2_0 
- x^2F^a_{\mu\nu}F_b^{\mu\nu}\rho^i\rho_i\D_a\D^b - 8
\(\tD_\sigma\M^{\sigma\nu}\)^i_\mu\(\tD^\rho M_{\rho\nu}\)^\mu_i,\eea 
where \bea \(\tD_\sigma\M^{\sigma\nu}\)^i_\mu\(\tD^\rho M_{\rho\nu}\)^\mu_i =
- {x^4\rho_i\rho^i\over128}\[\(F^a_{\mu\nu}F_a^{\mu\nu}\)^2 + 
\(F^a_{\mu\nu}\tF_a^{\mu\nu}\)^2 + \(F^a_{\mu\nu}F_b^{\mu\nu}\)^2 + 
\(F^a_{\mu\nu}\tF_b^{\mu\nu}\)^2\],\;\;\;\;\eea
giving
\bea {1\over2}\STr H^2_{\chi G} &=& {1\over2}\(\STr H^2_{\chi G}\)_0 
- 2e^{-K}\D\(A_i\aa^i + 
{\rm h.c.}\) - 4\D\(2M_\lambda^2 - \hV - e^{-K}a\aa \) \nonumber \\ & &
+ 2x^2\rho^i\rho_i\D^2 + {2\over x}\D_a\D_b\K^{ab} 
+ 2{e^{-K}\over x}\D^a\[(T_az)^iA_{ij}\A^j + {\rm h.c.}\] +
2xF^a_{\mu\nu}F_b^{\mu\nu}\K^a_b \nonumber \\ & &
+ 2i\({\D''_\mu\F_a^{\mu\rho}\over\sqrt{x}} -{\pp_\mu x\over x}F_a^{\mu\rho} \)
K_{i\m}\[\D_\rho\z^{\m}(T^az)^i - \D_\rho z^i(T^a\z)^{\m}\] \nonumber \\ & &
- 2F^a_{\mu\rho}\D''_\nu\F_a^{\nu\rho}{\pp^\mu x\over\sqrt{x}} - 
F^a_{\mu\nu}\tF_a^{\mu\nu}\({\nabla^2y\over2} - {\pp_\rho x\pp^\rho y\over x}\)
+ {2\pp_\mu y\pp^\nu y\over x}F_a^{\mu\rho}F^a_{\nu\rho}  \nonumber \\ & &
- F^a_{\mu\nu}F_a^{\mu\nu}\({\nabla^2x\over2} - {\pp_\rho x\pp^\rho x 
- \pp_\rho y\pp^\rho y\over x}\) 
- {3\over4}x^2\rho^i\rho_i\D_a\D^bF^a_{\mu\nu}F_b^{\mu\nu} 
\nonumber \\ & & + {x^4\rho^i\rho_i\over32}\[7\(F_a^{\mu\nu}F^a_{\mu\nu}\)^2 + 
7\(F_a^{\mu\nu}\tF^a_{\mu\nu}\)^2 + 6\(F_a^{\mu\nu}F^b_{\mu\nu}\)^2 +
6\(F_a^{\mu\nu}\tF^b_{\mu\nu}\)^2\] \nonumber \\ & & 
- x^2\lbr\rho_{ij}\D_\mu z^i\[\D^\mu z^jF^a_{\nu\rho}\(F_a^{\nu\rho} - {i\over2}
\tF_a^{\nu\rho}\) - 2\D^\nu z^j F^a_{\nu\rho} F_a^{\mu\rho}\] + {\rm h.c.}\rbr 
.\eea
The contribution to $T$ is 
\bea T^{\chi G} &=& T_4^{\chi G} + T_3^{\chi G}, \nonumber \\
T_3^{\chi G} &=& -{16\over3}\(\tD_\sigma\M^{\sigma\nu}\)^i_\mu
\(\tD^\rho M_{\rho\nu}\)^\mu_i,\nonumber \\  
T_4^{\chi G} &=& -4\[\(\m M^{\mu\nu}\)^i_\alpha\(\m M_{\mu\nu}\)^\alpha_i + 
{\rm h.c.}\] + {16\over3}\(\M^{\rho\sigma}M^{\mu\nu}\)^i_\alpha
\(\M_{\rho\sigma}M_{\mu\nu}\)^\alpha_i  \nonumber \\ 
&=& {x^2\rho_i\rho^i\over48}\[12F_a^{\mu\nu}F^b_{\mu\nu}\D^a\D_b
- \(F_a^{\mu\nu}F^b_{\mu\nu}\)^2 - \(F_a^{\mu\nu}\tF^b_{\mu\nu}\)^2\], \eea
which for future reference [see (C.59,62)] we write as 
\beq 2T_4^{\chi G} + T_3^{\chi G} = 
x^2\rho_i\rho^i\lbr{1\over24}\[\(F_a^{\mu\nu}F^a_{\mu\nu}\)^2 + 
\(F_a^{\mu\nu}\tF^a_{\mu\nu}\)^2\] + {1\over2}F_a^{\mu\nu}F^b_{\mu\nu}\D^a\D_b
\rbr. \eeq
The contribution to $\STr\hG^2$ is 
\bea \Tr\(\hG^\Phi_{z G}\)^2 &=&
x^4\rho_i\rho^i\[\(F^a_{\mu\nu}F_b^{\mu\nu}\)^2 + 
\(F^a_{\mu\nu}\tF_b^{\mu\nu}\)^2\],\nonumber \\
{1\over2}\btr\(\hG^\Theta_{\chi G}\)^2 &=& x^4\rho_i\rho^i
\[\(F^a_{\mu\nu}F_a^{\mu\nu}\)^2 + \(F^a_{\mu\nu}\tF_a^{\mu\nu}\)^2 + 
\(F^a_{\mu\nu}F_b^{\mu\nu}\)^2 + \(F^a_{\mu\nu}\tF_b^{\mu\nu}\)^2\], 
\nonumber \eea \beq {1\over12}\STr\hG_{\chi G}^2 =
{x^4\rho_i\rho^i\over24}\[\(F^a_{\mu\nu}F_b^{\mu\nu}\)^2 + 
\(F^a_{\mu\nu}\tF_b^{\mu\nu}\)^2 - \(F^a_{\mu\nu}F_a^{\mu\nu}\)^2 -
\(F^a_{\mu\nu}\tF_a^{\mu\nu}\)^2\]. \eeq

\subsubsection{Yang-Mills and gravity supertraces}

For the remaining bosonic contributions, we have $H^{g+G}_\Phi = X - N - K$;
we write $N_{ab} = N'_{ab} + \delta_{ab}n$, and evaluate separately in the next
subsection the terms that depend only on $n$ and are proportional to $N_G$, the
number of gauge degrees of freedom.  Then:
\bea \Tr X &=& \Tr X_0 - 20\D + 
xF_{\mu\nu}^aF_a^{\mu\nu}, \nonumber \\ 
\Tr X^2 &=& \Tr X^2_0 +40\D^2 + 80\D\hV - 8r\D + 
xF_{\mu\nu}^aF_a^{\mu\nu}\(r - 4V\) \nonumber \\ & &
+ 2r^\mu_\nu xF_{\mu\rho}^aF_a^{\nu\rho} 
- 6xr_{\mu\nu}^{\;\;\;\;\;\rho\sigma}F_a^{\mu\nu}F^a_{\rho\sigma} 
- {3x^2\over8}\(F_{\mu\nu}^aF_a^{\mu\nu}\)^2 \nonumber \\ & &
+ {x^2\over2}\(F_{\mu\nu}^a\tF_a^{\mu\nu}\)^2 
+ {29x^2\over8}\(F_{\mu\nu}^aF_b^{\mu\nu}\)^2 
+ {5x^2\over8}\(F_{\mu\nu}^a\tF_b^{\mu\nu}\)^2, \nonumber \\
\Tr N &=& 8\K^a_a + N_Gn, \nonumber \\
\Tr N^2 &=& 8(\K_{ab}\K^{ba} + \K_{ab}\K^{ab}) 
+ 4C_G^aF^a_{\mu\nu}F_a^{\mu\nu} + N_Gn^2 + 4r\K^a_a \nonumber \\ & &
+ 4x\(1 - {x^2\rho_i\rho^i\over2}\)\(r_\nu^\mu F^a_{\mu\rho}F_a^{\nu\rho} 
- {1\over 4}rF^a_{\mu\nu}F_a^{\mu\nu}\) + 
12xc_{abc}F^a_{\mu\nu}F_\rho^{b\;\;\mu}F^{c\rho\nu} \nonumber \\ & &
- 4\K^a_a\({\nabla^2x\over x} + {3\pp^\mu y\pp_\mu y\over 2x^2}\) 
- 2\(2 - x^2\rho_i\rho^i\){\pp^\mu x\over\sqrt{x}}F^a_{\mu\rho}
\D''_\nu\F_a^{\nu\rho} \nonumber \\ & &
+ \(1 - {x^2\rho_i\rho^i\over2}\)\[F_{\mu\nu}^aF_a^{\mu\nu}
\({\pp_\rho x\pp^\rho x\over x} - {\pp_\rho y\pp^\rho y\over 2x}\) 
+ 2F_{\mu\rho}^aF_a^{\nu\rho}{\pp_\nu y\pp^\mu y \over x}\]
\nonumber \\ & &
- \(\D^\nu z^i\rho_{ij}\f^j + {\rm h.c.}\)F_a^{\mu\rho}\(\pp_\mu x F^a_{\nu\rho}
- {\pp_\nu x\over4}F^a_{\mu\rho}\)
\nonumber \\ & &
+ {13x^2\over8}\[\(F^a_{\mu\nu}F_a^{\mu\nu}\)^2 
+ \(F^a_{\mu\nu}\tF_b^{\mu\nu}\)^2\] - {5x^2\over8}\[
\(F^a_{\mu\nu}F_b^{\mu\nu}\)^2 +
\(F^a_{\mu\nu}\tF_a^{\mu\nu}\)^2\] \nonumber \\ & &
 -{x^4\rho_i\rho^i\over2}\[\(F^a_{\mu\nu}F_b^{\mu\nu}\)^2
+ \(F^a_{\mu\nu}\tF_b^{\mu\nu}\)^2 + \(F^a_{\mu\nu}F_a^{\mu\nu}\)^2 +
\(F^a_{\mu\nu}\tF_a^{\mu\nu}\)^2\] \nonumber \\ & &
+ {x^6\(\rho_i\rho^i\)^2\over8}\[\(F^a_{\mu\nu}F_b^{\mu\nu}\)^2
+ \(F^a_{\mu\nu}\tF_b^{\mu\nu}\)^2 + \(F^a_{\mu\nu}F_a^{\mu\nu}\)^2 +
\(F^a_{\mu\nu}\tF_a^{\mu\nu}\)^2\], \nonumber \\
\Tr K^2 &=& 10\D''^\mu\F^a_{\mu\rho}\D''_\nu\F_a^{\nu\rho} 
+ 6r^\nu_\mu xF_a^{\mu\rho}F^a_{\nu\rho} 
- 3r_{\mu\nu}^{\;\;\;\;\;\rho\sigma}xF_a^{\mu\nu}F^a_{\rho\sigma} 
+ 6xc_{abc}F^a_{\mu\nu}F_\rho^{b\;\;\mu}F^{c\rho\nu} \nonumber \\ & &
- F_a^{\mu\nu}F^a_{\mu\nu}\({3\nabla^2x\over2} - {2\pp_\rho x\pp^\rho x\over
x} + {25\pp_\rho y\pp^\rho y\over 8x}\) + F_a^{\lambda\nu}F^a_{\rho\nu}\(
{5\pp_\lambda x\pp^\rho x\over 2x} + {6\pp_\lambda y\pp^\rho y\over x}\)
\nonumber \\ & &
- {1\over\sqrt{x}}\D''^\mu\F^a_{\mu\nu}\(10F_a^{\lambda\nu}\pp_\lambda x - 11
\tF_a^{\lambda\nu}\pp_\lambda y\) - {1\over4}F_a^{\mu\nu}\tF^a_{\mu\nu}\(3
\nabla^2y - {\pp_\rho y\pp^\rho x\over x}\).\eea
In writing these expressions we dropped total derivatives and used (B.10) 
and (B.12--B.14), as well as the Yang-Mills Bianchi identity.

Finally, writing $\(\hG^g_{\mu\nu}\)^a_b = \(\hG'_{\mu\nu}\)^a_b + 
\hgg_{\mu\nu}\delta^a_b$, we have 
\bea \Tr\(\hG^{g+G}_\Phi\)^2 &=& \Tr\(G^G_\Phi\)^2_0 
-4C^a_GF^a_{\mu\nu}F_a^{\mu\nu} + N_G\hgg^2 
+ \tF^a_{\mu\nu}F_a^{\mu\nu}\(3\nabla^2 y - 
{4\pp_\nu x\pp^\nu y\over x}\) \nonumber \\ & & 
+ x^3\rho_i\rho^i\(4r_\mu^\nu F_a^{\mu\rho}F^a_{\nu\rho} - 
rF^a_{\mu\nu}F_a^{\mu\nu}\) + 8\(\D''_\mu\F_a^{\mu\nu}\)^2 
+ 8xr^\nu_\mu F^a_{\nu\rho}F_a^{\mu\rho} \nonumber \\ & &
+ {1\over 2x}F_{\mu\nu}F^{\mu\nu}\(4\pp^\rho x\pp_\rho x -
\pp^\rho y\pp_\rho y\) - {2\over x}F_{\mu\rho}F^{\nu\rho}\(\pp^\mu x\pp_\nu x 
+ 2\pp^\mu y\pp_\nu y\)\nonumber \\ & &
+ {x^6\(\rho_i\rho^i\)^2\over4}\[\(F^a_{\mu\nu}F_b^{\mu\nu}\)^2
+ \(F^a_{\mu\nu}\tF_b^{\mu\nu}\)^2 - 3\(F^a_{\mu\nu}F_a^{\mu\nu}\)^2 - 3
\(F^a_{\mu\nu}\tF_a^{\mu\nu}\)^2\] \nonumber \\ & &
+ {x^4\rho_i\rho^i\over2}\[\(F^a_{\mu\nu}F_b^{\mu\nu}\)^2
+ \(F^a_{\mu\nu}\tF_b^{\mu\nu}\)^2 + \(F^a_{\mu\nu}F_a^{\mu\nu}\)^2 +
\(F^a_{\mu\nu}\tF_a^{\mu\nu}\)^2\] \nonumber \\ & &
- {3x^2\over2}\(F^a_{\mu\nu}F_b^{\mu\nu}\)^2 - 
x^2\(F^a_{\mu\nu}F_a^{\mu\nu}\)^2 
+ {5x^2\over4}\(F^a_{\mu\nu}\tF_a^{\mu\nu}\)^2 \nonumber \\ & &
- {x\rho_i\rho^i\over2}\(F_a^{\mu\nu}F^a_{\mu\nu}\pp_\rho y\pp^\rho y
- 4F_a^{\mu\rho}F^a_{\nu\rho}\pp_\mu y\pp^\nu y\) \nonumber \\ & &
- \(8\pp^\mu xF_{\mu\rho} 
- 4\pp^\mu y\tF_{\mu\rho}\){\D''_\nu\F^{\nu\rho}\over\sqrt{x}},\eea
where $C^a_G$ is the Casimir in the adjoint representation, and we used 
(C.10) and (B.12--14).

For the fermions we define $H^{g + G} = H^g + H^G + H^{gG}$, with
\bea H^g_{ab} &=& H'_{ab} + \delta_{ab}\pmatrix{1&0\cr0&1\cr}\otimes
\(h^g_1 + h^g_2 + h^g_3\), \nonumber \\
\(G^g_{\mu\nu}\)^a_b &=& \(G'_{\mu\nu}\)^a_b + \delta^a_b
\pmatrix{1&0\cr0&1\cr}\otimes\(\tgg_{\mu\nu} + i\gamma_5L_{\mu\nu}\), 
\nonumber \\
\(\hG^g_{\mu\nu}\)^a_b &=& \(\hG'^g_{\mu\nu}\)^a_b 
+ \delta^a_b\pmatrix{1&0\cr0&1\cr}\otimes\hgg^g_{\mu\nu} ,\eea
where $h_i,\tgg_{\mu\nu}$ and
$\hgg_{\mu\nu}$ are $4\times 4$ Dirac matrices.  Then we obtain:
\bea {1\over8}\btr H^{g+G}_1 &=& {1\over8}\btr\(H_1^G\)_0 
+ {N_G\over4}\btr h_1 + 2\K^a_a - \D\(2 - x^2\rho^i\rho_i\) 
+ {x\over2}F^a_{\mu\nu}F_a^{\mu\nu}, \nonumber \\
{1\over8}\btr\(H^{g+G}_1\)^2 &=& {1\over8}\btr\(H_1^G\)^2_0 +
{N_G\over4}\btr h^2_1 + 2M_\lambda^2\[2\K^a_a + \D\(3 + x^2\rho^i\rho_i\)\] 
\nonumber \\ & &
+ \D^2\[2 - 2x^2\rho_i\rho^i + \(x^2\rho_i\rho^i\)^2\] + 4\K^{ab}\K_{ba}
- {2\over x}\(1-x^2\rho^i\rho_i\)\K^{ab}\D_a\D_b 
\nonumber \\ & & - {1\over x^2}\(\pp^\mu x\pp_\mu x + \pp^\mu y\pp_\mu y\)\D
- {8\over x}\D_\mu z^i\D^\mu\z^{\m}(T_a z)^j(T^a\z)^{\n}K_{i\n}K_{j\m}
\nonumber \\ & & + {2\over x^2}\D_a\[\(\pp_\mu x + i\pp_\mu y\)\D^\mu\z^{\m}
(T^a z)^iK_{i\m} + {\rm h.c.}\] \nonumber \\ & & 
+ 2\D\(M_\psi^2 + \hV -4e^{-K}\aa a\)+
xF^a_{\mu\nu}F_a^{\mu\nu}\({1\over2}M_\lambda^2 + M_\psi^2 - e^{-K}\aa a\)
\nonumber \\ & & + {1\over 2}F^a_{\mu\nu}F_b^{\mu\nu}\D_a\D^b\(1 - 
{x^2\rho^i\rho_i\over2} + {\(x^2\rho^i\rho_i\)^2\over4}\) \nonumber \\ & & 
+ {x^2\over 16}\[\(F^a_{\mu\nu}F_b^{\mu\nu}\)^2 + \(F^a_{\mu\nu}F_a^{\mu\nu}\)^2
- \(F^a_{\mu\nu}\tF_b^{\mu\nu}\)^2 - \(F^a_{\mu\nu}\tF_a^{\mu\nu}\)^2\]
\nonumber \\ & & + {x^4\rho^i\rho_i\over 32}\[\(F^a_{\mu\nu}F_a^{\mu\nu}\)^2 
+ \(F^a_{\mu\nu}\tF_a^{\mu\nu}\)^2\], \nonumber \\
- {1\over8}\btr\(H^{g+G}_2\)^2 &=& - {1\over8}\btr\(H_2^G\)^2_0 -
{N_G\over4}\btr h^2_2 + \D_a\D^bF^a_{\mu\nu}F_b^{\mu\nu} - 
2xe^{-K}\aa aF^a_{\mu\nu}F_a^{\mu\nu}\nonumber \\ & & 
- \({\pp^\mu x\pp_\mu x\over2x^2} - {\pp^\mu y\pp_\mu y\over2x^2}\)\D + 
{\pp^\mu x\over x^2}\D_a\[\D_\mu z^iK_{i\m}(T^a\z)^{\m} + {\rm h.c.}\] 
\nonumber \\ & & - {1\over x}\lbr K_{i\n}K_{j\m}\D^\mu z^j
(T_a\z)^{\m}\[(T^az)^i\D_\mu\z^{\n} + (T^a\z)^{\n}\D_\mu z^i\] + {\rm h.c.}\rbr
\nonumber \\ & & - {x^2\over4}\(F^a_{\mu\nu}\tF_a^{\mu\nu}\)^2
+ {1\over4x}\(\pp^\mu x\pp_\nu x - \pp^\mu y\pp_\nu y\)
F^a_{\mu\rho}F_a^{\nu\rho}
\nonumber \\ & & - {1\over8x}\(\pp^\mu x\pp_\mu x - \pp^\mu y\pp_\mu y 
\)F^a_{\mu\nu}F_a^{\mu\nu} , \nonumber \\
{1\over8}\btr H^{g+G}_3 &=& {1\over8}\btr\(H_3^G\)_0 + 
{N_G\over4}\btr h_3, \nonumber \\
{1\over8}\btr\(H^{g+G}_3\)^2 &=& {N+5\over16}r^2 + {N_G\over4}\btr h^2_3 +
{\pp_\mu y\pp^\nu y\over2x^2}F_a^{\mu\rho}F^a_{\mu\rho}
- {1\over16}\btr\tG'^2_{g+G}, \nonumber \\
{1\over4}\btr\(H_1H_3\)^{g+G} &=& {1\over4}\btr\(H_1H_3\)^G_0 
+ {N_G\over2}\btr\(h_1h_3\) - {1\over4}F^a_{\mu\nu}\tF_a^{\mu\nu}\(\nabla^2y 
- {\pp^\rho x\pp_\rho y\over x}\) \nonumber \\ & & 
+ \(r - {2\pp_\rho y\pp^\rho y\over x^2}\)\(\K_a^a + {x^2\over2}\rho^i\rho_i\D
\) + \(r - {\pp_\rho y\pp^\rho y\over x^2}\)\({x\over4}F^a_{\mu\nu}F_a^{\mu\nu}
- \D\) 
\nonumber \\ & & - T_3^{gG} + i\(2 - {x^2\rho^i\rho_i\over2}\)
\D_\mu z^j\D_\nu\z^{\m}K_{i\m}\D^aF_a^{\mu\nu} , \nonumber \\
{1\over2}\btr\hG^2_{g+G} &=& {1\over2}\btr\tG'^2_{g+G} + N_G\btr\hgg^2
+ {5\over2}\(r^2 - 2r^{\mu\nu}r_{\mu\nu}\)
+ \tF^a_{\mu\nu}F_a^{\mu\nu}\(6\nabla^2y - 8{\pp_\nu x\pp^\nu y\over x}\) 
\nonumber \\ & & + x^3\rho^i\rho_i\(2r^\mu_\nu F^a_{\mu\rho} F_a^{\nu\rho} - 
{r\over2}F^a_{\mu\nu} F_a^{\mu\nu}\) - 8\(\D''_\mu\F_a^{\mu\nu}\)^2 - 
8xr^\nu_\mu F^a_{\nu\rho}F_a^{\mu\rho} \nonumber \\ & & 
- {2\over x}F_{\mu\nu}F^{\mu\nu}\(\pp^\rho x\pp_\rho x - 
\pp^\rho y\pp_\rho y\) + {2\over x}F_{\mu\rho}F^{\nu\rho}
\(\pp^\mu x\pp_\nu x - 7\pp^\mu y\pp_\nu y\) \nonumber \\ & &
- {5x\rho^i\rho_i\over4}\(F_{\mu\nu}F^{\mu\nu}\pp^\rho y\pp_\rho y - 4
F_{\mu\rho}F^{\nu\rho}\pp^\mu y\pp_\nu y\)
\nonumber \\ & &
+ 8\(\pp^\mu xF_{\mu\rho} + \pp^\mu y\tF_{\mu\rho}\)
{\D''_\nu\F^{\nu\rho}\over\sqrt{x}}
- {x^6\over4}\(\rho^i\rho_i\)^2\[\(F^a_{\mu\nu} F_b^{\mu\nu}\)^2 
+ \(F^a_{\mu\nu}\tF_b^{\mu\nu}\)^2\] \nonumber \\ & &
+ {x^4\rho^i\rho_i\over2}\[\(F^a_{\mu\nu}F_b^{\mu\nu}\)^2 
+ \(F^a_{\mu\nu}\tF_b^{\mu\nu}\)^2 - \(F^a_{\mu\nu}F_a^{\mu\nu}\)^2 -
\(F^a_{\mu\nu}\tF_a^{\mu\nu}\)^2\] \nonumber \\ & &
+ {x^2\over2}\[\(F^a_{\mu\nu}F_a^{\mu\nu}\)^2 - \(F^a_{\mu\nu}F_b^{\mu\nu}\)^2 +
5\(F^a_{\mu\nu}\tF_b^{\mu\nu}\)^2 - 5
\(F^a_{\mu\nu}\tF_a^{\mu\nu}\)^2\], \nonumber \\
{1\over2}\btr\tG'^2_{g+G} &=& 4\(r^2 - 4r^{\mu\nu}r_{\mu\nu}\) +
20\Gamma_{\mu\nu}\Gamma^{\mu\nu} 
-4C^a_GF^a_{\mu\nu}F_a^{\mu\nu} \nonumber \\ & &
+ 16\D''^\nu\F^a_{\rho\nu}\D''_\mu\F_a^{\rho\mu} -16{\pp^\nu x\over\sqrt{x}}
F^a_{\rho\nu}\D''_\mu\F_a^{\rho\mu} + 16xr^\mu_\nu F_a^{\rho\nu}
F^a_{\rho\mu} \nonumber \\ & &
+ {4\over x}\(\pp_\rho x\pp^\rho xF^a_{\mu\nu}F_a^{\mu\nu} 
- \pp_\mu x\pp^\nu xF_a^{\mu\rho}F^a_{\nu\rho}\) + 4x^2 
\(F^a_{\mu\nu}\tF_a^{\mu\nu}\)^2 
\nonumber \\ & & - 2x^2\[\(F^a_{\mu\nu}F_b^{\mu\nu}\)^2 +
\(F^a_{\mu\nu}\tF_b^{\mu\nu}\)^2 + \(F^a_{\mu\nu}F_a^{\mu\nu}\)^2\]. \eea
The nonvanishing contributions to $T^{g+G}$ are:
\bea T_3^{gG} &=& 2\[\tX_-^{\mu\nu}(M,\M)\]^a_\rho\(G'^+_{\mu\nu}\)_a^\rho - 2
\[\tX_-^{\mu\nu}(\M,M)\]^a_\rho\(G'^-_{\mu\nu}\)_a^\rho \nonumber \\ 
&=& -{1\over4}\[F^a_{\mu\nu}F_a^{\mu\nu}\nabla^2 x 
+ \tF^a_{\mu\nu}F_a^{\mu\nu}\(\nabla^2 y - {\pp_\rho x\pp^\rho y\over 2x}\)\] 
+ {\pp_\mu x\pp^\nu x\over 2x}F^a_{\nu\rho}F_a^{\mu\rho}, \nonumber \\ 
T_3^g &=& \[\tX_-^{\mu\nu}(M,\M)\]^a_b\(G'^+_{\mu\nu}\)_a^b - 
\[\tX_-^{\mu\nu}(\M,M)\]^a_b\(G'^-_{\mu\nu}\)_a^b \nonumber \\ & &
+ iN_G{\pp^\mu y\over x}\(\m_\lambda\tD_\mu m_\lambda - {\rm h.c}\) 
+ r^\mu_\nu\Tr\(\tM^{\nu\rho}\M_{\mu\rho} - M^{\nu\rho}\tbM_{\mu\rho}\)_i^j
\nonumber \\ &=& {i\over2}x^2\rho^i\rho_i\D_\mu z^j\D_\nu\z^{\m}K_{i\m}\D^a
F_a^{\mu\nu} + {1\over8}x^3\rho^i\rho_i\(r^\mu_\nu F_a^{\nu\rho}F^a_{\mu\rho} 
- {1\over4}rF_a^{\mu\nu}F^a_{\mu\nu}\) + N_Gt_3, \nonumber \\ 
T^{gG}_4 &=& - 4i\(m\M_{\mu\nu}\)^a_\alpha\(m\tbM^{\mu\nu}\)_a^\alpha + 
{\rm h.c.} = {x\over 2}F^a_{\mu\nu}F_a^{\mu\nu}M_\lambda^2 
+ ixF^a_{\mu\nu}\tF_a^{\mu\nu}e^{-K}\(\aa A - {\rm h.c.}\), \nonumber \\ 
T^g_4 &=& T^\chi_4 + T^{\chi G}_4 . \eea

Finally, for the ghost sector, defining $\(H_{gh}^g\)^a_b =
\(H'_{gh}\)^a_b + h_{gh}\delta^a_b$, we have 
\bea \Tr H^g_{gh} &=& 2\K^a_a - {x\over2}F^a_{\mu\nu}F_a^{\mu\nu} + N_Gh_{gh},
\;\;\;\;\Tr H^G_{gh} = \Tr \(H_{gh}\)_0 + {3x\over2}F^a_{\mu\nu},
\nonumber \\
\Tr\(H^g_{gh}\)^2 &=& N_Gh_{gh}^2 + 2(\K_{ab}\K^{ba} + \K_{ab}\K^{ab}) 
- 4\K^a_a\({\nabla^2 x\over2x} - {\pp_\mu x\pp^\mu x\over4x^2}\) 
\nonumber \\ & & + {1\over 2}F^a_{\mu\nu}F_a^{\mu\nu} 
\(\nabla^2 x - {\pp_\mu x\pp^\mu x\over2x}\) 
- 2xF^a_{\mu\nu}F_b^{\mu\nu}\K_{ba} + 
{x^2\over 4}\(F^a_{\mu\nu}F_b^{\mu\nu}\)^2, \nonumber \\ 
\Tr\(H^G_{gh}\)^2 &=& \(\Tr H_{gh}^2\)_0 + {9x^2\over16}\[\(F^a_{\mu\nu}
F_a^{\mu\nu}\)^2  + \(F^a_{\mu\nu}F_b^{\mu\nu}\)^2 + \(F^a_{\mu\nu}
\tF_b^{\mu\nu}\)^2\] \nonumber \\ & & 
- 3xF^a_{\mu\rho}F^{\nu\rho}_a\(2\D^\mu z^I\D_\nu z^JZ_{IJ} + 
r^\mu_\nu\), \nonumber \\ 
\Tr\(H^{gG}_{gh}\)^2 &=& - \(\D''^\nu\F_{a\mu\nu}\)^2 
- {1\over4x}\(F_{a\mu\nu}{\pp^\nu
x}\)^2 - 4\(q_{aI}\D_\mu z^I\)^2 \nonumber \\ & &
- \D''^\mu\F^a_{\nu\mu}\(F_a^{\nu\rho}{\pp_\rho x\over\sqrt{x}} + 
4q_{aI}\D^\nu z^I\) - 2F_a^{\mu\rho}{\pp_\rho x\over\sqrt{x}}q^a_I\D_\mu z^I,
\nonumber \\ 
\Tr\(\hG_{\mu\nu}\hG^{\mu\nu}\)^{gh} &=& \(\Tr G_{\mu\nu}G^{\mu\nu}\)^{gh}_0 
- {\pp_\nu x\pp^\mu x\over 2x}F^a_{\mu\rho}F_a^{\nu\rho} +
{\pp_\rho x\pp^\rho x\over 2x}F^a_{\mu\nu}F_a^{\mu\nu}
+ 2xr_\mu^\nu F_a^{\mu\rho}F^a_{\nu\rho} \nonumber \\ & &
+ 2\(\D''_\mu\F_a^{\mu\nu}\)^2 - 
2{\pp^\mu x\over\sqrt{x}}F^a_{\mu\rho}\D''_\nu\F_a^{\nu\rho}
- C^a_GF^a_{\mu\nu}F_a^{\mu\nu} + {x^2\over4}\(F^a_{\mu\nu}\tF_a^{\mu\nu}\)^2 
\nonumber \\ & & - {x^2\over8}\[\(F^a_{\mu\nu}F_a^{\mu\nu}\)^2 + 
\(F^a_{\mu\nu}F_b^{\mu\nu}\)^2 + \(F^a_{\mu\nu}\tF_b^{\mu\nu}\)^2\].
\eea 
For the ghostino, Tr$G_{\mu\nu}G^{\mu\nu}$ is given in (B.18) of I, and the 
remaining traces are modified with respect to that equation by\footnote{
The last term in the equation for $\Tr H^2_{Gh}$ should be 
$- 18\Gamma_{\mu\nu}\Gamma^{\mu\nu}.$}
\bea
\Tr H_{Gh} &=& \(\Tr H_{Gh}\)_0 + 4\D + xF^a_{\mu\nu}F_a^{\mu\nu}, 
\nonumber \\ 
\Tr H^2_{Gh} &=& \(\Tr H^2_{Gh}\)_0 + 4\D^2 + 2x\D F^a_{\mu\nu}F_a^{\mu\nu} -
24i\D^aF_a^{\mu\nu}\D_\mu z^iK_{i\m}\D_\nu\z^{\m} \nonumber \\ & &
+ 2\(4\D + xF^a_{\mu\nu}F_a^{\mu\nu}\)\(\hV + M_\psi^2 -
\D_\rho z^iK_{i\m}\D^\rho\z^{\m} - {r\over4}\) \nonumber \\ & &
+ 2\D_a\D^bF^a_{\mu\nu}F_b^{\mu\nu} + 
{x^2\over4}\[\(F^a_{\mu\nu}F_a^{\mu\nu}\)^2 - 
\(F^a_{\mu\nu}\tF_a^{\mu\nu}\)^2\] . \eea 
For the supertraces we obtain [see (B.17--20)] 
\bea \STr H^{g+G} &=& \STr H^G_0 + N_G\STr h^g -2\D\(4 + x^2\rho_i\rho^i\),
\nonumber \\
{1\over2}\STr H^2_{g+G} &=& {1\over2}\[\STr\(H^2_G\)_0 + 
\STr\(H^2_{\chi G}\)_0 - \STr H^2_{\chi G}
+ N_G\STr h^2_g \] - T_3^{g+G} - T_4^{g G} \nonumber \\ & &
- {7\over gx}\L_{a\mu}\L^{a\mu} + {4\over\sqrt{g}x}
\[4i\L^{a\mu}\(K_{i\m}\D_\mu\z^{\m}(T_az)^i - {\rm h.c.}\) + \D^a(T_az)^I
\L_I\] \nonumber \\ & & + {\L^\nu_a\over\sqrt{g}}\[\(7 + x^2\rho^i\rho_i\)
\pp^\mu xF^a_{\mu\nu} + 3\pp^\mu y\tF^a_{\mu\nu}\] 
- {12\over\sqrt{g}}\Delta_{\D}\L + r\K^a_a \nonumber \\ & & 
+ x\(2 - {7x^2\rho_i\rho^i\over8}\)r_\nu^\mu F^a_{\mu\rho}F_a^{\nu\rho} 
- {x\over 4}\(3 - {7x^2\rho_i\rho^i\over8}\)rF^a_{\mu\nu}F_a^{\mu\nu} 
\nonumber \\ & & + 4\K^a_bF^{\mu\nu}_aF_{\mu\nu}^b 
+ 3C^a_G\(\cW^a_a + \cbW^a_a\) + {2\over x^2}C^a_G\D^a\D_a 
- 2x\rho^i\rho_i\K^{ab}\D_a\D_b \nonumber \\ & &
- \(5 + {x^2\rho_i\rho^i\over2}\)r\D
- \K^a_a\({\pp^\mu x\pp_\mu x\over x^2} + {\pp^\mu y\pp_\mu y\over x^2}\) 
- 2x^2\rho^i\rho_i\D M_\lambda^2 \nonumber \\ & & 
- 4\K^a_aM_\lambda^2 - 2e^{-K}\(\D A_i\aa^i + {\rm h.c.}\)
+ 2\D\(31\hV + 29M_\psi^2 + 11M_\lambda^2\)\nonumber \\ & & 
-4e^{-K}a\aa\D + 2x\(\cW + \cbW\)\(2M_\psi^2 - M_\lambda^2\) 
- 4xe^{-K}\(\cW A\aa + {\rm h.c.}\) \nonumber \\ & & 
+ 2i\(2{\D''_\mu\F_a^{\mu\rho}\over\sqrt{x}} - {\pp_\mu x\over x}F_a^{\mu\rho}\)
K_{i\m}\[\D_\rho\z^{\m}(T^az)^i - \D_\rho z^i(T^a\z)^{\m}\]
\nonumber \\ & & - 2{\pp^\mu x\over x}F^a_{\mu\rho}\D''_\nu\F_a^{\nu\rho}
- {1\over x^2}\D_a\[\(3\pp^\mu x + 8i\pp^\mu y\)K_{i\m}(T^az)^i\D_\mu\z^{\m} 
+ {\rm h.c.}\] 
\nonumber \\ & & + {4\over x}\D^\mu z^i\D_\mu\z^{\m}K_{j\m}\D_aD_i(T^az)^j  
- i\(26 - x^2\rho^i\rho_i\)K_{j\m}\D_\mu z^j\D_\nu\z^{\m}\D^aF_a^{\mu\nu}
\nonumber \\ & & - {i\over x}\[\(5 + x^2\rho^i\rho_i\)\pp_\mu x F_a^{\mu\nu}
+ 5\pp_\mu y\tF_a^{\mu\nu}\]\(K_{j\m}\D_\nu\z^{\m}(T^az)^j - {\rm h.c.}\)
\nonumber \\ & & + {13\over 2x^2}\(\pp^\mu x\pp_\mu x + \pp^\mu y\pp_\mu y\)\D 
+ \({\pp^\rho x\pp_\rho x\over8x} + 
{\pp^\rho y\pp_\rho y\over4x}\)F^a_{\mu\nu}F_a^{\mu\nu} \nonumber \\ & & 
- {1\over 4x}F^a_{\mu\rho}F_a^{\nu\rho}\(6\pp^\mu x\pp_\nu x - \pp^\mu y
\pp_\nu y\) - {1\over x}F^a_{\mu\nu}\tF_a^{\mu\nu}\pp^\rho x\pp_\rho y 
\nonumber \\ & & - {x\rho^i\rho_i\over 4}\[F^a_{\mu\nu}F_a^{\mu\nu}
\(\pp^\rho x\pp_\rho x - {1\over2}\pp^\rho y\pp_\rho y\) +
F^a_{\mu\nu}\tF_a^{\mu\nu}\pp^\rho x\pp_\rho y +
2F^a_{\mu\nu}F_a^{\mu\rho}\pp^\nu y\pp_\rho y \]\nonumber \\ & &
+ \rho^i\rho_i\pp^\mu y\pp_\mu y\D 
- F_a^{\mu\rho}\({\pp_\mu x\over2}F^a_{\nu\rho} - {\pp_\nu x\over8}
F^a_{\mu\rho}\)\(\D^\nu z^i\rho_{ij}\f^j + {\rm h.c.}\) \nonumber \\ & &
+ x^2\[\(F^a_{\nu\rho}F_a^{\mu\rho} - iF^a_{\nu\rho}\tF_a^{\mu\rho}\)
\(2\D_\mu z^i\D^\nu z^j - g_\mu^\nu\D^\sigma z^i\D_\sigma
z^j\)\rho_{ij} + {\rm h.c.}\] \nonumber \\ & &
+ 6x^2\cW^{ab}\cbW_{ab} + 8\D_a\D_b\(\cW^{ab}+ \cbW^{ab}\)  + 2\D^2\(19
- x^2\rho^i\rho_i\) \nonumber \\ & &
+ x^3\rho^i\rho_i\(x^2\rho^i\rho_i - 1\)\[x\cW\cbW + x\cW^{ab}\cbW_{ab} + 
\D\(\cW + \cbW\) \] \nonumber \\ & & 
- x^2\rho^i\rho_i\({3\over2} - {x^2\rho^i\rho_i\over4}\)\D_a\D_b\(\cW^{ab}+ 
\cbW^{ab}\) \nonumber \\ & &
- \(8\D + 2xF^a_{\rho\sigma}F_a^{\rho\sigma}\)\D_\mu z^i\D^\mu\z^{\m}K_{i\m}
+ 12xF^a_{\rho\mu}F_a^{\rho\nu}\D_\nu z^i\D^\mu\z^{\m}K_{i\m},\nonumber \\
{1\over12}\STr\hG_{g+G}^2 &=& {1\over12}\STr\(G_{g+G}^2\)_0 + {1\over12}
N_G\STr\hgg^2 - {1\over12}\STr\hG_{\chi G}^2 - {1\over12}\STr\hG_{\chi}^2
\nonumber \\ & & - T^{\chi G} - T_4^g - T_4^\chi 
+ {x^3\rho_i\rho^i\over4}\(r_\mu^\nu F_a^{\mu\rho}F^a_{\nu\rho} 
- {r\over4}F^a_{\mu\nu}F_a^{\mu\nu}\)\nonumber \\ & & 
- {1\over8x}\(F_{\mu\nu}F^{\mu\nu}\pp^\rho y\pp_\rho y - 2
F_{\mu\rho}F^{\nu\rho}\pp^\mu y\pp_\nu y\) 
+ {i\over6}F^a_{\mu\nu}K_{j\m}\D^\mu z^j\D^\nu\z^{\m}D_i(T_az)^i
 \nonumber \\ & &
+ {x\rho^i\rho_i\over96}\(F_{\mu\nu}F^{\mu\nu}\pp^\rho y\pp_\rho y - 4
F_{\mu\rho}F^{\nu\rho}\pp^\mu y\pp_\nu y\)
+ x^4\rho_i\rho^i\(\cW^{ab}\cbW_{ab} + \cW\cbW \)\nonumber \\ & &
+ x^2\rho_i\rho^i\[{3\over2}\D_a\D_b\(\cW^{ab} + \cbW^{ab}\) + x\D\(\cW + 
\cbW\) + 6\D^2\] \nonumber \\ & & 
+ x^4\(\rho_i\rho^i\)^2\[x^2\(\cW^{ab}\cbW_{ab} - \cW\cbW\) - 2\D^2 
- x\D\(\cW + \cbW\)\] .\eea
The space-time curvature dependent terms in the supertraces evaluated in
sections C.4--7 give a contribution $\L_r$ of the form (2.23) of I with
\bea H_{\mu\nu} &=& H^0_{\mu\nu} + H^g_{\mu\nu} + 
\lll\[x\(4 - x^2\rho_i\rho^i\)F^a_{\mu\rho}F_{a\nu}^{\;\;\;\;\rho} - 
g_{\mu\nu}xF^a_{\rho\sigma} F_a^{\rho\sigma}\({3\over2} - {1\over4}
x^2\rho_i\rho^i\)\], 
\nonumber \\ \epsilon_0 &=& \(\epsilon_0\)_0 + \epsilon^g_0 
-\lll\lbr{22\over 3}\D + 2x^2\rho_i\rho^i\D + {2\over 3x}\D_aD_i(T^az)^i\rbr, 
\nonumber \\
\alpha &=& \alpha_0 + \alpha^g, \;\;\;\; \beta = \beta_0 + \beta^g, \eea 
where $\alpha^g,\;etc.$ are evaluated in section C.8.
The metric redefinition in \newline (2.24--25) of I gives (4.11), and we 
get a correction\footnote{Eq.(B23) of I should read:
\bea {1\over\sqrt{g}}\Delta_r\L
 &=&  \lll\Bigg[\lbr -2e^{-K}\(A_{ki}\A^{ik} - {2\over 
3}R^k_nA_k\A^n\) - (N+17)\hV - {4N+32\over 3}M^2_\psi \rbr \hV \nonumber \\ & & 
 + \[K_{i\m}\lbr {N+59\over 3}\hV + e^{-K}\(A_{ki}\A^{ik} - 
{2\over 3}R^k_nA_k\A^n\) + {2N+16\over 3}M^2_\psi \rbr 
+{4\over 3}R_{i\m}\hV\]\D_\rho z^i\D^\rho\z^{\m}  \nonumber \\
& &
 - \lbr\({2\over 3}R_{i\m} + 8K_{i\m}\)\D_\rho z^i\D^\rho\z^{\m}g_{\mu\nu} 
- {N+29\over 6}\(\D_\mu z^i\D_\nu\z^{\m} + \D_\nu z^i\D_\mu\z^{\m}\)K_{i\m}
\rbr\D^\mu z^j\D^\mu\z^{\n}K_{i\n} \Bigg]. \nonumber \eea } 
$\Delta_r\L$:
\bea \Delta_r\L &=& \(\Delta_r\L\)_0 + \Delta_{rg}\L 
+ \lll\Bigg\{{N-67\over3}\D^2 - {2N+118\over3}\D\hV - 
{4N+32\over3}\D M_\psi^2 \nonumber \\ & &
+\(\D_\mu z^i\D^\mu\z^{\m}K_{i\m} -2V\)\[2x^2\rho_i\rho^i\D
+ {2\over 3x}\D_aD_i(T^az)^i \] - 2xV\(\cW + \cbW\)
\nonumber \\ & &
-2\D e^{-K}\(A_{ij}\A^{ij} - {2\over 3}R^i_jA_i\A^j\) 
+ {1\over3}\D\D_\mu z^i\D^\mu\z^{\m}\[4R_{i\m} - (N-61)K_{i\m}\]
\nonumber \\ & & + \({N+29\over6} - x^2\rho_i\rho^i\)\[2x^2\cW^{ab}\cbW_{ab}
+ \(\cW^{ab} + \cbW^{ab}\)\D_a\D_b + 2\D^2\]
\nonumber \\ & & + \({N+35\over3} - x^2\rho_i\rho^i\){x\over4}
F^a_{\rho\sigma}F_a^{\rho\sigma}\D_\mu z^i\D^\mu\z^{\m}K_{i\m}
\nonumber \\ & & - \({N+29\over3} - x^2\rho_i\rho^i\)x
F^a_{\rho\mu}F_a^{\rho\nu}\D_\nu z^i\D^\mu\z^{\m}K_{i\m}
\Bigg\}, 
\eea 
where $\Delta_{rg}\L$ is given in (C.73) below.

\subsubsection{Order $N_G$ contributions}

The bosonic traces are 
\bea 
n &=& r - {\nabla^2x\over x}
- {3\pp^\mu y\pp_\mu y\over 2x^2}, \nonumber \\
n^2 &=& r_{\mu\nu}r^{\mu\nu} - r\[{\nabla^2 x\over x} 
-{\pp_\mu x\pp^\mu x\over 2x^2} +{\pp_\mu y\pp^\mu y\over x^2}\] 
+ r^{\mu\nu}\[{2\nabla_\nu\pp_\nu x\over x}
- {\pp_\mu x\pp_\nu x\over x^2} + {\pp_\mu y\pp_\nu y\over x^2}\] 
\nonumber \\ & &
+ \({\nabla^2x\over x}\)^2 - {3\pp_\mu x\pp^\mu x\nabla^2x\over 2x^3}
+ {\pp_\mu y\pp^\mu y\nabla^2x\over x^3} - {\pp_\mu y\pp^\mu x\nabla^2y\over
x^3} \nonumber \\ & &
+ {1\over 4x^4}\[3\(\pp_\mu x\pp^\mu x\)^2 + 
8\(\pp_\mu y\pp^\mu x\)^2 + 3\(\pp_\mu y\pp^\mu y\)^2 
- 5\pp_\mu x\pp^\mu x\pp_\nu y\pp^\nu y\], \eea
and
\bea \hgg^2 &=& {1\over x^2}\[3(\nabla^2y)^2 - 
6{\pp_\mu y\pp^\mu x\over x}\nabla^2y
+ {\(\pp_\mu y\pp^\mu x\)^2\over x^2} + 
{2\pp_\mu y\pp^\mu y\pp_\nu x\pp^\nu x\over x^2} - 
{3(\pp_\mu y\pp^\mu y)^2\over 4x^2}\] \nonumber \\ & & 
+ \(r^2 - 4r_{\mu\nu}r^{\mu\nu}\) - 2r^{\mu\nu}{\pp_\mu y\pp_\nu y\over x^2} 
+ r{\pp^\mu y\pp_\mu y\over x^2} .\eea
The fermion traces are (here $\btr$ includes the ordinary Dirac trace;
$\btr {\bf 1} = 4$):
\bea {1\over4}\btr h_1 &=& M_\lambda^2 , \;\;\;\; {1\over4}\btr h_1^2 = 
M_\lambda^4, \nonumber \\ 
-{1\over4}\btr h_2^2 &=& 
e^{-K}\D_\mu z^i\D^\mu\z^{\m}\[\(a_i - A_i\)\(\aa_{\m} - \A_{\m}\) 
+ x^2\rho_{ik}\A^k\rho_{\m}^jA_j + {\f_{\m}f_i\over4x^2}a\aa\]
\nonumber \\ & &
+ e^{-K}\[\D_\mu z^i\D^\mu z^j\(a_i - A_i\) 
\({f_j\over2x}\aa - x\rho_{jn}\A^n\)+ {\rm h.c.}\] \nonumber \\ & &
- {1\over 2}e^{-K}\D_\mu z^i\D^\mu\z^{\m}\(\f_{\m}\rho_{ik}a\A^k 
+ {\rm h.c.}\), \nonumber \\ 
\btr h_3 &=& r - 2{\pp_\mu y\pp^\mu y\over x^2}, 
\nonumber \\ 
\btr\(h_1h_3\) &=& \(r - {\pp_\mu y\pp^\mu y\over 2x^2}\)M_\lambda^2,
\nonumber \\ 
\btr h_3^2 &=& {1\over 4}r^2 - {1\over2}\btr\(\tgg_{\mu\nu}\tgg^{\mu\nu} 
- Z^{\mu\nu}Z_{\mu\nu}\) - {(\nabla^2y)^2\over x^2} 
\nonumber \\ & & + 2{\nabla^2y\pp_\mu x\pp^\mu y\over x^3}
- r{\pp_\mu y\pp^\mu y\over x^2} + {(\pp_\mu y\pp^\mu y)^2\over x^4} - 
{(\pp_\mu y\pp^\mu x)^2\over x^4}, \nonumber \\ 
\btr\hgg_{\mu\nu}\hgg^{\mu\nu} &=& \btr \tgg_{\mu\nu}\tgg^{\mu\nu}
+ 6{(\nabla^2y)^2\over x^2} - 12{\nabla^2y\pp_\mu x\pp^\mu y\over x^3}
+ 2r{\pp_\mu y\pp^\mu y\over x^2} 
- 4r^{\mu\nu}{\pp_\mu y\pp_\nu y\over x^2} \nonumber \\ & & 
- 6{(\pp_\mu y\pp^\mu y)^2\over x^4} + 2{(\pp_\mu x\pp^\mu y)^2\over x^4} 
+ 4{\pp_\mu y\pp^\mu y\pp_\nu x\pp^\nu x\over x^4}, \nonumber \\ 
\btr \tgg_{\mu\nu}\tgg^{\mu\nu} &=& 4\Gamma_{\mu\nu}\Gamma^{\mu\nu} +
\btr Z_{\mu\nu}Z^{\mu\nu}= 4\Gamma_{\mu\nu}\Gamma^{\mu\nu} +
{1\over 2}r^2 - 2r_{\mu\nu}r^{\mu\nu}.\eea
To evaluate $t_3$, Eq. (C.58), we write it as 
\bea t_3 &=& {\pp^\mu x + i\pp_\mu y\over2x}\m_\lambda\tD^\mu m_\lambda -
{\pp^\mu x - i\pp_\mu y\over2x}\m_\lambda\tD^\mu m_\lambda + {\rm h.c.} 
\nonumber \\ &=&  {e^{-K}\over2x}\(\f_{\m}\D^\mu\z^{\m} - f_i\D^\mu z^i\)
\(\aa - \A\)\[\D_\mu z^j\(a_j - A_j\) \right.\nonumber \\ & &
\left.- \D_\mu\z^{\n}\({\f_{\n}\over2x}a - x\rho^j_{\n}A_j\)\]
 + {\rm h.c.}. \eea 

The ghost traces are:
\bea \Tr h_{gh} &=& - {\nabla^2 x\over2x} + {\pp_\mu x\pp^\mu x\over4x^2},
\;\;\;\;\Tr h_{gh}^2 = \({\nabla^2 x\over2x} - 
{\pp_\mu x\pp^\mu x\over4x^2}\)^2 . \eea

The supertraces are 
\bea -{r\over6}\STr h &=& {r\over 3}M^2_\lambda 
- {r^2\over12} + r{\pp_\mu x\pp^\mu x\over12x^2} 
+ r{\pp^\mu y\pp_\mu y\over 12x^2}, \nonumber \\ 
{1\over2}\STr h^2 &=& - t_3 -{r^2\over16} + {1\over2}r_{\mu\nu}r^{\mu\nu} 
- M^2_\lambda\({r\over2} - {\pp_\mu y\pp^\mu y\over 4x^2}\) - M_\lambda^4 
+ {1\over2}\Gamma_{\mu\nu}\Gamma^{\mu\nu}\nonumber \\ & & 
- r\({\nabla^2 x\over 2x} -{\pp_\mu x\pp^\mu x\over 4x^2}
+{\pp_\mu y\pp^\mu y\over 4x^2}\) + r^{\mu\nu}\({\nabla_\nu\pp_\nu x\over x}
- {\pp_\mu x\pp_\nu x\over2x^2}+ {\pp_\mu y\pp_\nu y\over2x^2}\)
\nonumber \\ & & + {\(\nabla^2 x\)^2\over4x^2} + {\(\nabla^2y\)^2\over4x^2} 
-{1\over2x^3}\[\(\pp_\mu x\pp^\mu x - 
\pp_\mu y\pp^\mu y\)\nabla^2x + 2\pp_\mu x\pp^\mu y\nabla^2y\] \nonumber \\ & &
+ {5\over16x^4}\left|\(\pp_\mu x + i\pp_\mu y\)\(\pp^\mu x + 
i\pp^\mu y\)\right|^2 - 
{3\over16x^4}\(\pp_\mu y\pp^\mu y\)^2 \nonumber \\ & &
+ e^{-K}\D_\mu z^i\D^\mu\z^{\m}\[\(a_i - A_i\)\(\aa_{\m} - \A_{\m}\) 
+ x^2\rho_{ik}\A^k\rho_{\m}^jA_j - {\f_{\m}f_i\over4x^2}a\aa\]
\nonumber \\ & & + e^{-K}\bigg\{ \D_\mu z^i\D^\mu z^j\bigg[\(a_i - A_i\) 
\({f_j\over2x}\A - x\rho_{jn}\A^n\) + {f_if_j\over4x^2}a\aa\nonumber \\ & & 
\qquad - {f_j\over2}\(a-A\)\rho_{ik}\A^k\bigg] + {\rm h.c.}\bigg\}\nonumber \\ 
& & + {e^{-K}\over 2x}\lbr\D_\mu z^i\D^\mu\z^{\m}\f_{\m}
\[\(\aa - \A\)\(a_i - A_i\) - x\rho_{ik}A\A^k \]
+ {\rm h.c.}\rbr,   \nonumber \\
{1\over12}\STr\hgg^2 &=& {1\over16}\(r^2 - 4r_{\mu\nu}r^{\mu\nu}\) 
+ {3\over 16x^4}\(\pp_\mu y\pp^\mu y\)^2
- {1\over6}\Gamma_{\mu\nu}\Gamma^{\mu\nu}.\eea 
Dropping the total derivative
$$ \pp_\mu\({\pp^\mu x\over x}M_\lambda^2\) = {\nabla^2x\over x}M_\lambda^2
- {\pp^\mu x\pp_\mu x\over x}M_\lambda^2 + {\pp^\mu x\over x}\D_\mu M_\lambda^2
$$ and using the equations of motion (B.18), we can write
\bea & & N_G\lll\[\STr\(h^2 - {r\over6}h + {1\over12}\hgg^2\) + t_3\] = 
\L_{rg} + N_G\lll\[{x^2\rho^i\rho_j\over\sqrt{g}}\L_i\L^j + \(X_g^i\L_i
+ {\rm h.c.}\)\] \nonumber \\ & & \qquad +
N_G\sqrt{g}\lll\Bigg\{ x^6\(\rho^i\rho_i\)^2\cW\cbW - 2M_\lambda^4 + 3M_\psi^2
- 2M_\psi^2M_\lambda^2 + \hV^2  \nonumber \\ & & \qquad 
+ 6e^{-K}a\aa M_\psi^2 - e^{-K}\(\aa^iA_i + {\rm h.c.}\)\(\hV + M_\psi^2\) 
\nonumber \\ & & \qquad + e^{-2K}a_i\A^i\aa^jA_j - 2e^{-2K}\(\aa^iA_ia\A + 
{\rm h.c.}\) + 2\hV\(2M_\psi^2 - 2M_\lambda^2 + e^{-K}a\aa\)\nonumber \\ & &
\qquad + e^{-K}\D_\mu z^i\D^\mu\z^{\m}\[\(a_i - A_i\)\(\aa_{\m} - \A_{\m}\) 
+ x^2\rho_{ik}\A^k\rho_{\m}^jA_j + {\f_{\m}f_i\over4x^2}a\aa\]
\nonumber \\ & & \qquad + e^{-K}\lbr\D_\mu z^i\D^\mu z^j\[\(a_i - A_i\) 
\({f_j\over2x}\A - x\rho_{jn}\A^n\) - {f_i\over2x}a_i\A - 
f_i\(a - A\)\rho_{ik}\A^k\] + {\rm h.c.}\rbr \nonumber \\ & & \qquad 
+ {e^{-K}\over 2x}\lbr\D_\mu z^i\D^\mu\z^{\m}\f_{\m}\[2\aa a_i 
- x\rho_{ik}\(a - A\)\A^k \] + {f_if_j\over2x^2}\aa\(2a - A\)
\D_\mu z^i\D^\mu z^j + {\rm h.c.}\rbr \nonumber \\ & & \qquad 
+ x\(\rho_{ij}\D_\mu z^i\D^\mu z^j + {\rm h.c.}\)\(M_\psi^2 - \hV\) + 
e^{-K}\[x\rho_{ij}\D_\mu z^i\D^\mu z^j\(a_k\A^k 
- 2\A a\) + {\rm h.c.}\] \nonumber \\ & & \qquad 
{1\over16x^4}\left|\(\pp_\mu x + i\pp_\mu y\)\(\pp^\mu x + \pp^\mu y\)\right|^2
- x^3\rho^i\rho_i\(\cW + \cbW\)\(M_\psi^2 + \hV\) \nonumber \\ & & \qquad 
+ x^3\rho^k\rho_k\[\cW\(x\rho_{ij}\D_\mu z^i\D^\mu z^j + 
e^{-K}A_i\aa^i - 2e^{-K}\aa A\) + {\rm h.c.}\] \nonumber \\ & & \qquad
+ {1\over3}\Gamma_{\mu\nu}\Gamma^{\mu\nu} 
+ x^2\rho_{ij}\D_\mu z^i\D^\mu z^j\rho_{\n\m}\D_\nu\z^{\m}\D^\nu\z^{n} 
+ {\rm total\;derivative} .\eea 
where 
\beq X_g^i = (X_g^{\ibar})^* = {\f^i\over2x}\[x^3\rho^j\rho_j\cbW + 
x\rho_{jk}\D_\mu z^j\D^\mu z^k + e^{-K}\(\aa^jA_j - 2\aa A\) - V - M_\psi^2
\], \eeq
and $\L_{rg}$ is of the form (2.23) of I with
\bea 
\alpha^g &=& - {N_G\over 6}\lll, \;\;\;\; \beta^g = {N_G\over 2}\lll, \;\;\;\;
\epsilon^g_0 = -\lll{N_G\over3}M^2_\lambda, \nonumber \\
H^g_{\mu\nu} &=& N_G\lll\bigg\{g_{\mu\nu}\(-{\nabla^2x\over x} 
+ {2\pp_\rho x\pp^\rho x\over3x^2} - {\pp_\rho y\pp^\rho y\over3x^2}\)
\nonumber \\ & &  
+ 2{\nabla_\mu\pp_\nu x\over x} - {\pp_\mu x\pp_\nu x\over x^2} 
+ {\pp_\mu y\pp_\nu y\over x^2} \bigg\}. \eea 
Finally, using the equations of motion (B.17--18) we obtain [see (C.62)]:
\bea 
\Delta_{rg}\L &=& - N_G\lll{\pp_\rho x\over x}\[F_a^{\rho\mu}\L_\mu^a +
\(\D^\rho z^i\L_i + {\rm h.c.}\)\]\nonumber \\ & &
+ \sqrt{g}N_G\lll\Bigg\{ - {1\over3}V^2 +
{1\over3}M_\lambda^2\(\D_\mu z^i\D^\mu\z^{\m}K_{i\m} -2V\)
+{1\over3}V\D_\mu z^i\D^\mu\z^{\m}K_{i\m} \nonumber \\ & &
+\({\pp_\nu x\pp^\nu x\over x^2}  + {\pp_\nu y\pp^\nu y\over x^2}\)
\({1\over3}\D_\mu z^i\D^\mu\z^{\m}K_{i\m} - {1\over6}V - 
{x\over8}F_a^{\mu\rho}F^a_{\mu\rho}\) \nonumber \\ & &
+ \({x\over2}F_a^{\mu\rho}F^a_{\nu\rho} - \D_\nu z^i\D^\mu\z^{\m}K_{i\m}\)
\({\pp_\mu x\pp^\nu x\over x} +
{\pp_\mu y\pp^\nu y\over x}\) -{1\over3}\(\D_\mu z^i\D^\mu\z^{\m}K_{i\m}\)^2
\nonumber \\ & & + {1\over2}K_{i\m}K_{j\n}\(\D_\mu z^i\D^\mu z^j\D_\nu\z^{\m}
\D^\nu\z^{\n} + \D_\mu z^i\D^\mu\z^{\n}\D_\nu\z^{\m}\D^\nu z^j\)
\nonumber \\ & & 
+ {1\over4}xF^a_{\rho\sigma}F_a^{\rho\sigma}\D_\mu z^i\D^\mu\z^{\m}K_{i\m}
- xF^a_{\rho\mu}F_a^{\rho\nu}\D_\nu z^i\D^\mu\z^{\m}K_{i\m}\nonumber \\ & &
+ {x^2\over16}\[\(F^a_{\rho\sigma}F_b^{\rho\sigma}\)^2
+ \(F^a_{\rho\sigma}\tF_b^{\rho\sigma}\)^2\]\Bigg\}.
\eea 

Combining the results of sections C.4--8, and
evaluating $\L_1 - \L_r + \Delta_r\L - \Delta_K\L -
\Delta_x\L - \L_AX^A - \L_A\L_BX^{AB}$, with
$$ -4e^{-K}\A^iA\L_i + {\rm h.c.} =  - \(\L_IX^I\)_0 - 4\sqrt{g}\[x\(\cW + \cbW
\)M_\psi^2 - 4\D M_\psi^2 - xe^{-K}\(\cW\aa A + {\rm h.c.}\)\],$$ 
yields the result given in (4.6--12), where we used gauge invariance to write
\bea 0 &=&\delta_a\(\D_\mu z^i\D^\mu z^j\rho_{ij}\) = 
\D_\mu z^i\D^\mu z^j\[(T_az)^k
\rho_{ijk} + 2\rho_{ik}D_j(T_az)^k - (T_a\z)^{\m}\rho_{\m ij}\] \nonumber \\
&=& \D_\mu z^i\(\D^\mu\[(T_az)^j\rho_{ij}\] + \[\rho_{ik}D_j(T_az)^k - 
(T_a\z)^{\m}\rho_{\m ij}\]\D_\mu z^j - (T_az)^j\rho_{\m ij}\D^\mu\z^{\m}\),
\nonumber \\ 0 &=& \delta_a\(\rho_i\A^i\) = \rho_{ij}(T_az)^j\A^i - 
\rho_i(T_a\z)^{\m}\A^i_{\m} + \D_a\rho_i\A^i, \nonumber \\ 
0 &=& \delta_a\(f_i\f^i\) = 2x\[\rho^i_{\m}(T_a\z)^{\m}f_i -
\rho_{ij}(T_az)^j\f^i\] \nonumber \\ 
0 &=& \delta_a\(\rho_i\hV^i\) = \rho_{ij}(T_az)^j\hV^i - {e^{-K}\over x}
\(a_j\A - a_{ij}\A^i + A_{ij}\aa^i- x\rho_{\m ij}A^{\m}\A^i\)(T^az)^j 
\nonumber \\ & & - {e^{-K}\over x}\[\D_a(a\aa -A\aa - a\A) + 
2x\rho_{\m}^i(T_a\z)^{\m}A_i(\aa-\A)\]. \eea
We also used the following identities, that hold up to a total derivative:
\bea 0 &=& \D_\mu z^i\D^\mu\(\D_a\rho_{ij}(T^az)^j\) -\rho_{ij}\D_a(T^az)^j
\[g^{-{1\over2}}\L^i + \hV^i + {1\over x}\D_b(T^bz)^i + {1\over 2}\f^i\cbW\], 
\nonumber \\ 0 &=& - {\pp^\mu y\over x}K_{i\m}
\[(T_az)^i\D^\nu\z^{\m} + (T_a\z)^{\m}\D^\nu z^i\]F^a_{\mu\nu}\nonumber \\ & & 
+ \D_a{\pp^\mu y\over x^2}\[{1\over\sqrt{g}}\L_\mu^a 
+ 2\pp^\nu xF^a_{\mu\nu} - iK_{i\m}\(\D_\mu\z^{\m}(T^az)^i - 
\D_\mu z^i(T^a\z)^{\m}\)\] , \nonumber \\ 
0 &=& -{\pp^\mu x\over x}K_{i\m}
\[(T_az)^i\D^\nu\z^{\m} + (T_a\z)^{\m}\D^\nu z^i\]\tF^a_{\mu\nu}, \nonumber \\ 
0 &=& - {\pp_\mu x\over x^2}\D^aK_{i\m}
\[(T_az)^i\D^\mu\z^{\m} + (T_a\z)^{\m}\D^\mu z^i\]  \nonumber \\ & & + 
\D\Bigg[{1\over x^2}\(\pp_\mu x\pp^\mu x + \pp_\mu y\pp^\mu y\) -
4\(M_\psi^2 - M_\lambda^2\) - 2\(\hV + e^{-K}a\aa\)\quad  \nonumber \\ & & 
+ \({\f^i\L_i\over2x\sqrt{g}} + e^{-K}a_i\A^i + x\rho_{ij}\D_\mu z^i\D^\mu
z^j + x^3\rho^i\rho_i\cW + {\rm h.c.}\)\Bigg], \nonumber \\
0 &=& - 2x^2\rho^i\rho_i\D_\nu A^\nu + \(\f^i\rho_{ij}\D_\nu z^j + {\rm h.c.}\)
A^\nu,\eea
where $ - \D_\nu A^\nu$ is given by the right hand sides of the second and third
equations in (C.76), with $ A_\nu = (\pp^\mu y/x)\D_aF^a_{\mu\nu},\;
(\pp^\mu x/x)\D_a\tF^a_{\mu\nu},$ respectively.

\end{document}